\documentclass[oldversion]{aa} % double column 
\usepackage{graphicx}
\usepackage{aalongtable}
\usepackage{txfonts}
\usepackage{rotating}
\usepackage{lscape}
\usepackage{placeins}
\newcommand{\gsim}{\;\lower.6ex\hbox{$\sim$}\kern-7.75pt\raise.65ex\hbox{$>$}\;}
\newcommand{\lsim}{\;\lower.6ex\hbox{$\sim$}\kern-7.75pt\raise.65ex\hbox{$<$}\;}

\begin{document}
\title{Homogeneous abundances of Mg, Si, Ca, and Ti for about 1500 red giants 
in 16 globular clusters from FLAMES spectra\thanks{Based on observations  collected at ESO telescopes under
programmes 072.D-507 and 073.D-0211}
\thanks{Abundances for individual stars are only available at the CDS
via anonymous ftp to cdsarc.u-strasbg.fr (130.79.128.5) or via http://cdsarc.
u-strasbg.fr/viz-bin/cat/J/A+A/??/??}
 }

\author{
Eugenio Carretta\inst{1}
}

\authorrunning{E. Carretta}
\titlerunning{Abundances of alpha-elements}

\offprints{E. Carretta, eugenio.carretta@inaf.it}

\institute{
INAF-Osservatorio di Astrofisica e Scienza dello Spazio di Bologna, Via P. Gobetti
 93/3, I-40129 Bologna, Italy}

\date{}

\abstract{The FLAMES survey ``Na-O anti-correlation and HB" uncovered the modern
standard for globular clusters (GCs), that is their ubiquitous multiple stellar
populations (MPs) distinct by the abundance of proton-capture elements.  That
survey can still be mined to extract a wealth of data. We derive new abundances
of Mg, Si, Ca, and Ti for 948, 954, 1542, and 1350 red giant branch stars in 16
GCs, both formed in situ or accreted in the Milky Way. The program GCs cover the
metallicity range from [Fe/H]=-2.35 dex to [Fe/H]=-0.74 dex. Both the halo and
disc GCs show a clear overabundance of $\alpha-$elements with the modulation in
Mg and Si due to the MPs phenomenon in different clusters. We found star to star
variations in Si abundance correlated to changes in Na  in more than half of our
sample, implying that temperatures in excess of about  65 MK were achieved in
the polluters responsible for the enrichment. We confirm with an enlarged sample
the previous result that significant variations in Mg are observed in GCs that
are metal-poor, massive or both. Evidence of excess of Ca with respect to
reference unpolluted field stars are found in NGC~6752 and NGC~7078, indicating
the action of proton-capture reactions at very high temperature regime in these
GCs. These excesses fit very well in a previously found relation as a function
of a combination of cluster mass and metallicity shown by other typical
signatures of MPs. At odds with previous results based on the Si abundance from
APOGEE, we found that the average abundance of $\alpha-$elements is not an
efficient discriminating factor between in situ and accreted GCs. }

\maketitle

\section{Introduction}

Globular clusters (GCs) represent a major component of old stellar populations
in galaxies. Being bright they are accessible even at large distances, providing
a useful link between the local population in the Milky Way and external
galaxies, including the high-redshift objects presently being studied with new
generation space telescopes (e.g. Adamo et al. 2024).
For some decades, GCs were considered as the first stellar aggregates originated
in galaxies and were thought as the best specimen of simple stellar populations
(coeval groups of stars with homogeneous initial chemical composition). In
relatively recent times both these statements underwent severe revisions. Since
the seminal paper by Searle and Zinn (1978) it was recognised that only a part
of GCs originated in the main body (bulge or disc) of the proto-galaxy, within a
dissipational collapse as in the scenario  devised by Eggen et al. (1962). About
half of the GCs population formed in individual fragments, later accreted in the
Galaxy. It is well known that the luminosity function of the outer halo GCs is
similar to that of GCs in dwarf spheroidal galaxies (e.g. Carretta et al. 2010a).

Second, the pioneering efforts of the Lick-Texas group (Kraft 1994, Sneden 2000)
to study the detailed chemistry of GCs was given a boost thanks the capability
offered by the FLAMES multi-object spectrograph at the ESO VLT. Our project
``Na-O anti-correlation and HB" (Carretta et al. 2006,2009a,b,c and further 
individual extensions) collected a large and homogeneous
database of abundances in GCs, unambiguously showing that these systems are
composed by multiple stellar populations (MPs) differing by their content of
light elements (mainly from He to Si, with extreme cases touching also heavier
elements like K, Ca, Sc). The MPs are the intrinsic nature of GCs, so that a
general definition of a genuine GC is that of a cluster where a Na-O
anti-correlation is observed (Carretta et al. 2010a).

Although the nucleosynthesis of the observed anti-correlations (C-N, Na-O, Mg-Al,
Mg-Si) and correlations (Na-Al, Mg-O, Si-Al, Si-Na) can be reconducted to an
unique mechanism, proton-capture reactions in H-burning at high temperature
(Langer et al. 1993, see also the review by Gratton et al. 2004, 2012, 2019),
the exact stellar site of this mechanism is still debated (e.g. Bastian and
Lardo 2018). An unambiguous quantitative agreement between observations and
different models and proposed scenarios for the self-enrichment of GCs in light
elements is still lacking.

It is then crucial to gather precise and detailed abundance data concerning as
many elements as possible. Proton-capture reactions produce and destroy different
species at different temperature thresholds. With an homogeneous dataset in
various GCs of different global parameters the aim of a self-consistent scenario
for the formation of GCs may hopefully be reached.

The study by Carretta et al. (2009a) was focused on Na and O abundances for more
than 1400 giants in 15 GCs derived from high-resolution GIRAFFE spectra, whereas
in Carretta et al. (2009b) we used UVES spectra to provide information also
about Mg, Al, Si but only for 215 stars targeted with the UVES Red Arm fibers.
In the present paper we extend the study of $\alpha-$elements Mg, Si, Ca, and Ti
to a larger sample, in principle all the stars with a Na or O abundance measured
in GIRAFFE and UVES spectra. Oxygen is the most abundant $\alpha-$element, being
about 10 times more abundant than Mg, Si, Ca, and Ti put together. However, in
almost all the GCs all these elements are enhanced above the solar level and
thus they represent an important ingredient of the cluster chemistry.
The level of $\alpha-$elements in a stellar system is a probe of how the system
was able to produce and retain the ejecta of massive stars before the onset of
strong contribution of iron from type Ia supernovae (SNe). 

Star to star variations in Mg content among MPs are tracers of proton-capture
reactions at temperatures higher than those involved in the Na-O anti-correlation
(e.g. Gratton et al. 2019), and are observed only in massive and/or metal-poor
GCs (Carretta et al. 2009b, see also M\'esz\'aros et al. 2020). Small variations
in Si are detected, as expected due to its large primordial abundance in GCs and
to the fact that the changes from the level provided by SNe is due to a
``leakage" from the Mg-Al cycle on $^{28}$Si (see Yong et al. 2005).
Significant variations in Si occur only when the reaction 
$^{27}$Al($p,\gamma$)$^{28}$Si takes over the reaction
$^{27}$Al($p,\alpha$)$^{24}$Mg  in the Mg-Al chain when the temperature exceeds $T_6 \sim 65$ K
(see Fig. 8 in Arnould et al. 1999).
As the temperature of putative polluters increases, the addition of protons may
overcome higher Coulomb barriers and abundance variations in heavier elements
may be observable, like K (Cohen and Kirby 2012 and Mucciarelli et al. 2012 in
NGC~2419; Carretta 2021 in NGC~4833; Carretta 2022 in NGC~6715). At very high
temperature regimes, significant excesses of Ca may be revealed (Carretta and
Bragaglia 2021).

Finally $\alpha-$elements (or a subset of them) were used to try to detect
signatures discriminating the accreted or in situ origin of GCs. However,
contrasting results were reached (see e.g. Horta et al. 2020 and Carretta and
Bragaglia 2023).

In the present paper, we analyse the GIRAFFE and UVES spectra of red giant
branch (RGB) stars in 16 GCs. The sample is the one analysed in Carretta et al.
(2009a), except  NGC~2808, NGC~6388, and NGC~6441, studied elsewhere. Part of
the abundances for Ca, Mg, and Si (those derived from UVES spectra, hereinafter
the UVES stars) were already given in Carretta et al.  (2010b) and Carretta et
al. (2009b). In a few cases, abundances of Mg and Si were extracted also from
GIRAFFE spectra of a large number of stars (hereinafter the GIRAFFE stars), like
in NGC~104 and NGC~6121 (Carretta et al. 2013a) and NGC~6752 (Carretta et al.
2012). These data are employed in the present paper, whenever useful. In the
present work we complete the  homogeneous derivation of abundances of Mg, Si,
and Ca for the remaining GIRAFFE stars, and of Ti for both GIRAFFE and UVES
stars. Since there are no significant offsets between abundances derived from
UVES and GIRAFFE in our analysis, we can safely merge the two samples in each
GC, so that we obtain the abundances of $\alpha-$elements in more than 1500 RGB
stars in 16 GCs. The list of studied GCs is given in Table~\ref{t:tableA}.
Adding previous studies of individual GCs obtained with the same setups and
identical methods of analysis, we gathered a total of about 2600 red giants in 
26 GCs with strictly homogeneous abundances of $\alpha-$elements from high
resolution optical spectra.

\begin{figure}[t]
\centering
\includegraphics[scale=0.40]{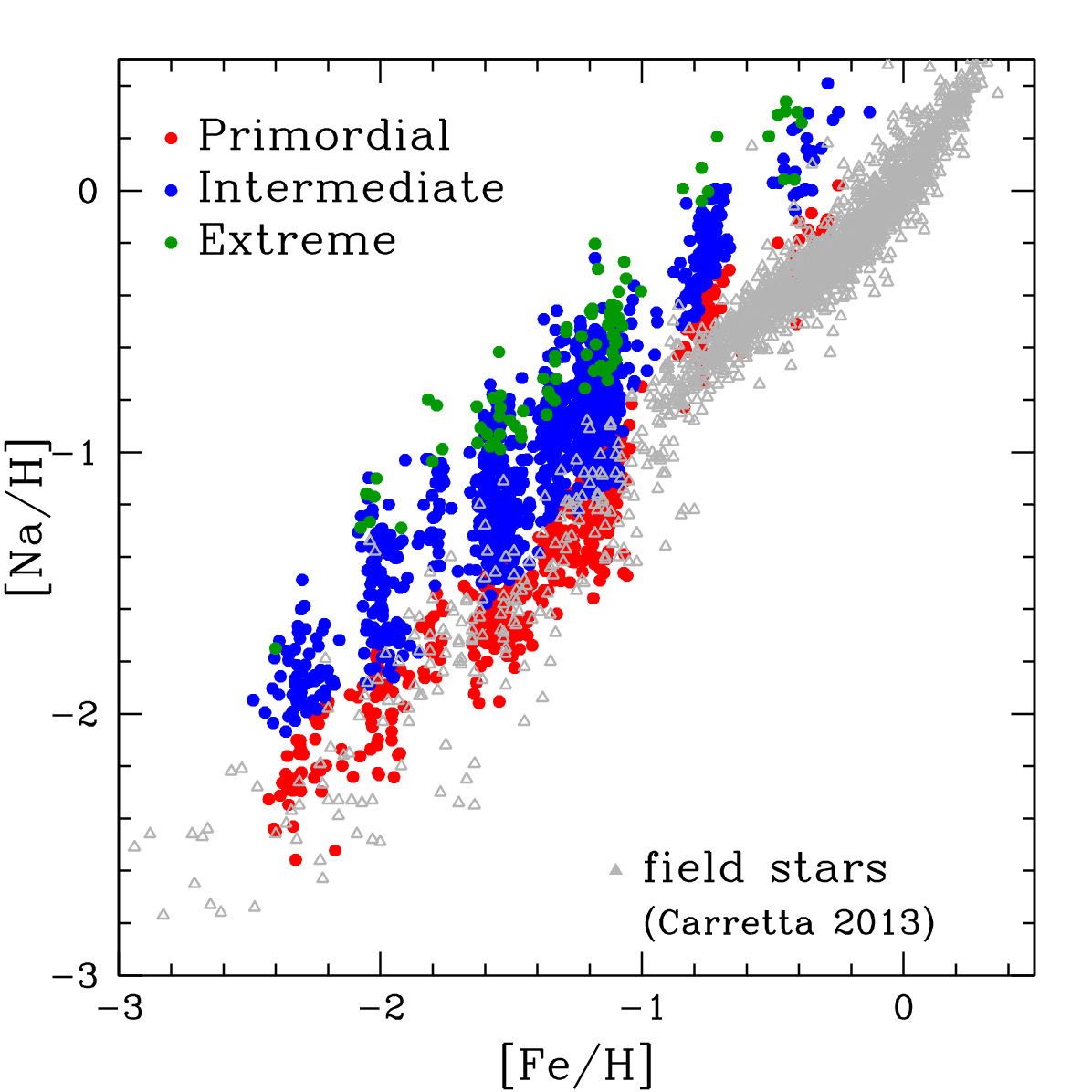}
\caption{Fractions of stars with primordial (P, red circles), intermediate
(I, blue circles), and extreme (E, green circles) composition in our program
GCs, superimposed to a reference sample of unpolluted field stars from Carretta
(2013: grey open triangles).}
\label{f:nahtot}
\end{figure}

In all GCs for the MPs we adopted the classification given in Carretta et al.
(2009a). The primordial (unpolluted) component P includes stars with 
[Na/Fe]\footnote{We adopt the usual spectroscopic notation, $i.e.$ 
[X]= log(X)$_{\rm star} -$ log(X)$_\odot$ for any abundance quantity X, and 
log $\epsilon$(X) = log (N$_{\rm X}$/N$_{\rm H}$) + 12.0 for absolute number
density abundances.}
between the minimum [Na/Fe] observed in each GC and 
[Na/Fe]$_{\rm min}+4\sigma$. The minimum sodium abundance is determined on the
Na-O anti-correlation in each GC.
These GCs stars show the composition of field stars of similar metallicity,
inherited essentially from SNe nucleosynthesis. The polluted stars are separated
into fractions with intermediate (I) and extreme (E) composition according to
their position along the Na-O anti-correlation ([O/Na]$>-0.9$ and [O/Na]$<-0.9$
dex respectively). The P stars in GCs well match  the reference sample of
unpolluted field stars of similar metallicity (see Fig.~\ref{f:nahtot}, adapted
from Carretta 2013).

The paper is organised as follows: the abundance analysis and the comparison
with other spectroscopic surveys are in Section 2, whereas the global pattern of
$\alpha-$elements is discussed in Section 3. The variations in $\alpha-$elements
in MPs are described in Section 4, and a summary of the results from the
present survey is given in Section 5.

\section{Analysis}

The spectroscopic material is described in Carretta et al. (2009a,b). Briefly,
for GIRAFFE spectra we used the two high-resolution setups HR11 (spectral
resolution $R=24,200$, centred on 5728~\AA) and HR13 ($R=22,500$, centred at
6273~\AA). The UVES Red Arm fed by FLAMES fibres (up to a maximum of 7 per
fibre positioning configuration) cover a spectral range from 4800 to 6800~\AA\
with a resolution $R\simeq 45,000$.
All targets were selected among isolated stars near the RGB ridge line in each
GC.

We adopted the atmospheric parameters derived in Carretta et al. (2009a,b) 
and the iron ratios [Fe/H] from Carretta et al. (2009c) for
all the analysed stars. Line lists, atomic parameters, and solar reference 
abundances (from Gratton et al. 2003) are strictly homogeneous with all the
previous studies from our group.

Abundances of Mg, Si, Ca, and Ti were derived  from equivalent widths (EWs)
measured with the package ROSA (Gratton 1988) as described in detail in
Bragaglia et al. (2001). Following the procedure adopted in Carretta et al.
(2009a) stars observed in each GC with both spectrographs were used to register
the EWs from GIRAFFE to those measured on the higher resolution UVES spectra.

For titanium, neutral transitions are available both in UVES and GIRAFFE spectra.
In addition, also singly ionised transitions could be measured for the few UVES
stars in each GC of the sample.
We found on average
[Ti/Fe]~{\sc i} - [Ti/Fe]~{\sc ii} $=-0.008\pm 0.004$ dex ($\sigma=0.050$ 
dex, 188 stars). The excellent match between [Ti/Fe]~{\sc i} and [Ti/Fe]~{\sc ii}
is a safety check on the reliability of the analysis and on the adopted scale of
atmospheric parameters, in particular the surface gravity. If not stated
otherwise, in the following for titanium we use the average abundance of
[Ti/Fe]~{\sc i}, derived from much larger samples of stars in each GC.

A second critical check consists in looking for trends of the abundances as a
function of temperature. Several signatures of MPs in GCs involve correlations
and anti-correlations among different species produced and destroyed in
proton-capture reactions. Hence, it is crucial to verify that in the data there
are no spurious trends due to some temperature-dependent errors from
the abundance analysis.

In Fig.~\ref{f:alphateff} we plot the abundances of Mg, Si, Ca,  [Ti/Fe]~{\sc
i}, and [Ti/Fe]~{\sc ii} for all stars analysed in the 16 program GCs as a
function of the effective temperature (T$_{\rm eff}$). The vertical scale is set
by the specie with the largest dispersion observed (Mg in NGC~7078). Blue
squares are the UVES stars and red circles are for GIRAFFE stars. 
For each abundance ratio in each GC we computed the linear regression as a 
function of temperature. Out of 80 cases, the relation is statistically
significant only for Mg and Si for NGC~6752 (from Carretta et al. 2012), and
for Si in NGC~6171 and NGC~6809 from the present work. These trends may help to
explain the larger offset in the abundances from UVES and GIRAFFE in NGC~6752
(see Table~\ref{t:tableA}). However, the average ratios in different temperature bins 
differ by only a few hundredths of dex even in these cases (with a maximum of
0.07 and 0.08 dex in NGC~6752), hence do not affect significantly our following
results.
Since there
are no significant gradients with temperature, this is a rather robust safety check on
our analysis. Since there are no significant offsets, we are entitled to safely
merge the UVES and GIRAFFE samples. For stars studied in both samples we however
privileged  the UVES abundances, which generally are based on a larger number of
lines.

The average results are listed in Tab.~\ref{t:tableA}. In the first line for
each  element in each GC we report the average abundance and the rms scatter
from the UVES sample, together with the cluster metallicity from Carretta et al.
(2009c). The second line is for the abundances from GIRAFFE, and in the third
line we  list the results of the merging of the two previous set, that is the
unique stars in each GC. In the present paper we provide a measurement of some
$\alpha-$elements (not previously published) for slightly more than 1500 RGB 
stars. 

\begin{table*}
\centering
\caption{Average abundances of $\alpha-$elements}
\begin{tabular}{lrrrrrrl}
\hline

GC   &    [Mg/Fe]      &   [Si/Fe]       &   [Ca/Fe]       & [Ti/Fe]~{\sc i}& [Ti/Fe]~{\sc ii} & [Fe/H] &ref.\\
     & n~~~ mean~ rms  & n~~~ mean~ rms  & n~~~ mean~ rms  & n~~~ mean~ rms & n~~~ mean~ rms   & mean~ rms   \\
\hline
0104 & 11 +0.521 0.030 & 11 +0.399 0.022 & 11 +0.315 0.013 & 11 +0.398 0.007 & 11 +0.379 0.015 &-0.768 0.054 &1,1,2,3,3,6 \\
     &147 +0.532 0.079 &147 +0.440 0.065 &147 +0.316 0.025 &146 +0.395 0.039 &                 & &4,4,3,3   \\
     &147 +0.530 0.076 &147 +0.433 0.061 &147 +0.315 0.024 &146 +0.394 0.036 &                 & &3,3,3,3   \\
\hline     
0288 & 10 +0.455 0.029 & 10 +0.372 0.030 & 10 +0.409 0.023 & 10 +0.271 0.034 & 10 +0.282 0.038 &-1.305 0.054 &1,1,2,3,3,6 \\
     &104 +0.472 0.046 &109 +0.390 0.029 &108 +0.402 0.046 &103 +0.285 0.041 &                 & &3,3,3,3   \\
     &107 +0.469 0.045 &112 +0.389 0.029 &111 +0.402 0.045 &106 +0.283 0.040 &                 & &3,3,3,3   \\
\hline 
1904 & 10 +0.279 0.061 & 10 +0.293 0.032 & 10 +0.278 0.007 & 10 +0.123 0.038 & 10 +0.146 0.025 &-1.579 0.033 &1,1,2,3,3,6 \\
     & 47 +0.271 0.035 & 58 +0.306 0.037 & 57 +0.276 0.040 & 56 +0.133 0.043 &                 & &3,3,3,3   \\
     & 57 +0.272 0.040 & 68 +0.304 0.036 & 67 +0.276 0.037 & 66 +0.131 0.042 &                 & &3,3,3,3   \\
\hline 
3201 & 13 +0.343 0.040 & 13 +0.298 0.046 & 13 +0.299 0.028 & 13 +0.086 0.038 & 13 +0.094 0.038 &-1.512 0.065 &1,1,2,3,3,6 \\
     &130 +0.339 0.051 &145 +0.301 0.038 &149 +0.303 0.043 &117 +0.082 0.035 &                 & &3,3,3,3   \\
     &131 +0.341 0.049 &146 +0.299 0.038 &150 +0.306 0.039 &118 +0.082 0.037 &                 & &3,3,3,3   \\
\hline      
4590 & 13 +0.350 0.057 & 13 +0.400 0.054 & 13 +0.263 0.035 & 10 +0.150 0.057 & 13 +0.148 0.040 &-2.265 0.047 &1,1,2,3,3,6 \\
     & 54 +0.372 0.070 & 47 +0.425 0.058 &117 +0.282 0.042 & 13 +0.168 0.032 &                 & &3,3,3,3   \\
     & 62 +0.364 0.066 & 54 +0.423 0.057 &120 +0.280 0.042 & 18 +0.162 0.048 &                 & &3,3,3,3   \\
\hline      
5904 & 14 +0.409 0.072 & 14 +0.304 0.049 & 14 +0.376 0.020 & 14 +0.174 0.031 & 14 +0.192 0.024 &-1.340 0.052 &1,1,2,3,3,6 \\
     &133 +0.422 0.049 &135 +0.321 0.031 &136 +0.368 0.034 &131 +0.181 0.027 &                 & &3,3,3,3   \\
     &135 +0.422 0.052 &137 +0.320 0.034 &138 +0.371 0.028 &133 +0.181 0.027 &                 & &3,3,3,3   \\
\hline      
6121 & 14 +0.552 0.030 & 14 +0.524 0.065 & 14 +0.415 0.033 & 14 +0.270 0.039 & 14 +0.237 0.038 &-1.168 0.046 &1,1,2,3,3,6 \\
     &103 +0.541 0.048 &103 +0.549 0.034 &103 +0.416 0.034 &103 +0.284 0.032 &                 & &4,4,3,3   \\
     &104 +0.545 0.045 &104 +0.538 0.039 &104 +0.415 0.035 &104 +0.281 0.033 &                 & &3,3,3,3   \\
\hline      
6171 &  5 +0.514 0.040 &  5 +0.535 0.080 &  5 +0.405 0.024 &  5 +0.164 0.031 &  5 +0.200 0.044 &-1.033 0.064 &1,1,2,3,3,6 \\
     & 33 +0.520 0.048 & 33 +0.522 0.042 & 33 +0.415 0.037 & 33 +0.178 0.039 &                 & &3,3,3,3   \\
     & 33 +0.522 0.044 & 33 +0.524 0.049 &108 +0.415 0.032 & 33 +0.174 0.038 &                 & &3,3,3,3   \\
\hline      
6218 & 11 +0.524 0.039 & 11 +0.350 0.060 & 11 +0.424 0.027 & 11 +0.249 0.021 & 11 +0.216 0.019 &-1.330 0.042 &1,1,2,3,3,6 \\
     & 79 +0.538 0.038 & 79 +0.359 0.042 & 81 +0.418 0.041 & 78 +0.250 0.026 &                 & &3,3,3,3   \\
     & 81 +0.537 0.037 & 81 +0.356 0.044 &108 +0.420 0.038 & 80 +0.251 0.024 &                 & &3,3,3,3   \\
 \hline     
6254 & 14 +0.487 0.043 & 14 +0.284 0.048 & 14 +0.341 0.038 & 14 +0.164 0.059 & 14 +0.183 0.030 &-1.575 0.059 &1,1,2,3,3,6 \\
     &122 +0.480 0.067 &137 +0.318 0.056 &144 +0.341 0.037 &123 +0.178 0.041 &                 & &3,3,3,3   \\
     &129 +0.481 0.065 &144 +0.314 0.057 &152 +0.342 0.037 &130 +0.176 0.043 &                 & &3,3,3,3   \\
\hline      
6397 & 13 +0.458 0.037 & 13 +0.337 0.047 & 13 +0.281 0.030 & 13 +0.173 0.030 & 13 +0.179 0.028 &-1.988 0.044 &1,1,2,3,3,6 \\
     & 88 +0.455 0.041 & 40 +0.335 0.034 &140 +0.292 0.036 & 27 +0.180 0.040 &                 & &3,3,3,3   \\
     & 97 +0.454 0.041 & 49 +0.337 0.037 &147 +0.290 0.035 & 38 +0.177 0.037 &                 & &3,3,3,3   \\
\hline      
6752 & 14 +0.501 0.048 & 14 +0.377 0.060 & 14 +0.396 0.032 & 14 +0.192 0.022 & 14 +0.207 0.013 &-1.555 0.051 &1,1,2,3,3,6 \\
     &119 +0.371 0.108 &131 +0.487 0.051 &122 +0.397 0.034 &128 +0.181 0.040 &                 & &5,5,3,3   \\
     &125 +0.387 0.107 &137 +0.476 0.062 &130 +0.397 0.033 &134 +0.181 0.039 &                 & &3,3,3,3   \\
\hline      
6809 & 14 +0.469 0.098 & 14 +0.376 0.057 & 14 +0.356 0.035 & 13 +0.130 0.060 & 14 +0.211 0.072 &-1.934 0.063 &1,1,2,3,3,6 \\
     &124 +0.487 0.061 &145 +0.383 0.045 &151 +0.363 0.044 &131 +0.150 0.056 &                 & &3,3,3,3   \\
     &129 +0.483 0.061 &146 +0.380 0.044 &151 +0.363 0.042 &133 +0.147 0.056 &                 & &3,3,3,3   \\
\hline      
6838 & 12 +0.493 0.038 & 12 +0.381 0.059 & 12 +0.307 0.056 & 12 +0.384 0.068 & 12 +0.342 0.023 &-0.832 0.061 &1,1,2,3,3,6 \\
     & 39 +0.496 0.044 & 39 +0.395 0.049 & 39 +0.314 0.049 & 39 +0.360 0.070 &                 & &3,3,3,3   \\
     & 51 +0.496 0.043 & 51 +0.391 0.051 & 51 +0.312 0.050 & 51 +0.366 0.070 &                 & &3,3,3,3   \\
\hline      
7078 & 13 +0.446 0.191 & 10 +0.429 0.104 & 13 +0.250 0.053 & 13 +0.227 0.034 & 13 +0.262 0.039 &-2.320 0.057 &1,1,2,3,3,6 \\
     & 51 +0.435 0.164 & 59 +0.488 0.088 & 82 +0.279 0.054 & 28 +0.217 0.042 &                 & &3,3,3,3   \\
     & 55 +0.420 0.175 & 59 +0.472 0.093 & 82 +0.273 0.054 & 36 +0.219 0.040 &                 & &3,3,3,3   \\
\hline      
7099 & 10 +0.508 0.036 &  9 +0.342 0.069 & 10 +0.309 0.035 & 10 +0.237 0.024 & 10 +0.243 0.020 &-2.344 0.049 &1,1,2,3,3,6 \\
     & 26 +0.505 0.054 & 16 +0.382 0.046 & 62 +0.324 0.053 & 16 +0.241 0.036 &                 & &3,3,3,3   \\
     & 33 +0.505 0.044 & 22 +0.367 0.058 & 68 +0.323 0.051 & 24 +0.240 0.032 &                 & &3,3,3,3   \\
     
\hline
\end{tabular}

\label{t:tableA}
\begin{list}{}{}
\item[1-] Carretta et al. (2009b);[2-] Carretta et al. (2010b); [3-] this work; 
[4-] Carretta et al. (2013a); [5-] Carretta et al. (2012); [6-] Carretta et al. (2009c)
\end{list}

\end{table*}

The abundances of $\alpha-$elements for individual stars are available in
electronic form at CDS. We provide star identification, coordinates, number of lines,
average abundance, line-to-line scatter, and a flag (UVES or GIRAFFE) for each
element.

\subsection{Error budget}

The error budget was estimated as described in detail in Carretta et al. 
(2009b) and Carretta et al. (2009a) for abundances derived from UVES and GIRAFFE
spectra, respectively. 
To obtain the star to star errors, two ingredients are necessary: a fair
estimate of the internal errors in the atmospheric parameters (as well as in
the EW measurement) and the sensitivity, that is the change in the abundance
ensuing a variation in one of the parameters. 
Internal errors are estimated and tabulated in Carretta et al. (2009a) and
Carretta et al. (2009b) for the GIRAFFE and the UVES samples, respectively.

To derive the sensitivity of the derived abundances on the adopted
atmospheric parameters we repeated the abundance analysis by changing only one
parameter (effective temperature, surface gravity, etc) each time for all stars 
in a GC. The sensitivity to each parameters was adopted as the average of all
the stars over the sampled temperature range. This procedure was repeated
separately for the UVES and GIRAFFE samples in each program GC.

The sensitivity of Mg, Si, Ca, [Ti/Fe]~{\sc i}, and [Ti/Fe]~{\sc ii} in
response to the amount of changes in the parameters listed in the table header
are reported in Table~\ref{t:sensAuves} and Table~\ref{t:sensAgira} for
UVES and GIRAFFE, respectively. By multiplying the sensitivity in each element
for the internal errors in the parameters, we obtained the internal errors in
abundances due to internal errors in the adopted parameters.

The resulting errors, the average number of lines for each species, and the
error estimates due to measurement errors in the EWs are listed for all GCs in
Table~\ref{t:errabuTOT}. Summing in quadrature all the
contributions we obtained the estimated internal errors listed in the table.

\subsection{Comparison with other spectroscopic surveys}

We compared our results with other large surveys (Gaia-ESO, APOGEE, GALAH) using
stars in common.

With the Gaia-ESO (GES) survey final data release DR5.1 (Gilmore et al. 2022,
Randich et al. 2022) we have in common 232 stars in six GCs and  68 stars when
considering only the UVES U580 setup, with large spectral coverage, in GES. 
The
mean offset in temperature (this work minus GES) is -65.1 K ($\sigma=160.7$ K,
total sample) or -183.5 K ($\sigma=138.2$ K, UVES). The corresponding offset in
surface gravity are +0.12 dex ($\sigma=0.49$ dex, total) and -0.15 dex
($\sigma=0.36$ dex, UVES). After bringing the GES data to our scale of solar
abundances, the offsets in [Fe/H] are +0.095 dex ($\sigma=0.136$ dex),
decreasing to +0.071 dex ($\sigma=0.105$ dex), when using only the U580
setup in GES.

The comparison for all the analysed elements is shown as a function of the
effective temperature in Fig.~\ref{f:ges} for the stars in common in the six GCs.
All the abundance ratios from GES were shifted to our scale. Filled points for GES
represent the U580 setup. If we restrict to the four GCs with sufficient stars
observed with the U580 setup, our abundances have an offset on average within
$\pm 0.1$ dex from the mean abundances from GES, except for Ca. However, the
star to star scatter is  noticeably larger for all the analysed species in GES
data. On average, in NGC~104, NGC~1904, NGC~6752, and NGC~7078 the rms scatter
of the mean is 0.147 dex, 0.105 dex, 0.133 dex, and 0.189 dex for Mg, Si, Ca,
and Ti in GES to be compared to the averages 0.096 dex, 0.062 dex, 0.037 dex,
and 0.034 dex for the corresponding stars in the present work. By looking at
Fig.~\ref{f:ges} this is probably because of the significant trends as a function
of the temperature present in GES data in several cases, even when only the
more reliable U580 dataset is considered, suggesting problems in the set of
atmospheric parameters or in the procedure used in the GES analysis. 

We have 321 stars in common in nine GCs with the GALAH survey DR4 (Buder et al.
2025). After excluding unreliable stars using the prescripted criteria from the
GALAH web site,\footnote{With these criteria all the 15 stars in common for
NGC~3201 are eliminated and we are left with only eight GCs.} to retain only
stars with reliable spectroscopic analysis and abundances, the mean offsets in
temperature and gravity are -67.3 K ($\sigma=95.3$ K) and -0.054 dex
($\sigma=0.060$ dex) for 158 stars. We used the values in table C.1 by Buder et
al. (2025) to correct the GALAH abundances to our scale. The comparison of the
abundances as a function of the temperature for the stars in common is shown in
Fig.~\ref{f:gal}. Overall, there is a reasonable agreement both in mean value
and scatter. The mean offsets (this work minus GALAH) are -0.016 dex, -0.003
dex, -0.124 dex, and -0.065 dex for Mg, Si, Ca and Ti. The average rms scatter
in GALAH are 0.059 dex, 0.041 dex, 0.100 dex, and 0.101 dex, respectively, to be
compared with 0.056 dex, 0.041 dex, 0.034 dex, and 0.038 dex for stars in 
common with the present work. 

\begin{figure}
\centering
\includegraphics[scale=0.40]{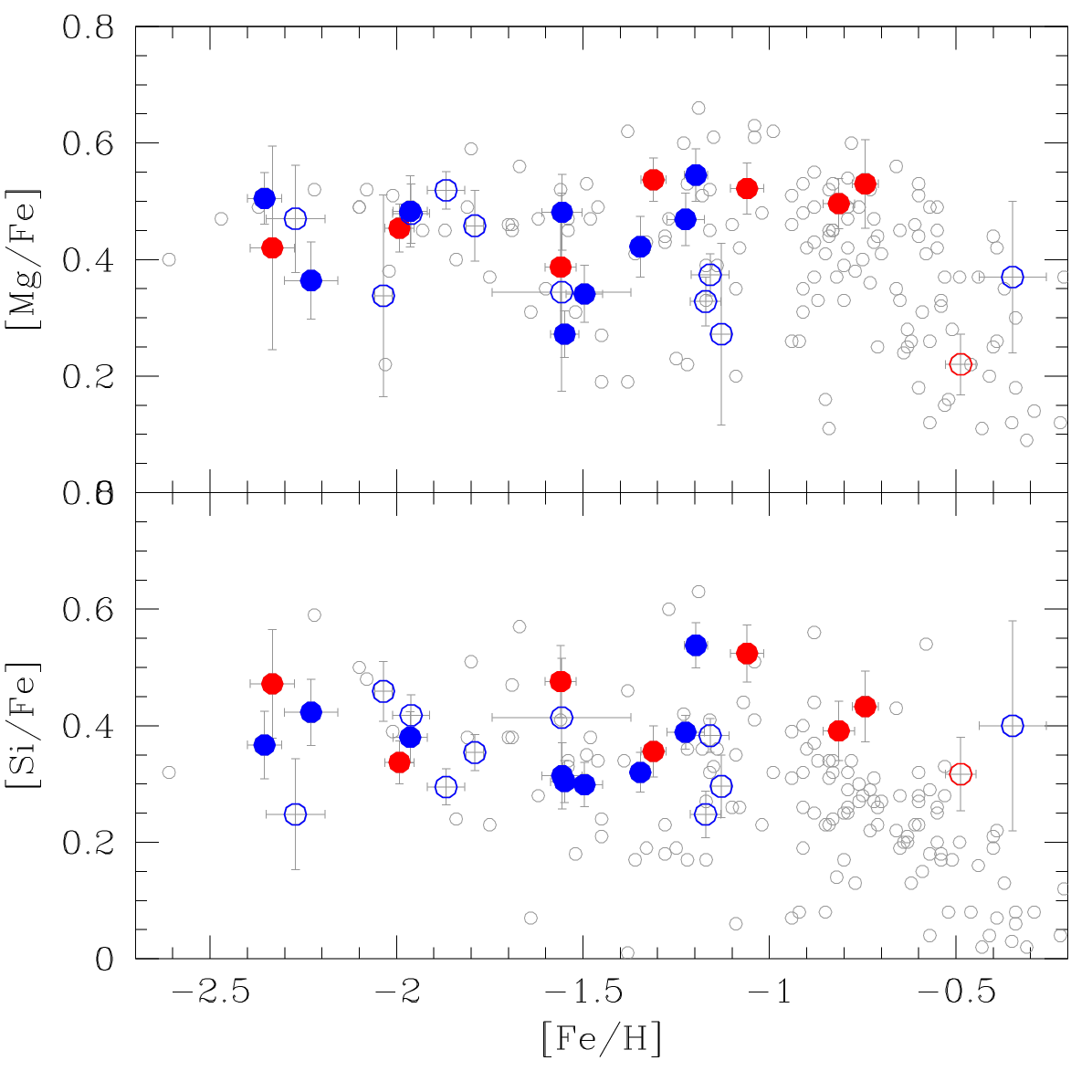}
\includegraphics[scale=0.40]{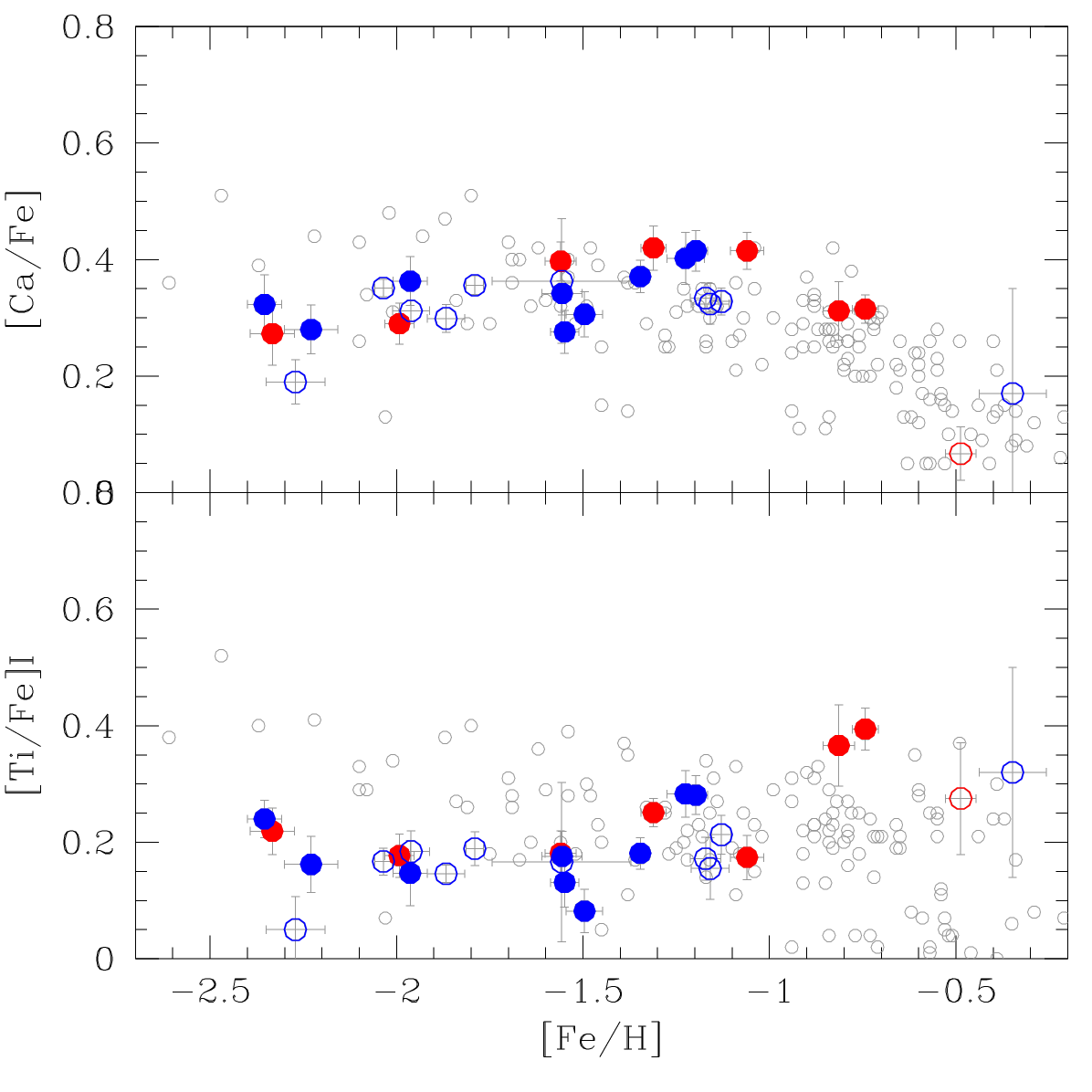}
\caption{Average abundance ratios [Mg/Fe], [Si/Fe], [Ca/Fe], [Ti/Fe]~{\sc i} as a
function of metallicity [Fe/H], with the 1$\sigma$ spread represented as error
bars. Filled symbols are for the 16 GCs of the present work, empty symbols
indicates GCs from previous, individual studies. Red and blue colour are for in
situ and ex situ GCs, respectively. Grey points are field stars from Gratton et
al. (2003).}
\label{f:mediepfield}
\end{figure}

With the recommended selection criteria (STARFLAG=0,ASPCAPFLAG=0) we found 373
stars in common in 15 GCs with APOGEE DR17 (Abdurro'uf et al. 2022). Mean
offsets are -162.2 K ($\sigma=107.0$ K) in temperature, -0.055 dex
($\sigma=0.146$ dex) in gravity, and -0.115 dex ($\sigma=0.103$ dex) in [Fe/H].
The comparison of the abundances, in Fig.~\ref{f:apo}, is done using the
original APOGEE abundances since no solar abundances are explicitly stated for that survey.
Our abundances of $\alpha-$elements from optical spectra are systematically
larger than those derived by APOGEE near-infrared spectra. On average we found 
offsets (this work minus APOGEE) of +0.212 dex ($\sigma=0.081$ dex, 358 stars) for
Mg, +0.157 dex ($\sigma=0.086$ dex, 358 stars) for Si, 
+0.127 dex ($\sigma=0.121$ dex, 358 stars) for Ca, and
+0.172 dex ($\sigma=0.127$ dex, 333 stars) for Ti~{\sc i}. The best agreement is
found for Ti~{\sc ii} (-0.007 dex), although with large scatter ($\sigma=0.171$
dex, 75 stars). The average rms scatter are 0.064 dex, 0.042 dex, 0.121 dex,
0.132 dex, and 0.136 dex for Mg, Si, Ca, Ti~{\sc i}, and Ti~{\sc ii} in APOGEE
to be compared with average values 0.067 dex, 0.043 dex, 0.038 dex, 0.039 dex,
and 0.025 dex in the present work.

\begin{table}
\centering
\caption{GCs from previous studies}
\begin{tabular}{lcl}
\hline

GC       &    [Fe/H]  &   ref.                       \\
\hline
NGC 362  & $-$1.166  & Carretta et al. (2013b)       \\
NGC 1851 & $-$1.185  & Carretta et al. (2011)       \\
NGC 2808 & $-$1.129  & Carretta (2015)              \\
NGC 4833 & $-$2.015  & Carretta et al. (2014a)      \\
NGC 5634 & $-$1.867  & Carretta et al. (2017)       \\
NGC 6093 & $-$1.792  & Carretta et al. (2015)       \\
NGC 6388 & $-$0.480  & Carretta and Bragaglia (2023)\\
NGC 6441 & $-$0.348  & Gratton et al. (2006,2007)   \\
NGC 6535 & $-$1.952  & Bragaglia et al. (2017)      \\
NGC 6715 & $-$1.505  & Carretta et al. (2010c)      \\
Terzan 8 & $-$2.271  & Carretta et al. (2014b)      \\
\hline
     
\hline
\end{tabular}
\label{t:tabprev}
\end{table}

\section{Results: the general pattern}

Average abundances of individual $\alpha-$elements are shown in Fig.~\ref{f:mediepfield}
as a function of the metallicity [Fe/H]. Filled points are the 16 program
GCs from the present paper, whereas empty circles are GCs from previous studies
of individual GCs, summarised in Table~\ref{t:tabprev} together with the mean
metallicity from UVES to locate them in the plots. The methodology of the
abundance analysis being identical, this addition provides an extended sample
which is very homogeneous, improving the statistics.
Red colour is used for in situ GCs and blue colour indicates accreted GCs. For
this classification we follow the associations given in Massari et al. (2019).
For GCs with uncertain progenitors (NGC~3201, NGC~5904, NGC~5634, and NGC~6535)
we adopt the associations for which both Forbes (2020) and Callingham et al.
(2022) are in agreement, that is Sequoia for NGC~3201 and NGC~6535, and Helmi
streams for NGC~5904 and NGC~5634. Empty grey circles are Milky Way field stars
used as a comparison. We adopted the sample from Gratton et al. (2003), which is
a mixture of the dissipational and accreted components in the  Galaxy, because
that study used the same line list and solar abundances as the present work.

In Fig.~\ref{f:mediepfield} we plot the average abundances of
$\alpha-$elements as a function of the metallicity. The ratios [Mg/Fe] and
[Si/Fe] are roughly constant up to the metallicity of our most metal-rich object,
NGC~6441. This pattern is consistent with a constant level of Mg shared
also by disk and bulge GCs with abundances derived from optical spectroscopy in
the literature (e.g. Mu\~{n}oz et al. 2017, 2018, 2020, Mura-Guzm\'{a}n et al.
2018, Puls et al. 2018). For Si, a decrease after the knee at [Fe/H]$\sim -1$
dex is not found in our data, and is apparent only in low mass bulge GCs 
(Mu\~{n}oz et al. 2018, 2020) from optical spectroscopy, whereas in APOGEE data
Si is clearly decreasing (e.g. H20). In the present work, a decrease in the
average abundances is observed only for Ca.
The average abundances of GCs are mostly comparable to those of field stars,
with the exception of Si in NGC~6441.

Moreover, for most elements we found no significant differences between
the distribution of ex situ and that of in situ GCs. Comparing the 19 ex situ
GCs and the 8 in situ GCs with a Kolmogorov-Smirnov (K-S) test we could not
safely reject the null hypothesis that the two distribution are extracted from
the same parent population looking at the average abundances of Mg, Si, Ca, and
Ti (p-value 0.41, 0.27, 0.74, and 0.06, respectively, with the usual 0.05
threshold for significance).

Our findings do not agree with the conclusions by Horta et al. (2020: H20).
Using the average Si abundances from APOGEE DR17 for 46 GCs, they claim that a
distinction exist between in situ and accreted GCs, at least in the metallicity
range from [Fe/H]=-1.5 to -1 dex, with the in situ subgroup displaying a higher
[Si/Fe] average abundance. In our data a K-S test for  [Si/Fe] over the
metallicity -1.5 $<$ [Fe/H] $<$ -1, where according to H20 the difference
between in situ and accreted GCs is most evident, returns a two-tailed
probability p=0.49 .

Their work is the only one, to our knowledge, to provide tabulated average
[Si/Fe] abundances from APOGEE for the analysed GCs, and we have 20 GCs in
common  between our sample (enlarged with the addition of previous homogeneous
studies in Tab.~\ref{t:tabprev}) and H20.

\begin{figure}[t]
\centering
\includegraphics[scale=0.40]{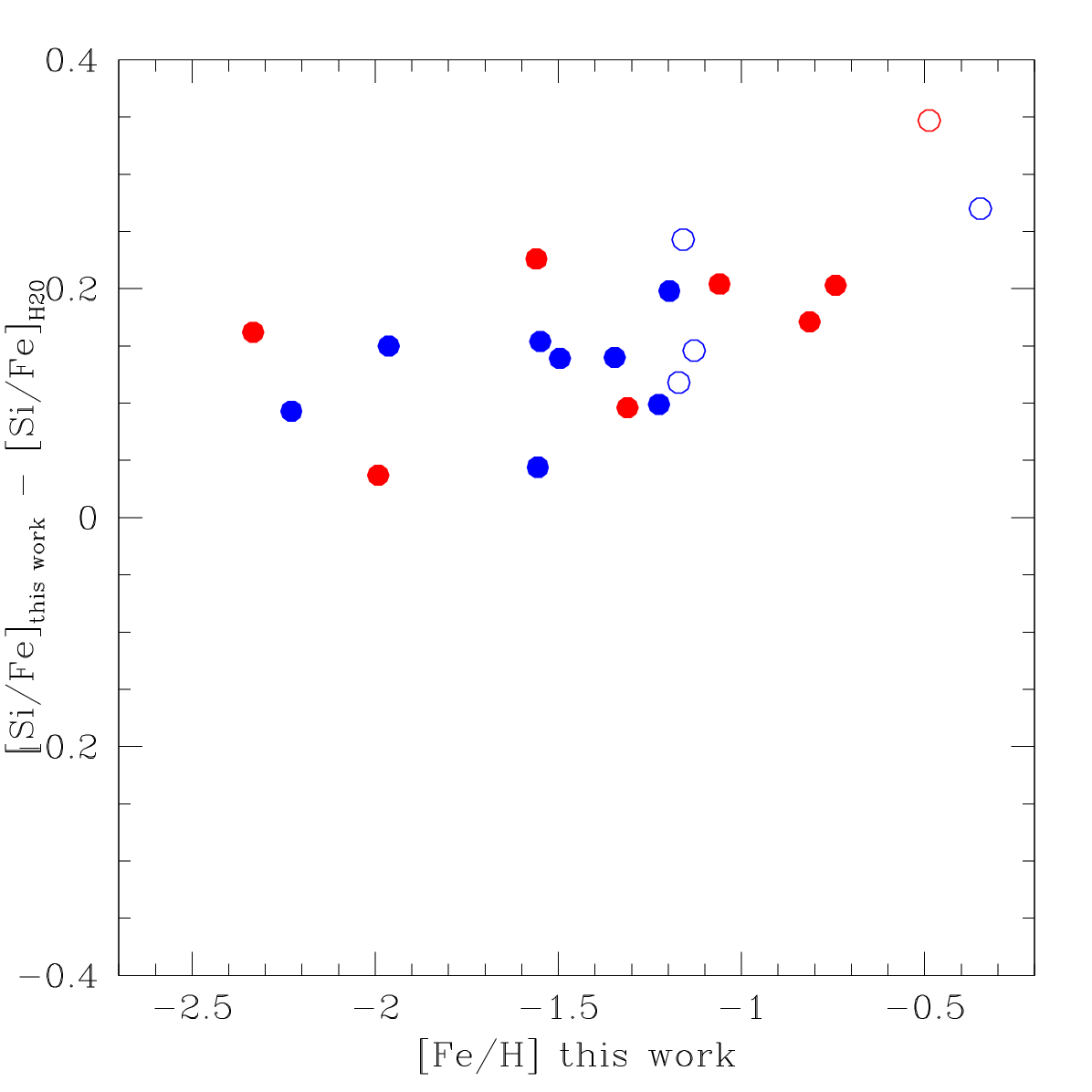}
\caption{Differences of the average [Si/Fe] abundance from this work (extended
sample) and APOGEE (from Horta et al. 2020) for 20 GCs in common. Symbols
are as in Fig.~\ref{f:mediepfield}.}
\label{f:horta}
\end{figure}

The offset in metallicity is scarcely significant. We found on average that
[Fe/H]$_{this work} -$ [Fe/H]$_{APOGEE} = -0.063\pm 0.016$ dex with 
$\sigma=$0.073 dex (20 GCs). The actual difference lies in the Si abundances and
the comparison is shown in  Fig.~\ref{f:horta}.

Computed over the 20 GCs, the abundance [Si/Fe] derived in the present work is
on average higher that those used by H20 by 0.162 dex ($\sigma=$0.075 dex), in
agreement with what found in the section 2.2 . However, more
importantly, this offset is not constant but it varies as a function of
metallicity, with a clear trend to increase as [Fe/H] increases.

Since the GCs formed in situ (either in the main disc or main bulge of the
progenitor Galaxy) are (on average) more metal-rich than those 
formed ex situ,  the discrepancy in the Si abundances is increased when the
majority of GCs is not accreted, at the metallicity regime $\gtrsim -1.2$ dex. 
This trend may explain the different conclusions reached by H20 and the 
present study, as well the large discrepancy between the APOGEE [Si/Fe] average
for NGC~6388 and the abundance given by Carretta and Bragaglia
(2023)\footnote{The in situ origin for NGC~6388 is corroborated also by Fe-peak
elements (Carretta and Bragaglia 2022) in a large number of stars, not only by
the $\alpha-$elements.}. However, internal errors in the [Si/Fe] ratio in
both the present work and in APOGEE are close to the claimed difference from
accreted and in situ GCs (about 0.1 dex). This evidence, coupled to possible
systematic offsets, would require even more precise abundances for a definitive
answer to this issue.

\begin{figure*}
\centering
\includegraphics[scale=0.23]{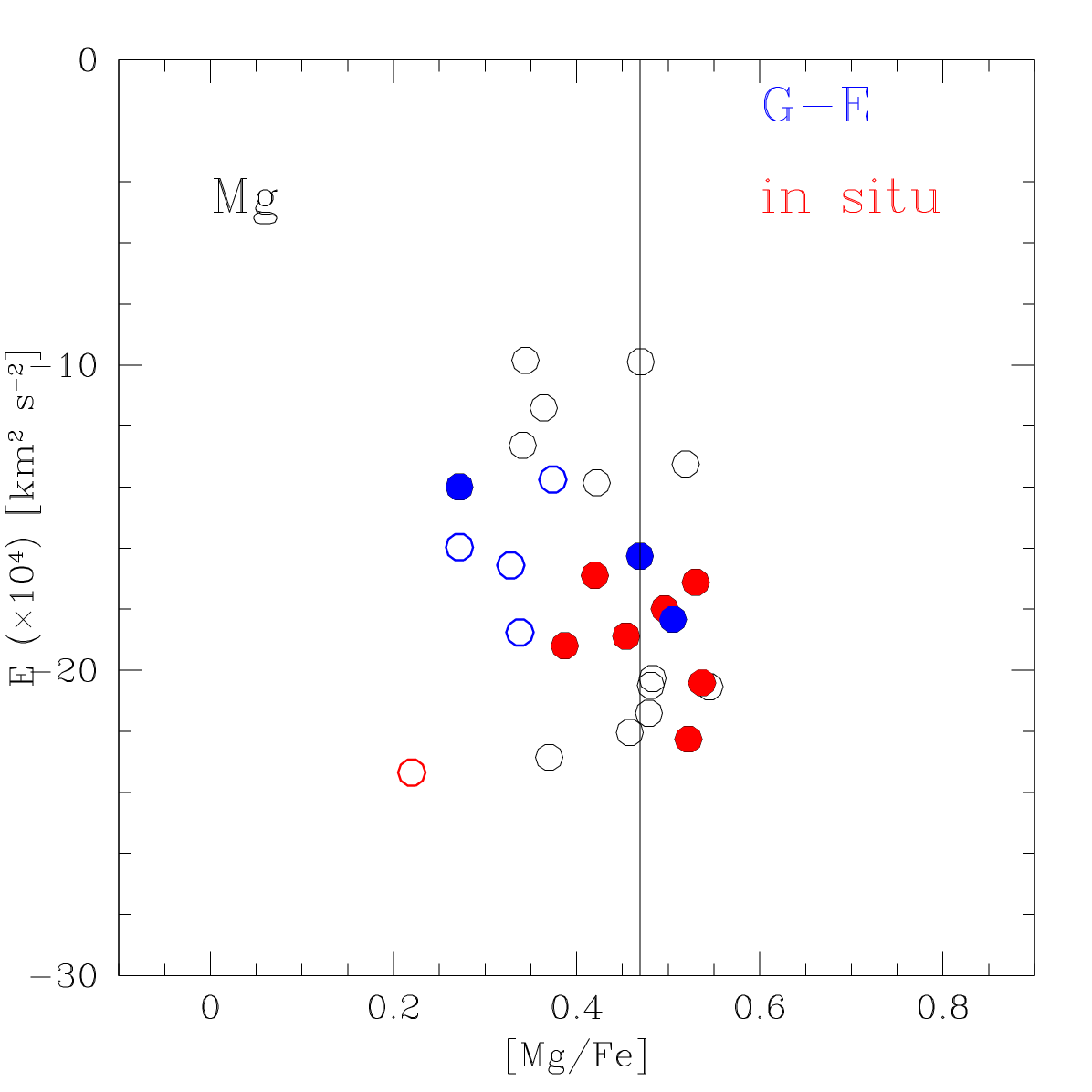}\includegraphics[scale=0.23]{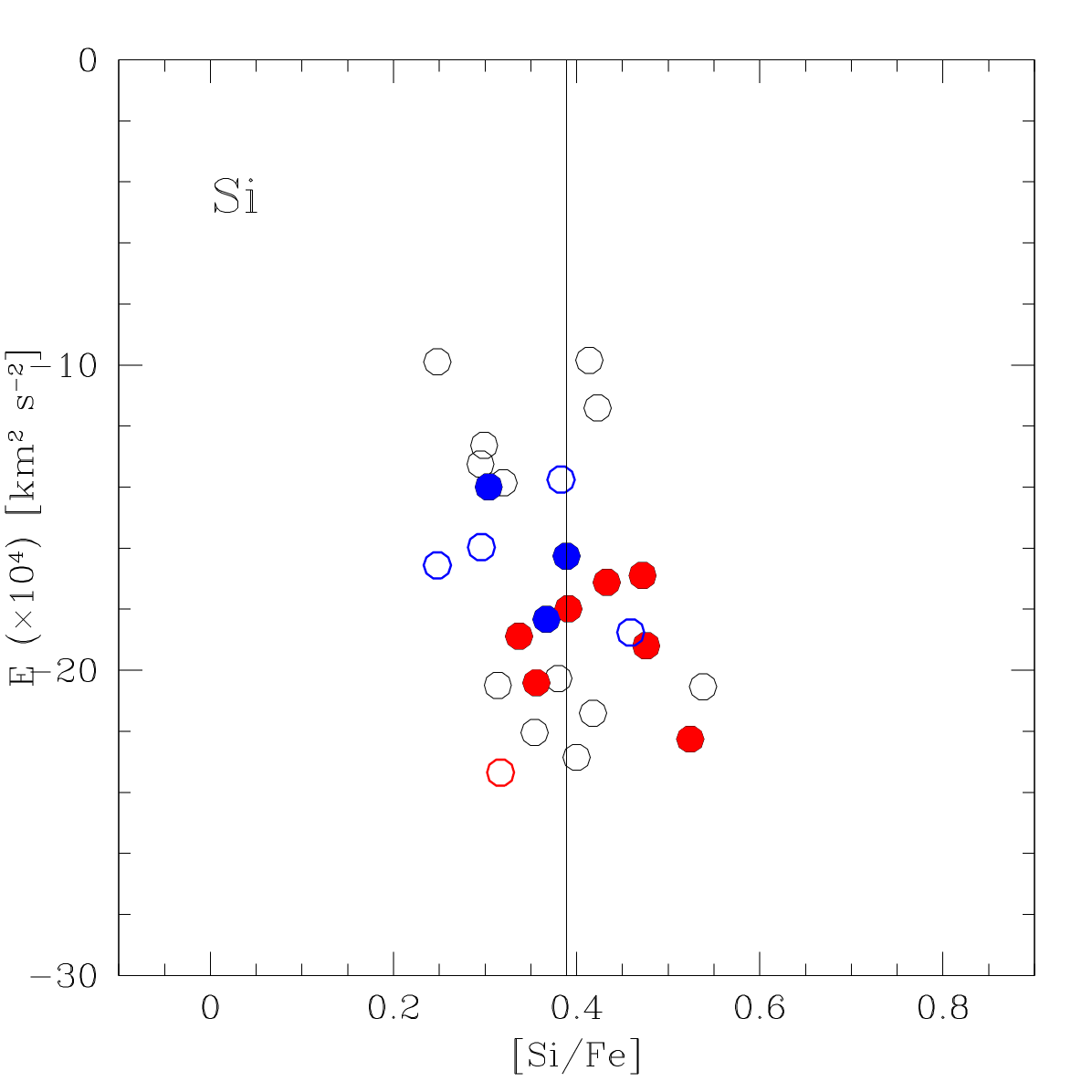}\includegraphics[scale=0.23]{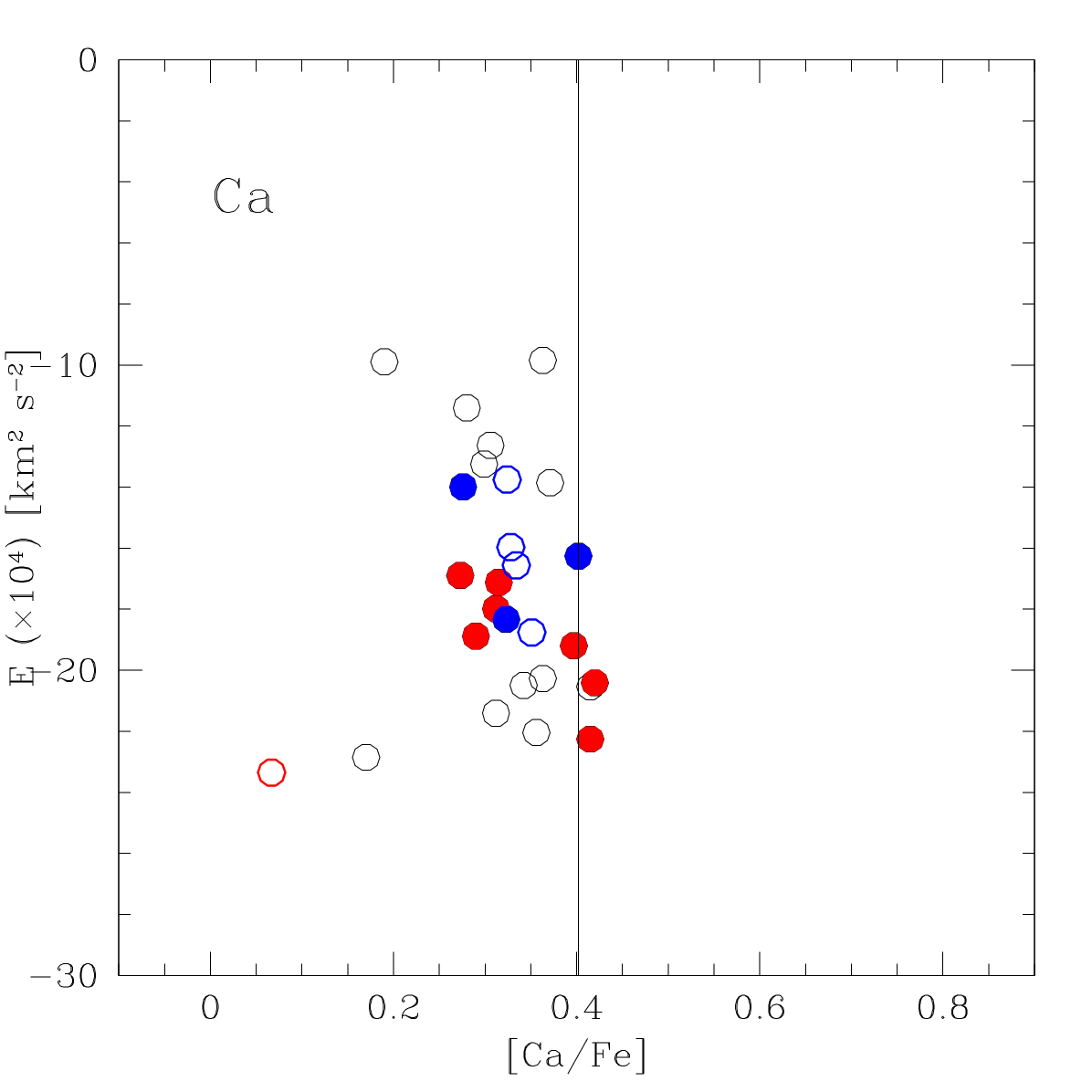}
\caption{Abundance ratios [Mg/Fe], [Si/Fe], and [Ca/Fe] as a function of the
orbital energy of the GCs, from Savino and Posti (2019). Red and blue symbols
are for GCs formed in situ and associated to Gaia-Enceladus. Filled and
empty circles refer to the present and previous works, respectively. In each panel, the
abundance level of NGC~288 is indicated by the vertical line.}
\label{f:eorb288}
\end{figure*}

Finally, our data seem not to confirm a tension between the kinematics of 
NGC~288 and its chemistry, as advocated by H20 from the
abundances in APOGEE. While the orbital parameters unambiguously associate
NGC~288 to the Gaia-Enceladus (GE) accretion event, H20 find the
level of $\alpha-$elements in this GC to be much higher than in other GE
candidate GCs and more in agreement with other GCs originated in situ. In our
analysis, the [Mg/Fe] ratio of NGC~288 is in agreement with that of 
other GCs associated to GE (blue points in 
Fig.~\ref{f:eorb288}), but also compatible with the abundance of in situ
GCs. The same is even more true concerning the Si abundance, whereas the Ca
abundance in NGC~288 is marginally higher than in other clusters from GE. We
must conclude that in both the present work and in APOGEE there is no strong
evidence favouring the attribution of NGC~288 to in situ or to GE groups 
from the chemical abundances alone.

\section{Temperature tomography of multiple populations}

With abundances of $\alpha-$elements in several hundreds of stars we have a
sample well suited to investigate the different temperature regimes in which 
proton-capture reactions were activated in GCs. More or less energetic reactions
being active in stars of different mass ranges, this study may provide useful
constraints to tune models of cluster formation and evolution.

\subsection{The high temperature regime: Mg and Si}

A complete survey of O, Na, Mg, Al, and Si and their inter-related abundances
was done in Carretta et al. (2009b), but only for a limited sample in each GC (a
maximum of 14 stars, those observed with the UVES Red-arm fibres). Despite the
small sample size, to that study dates the discovery that a clear Mg-Al
anti-correlation (produced at temperatures $\gtrsim 70$ MK, see e.g. Prantzos et
al. 2017, Gratton et al. 2019) is only detectable in metal-poor or massive GCs.
In the present study we do not have abundances of Al available for all stars, but
we are able to test variations in  Mg and Si.

\begin{figure*}[t]
\centering
\includegraphics[scale=0.30]{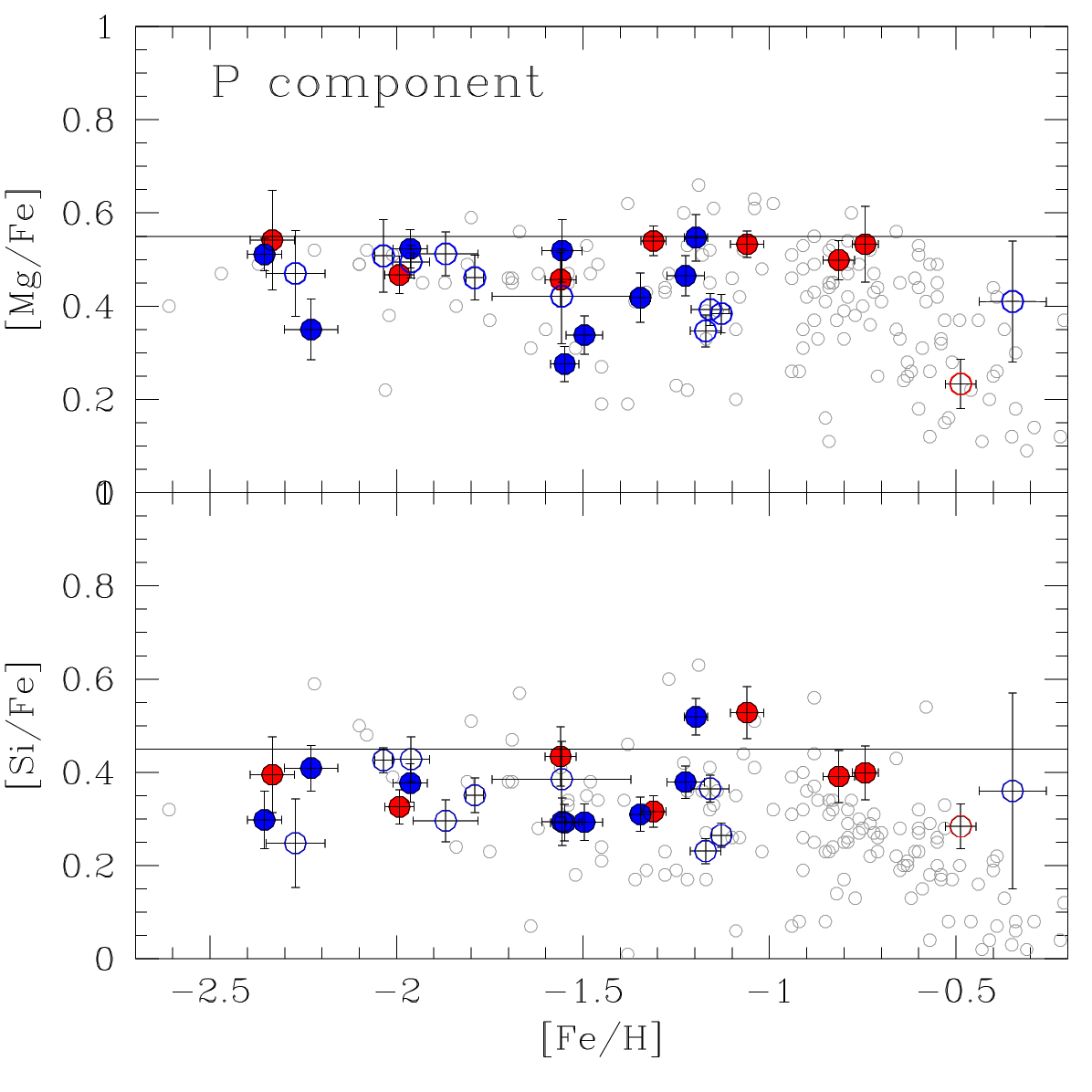}\includegraphics[scale=0.30]{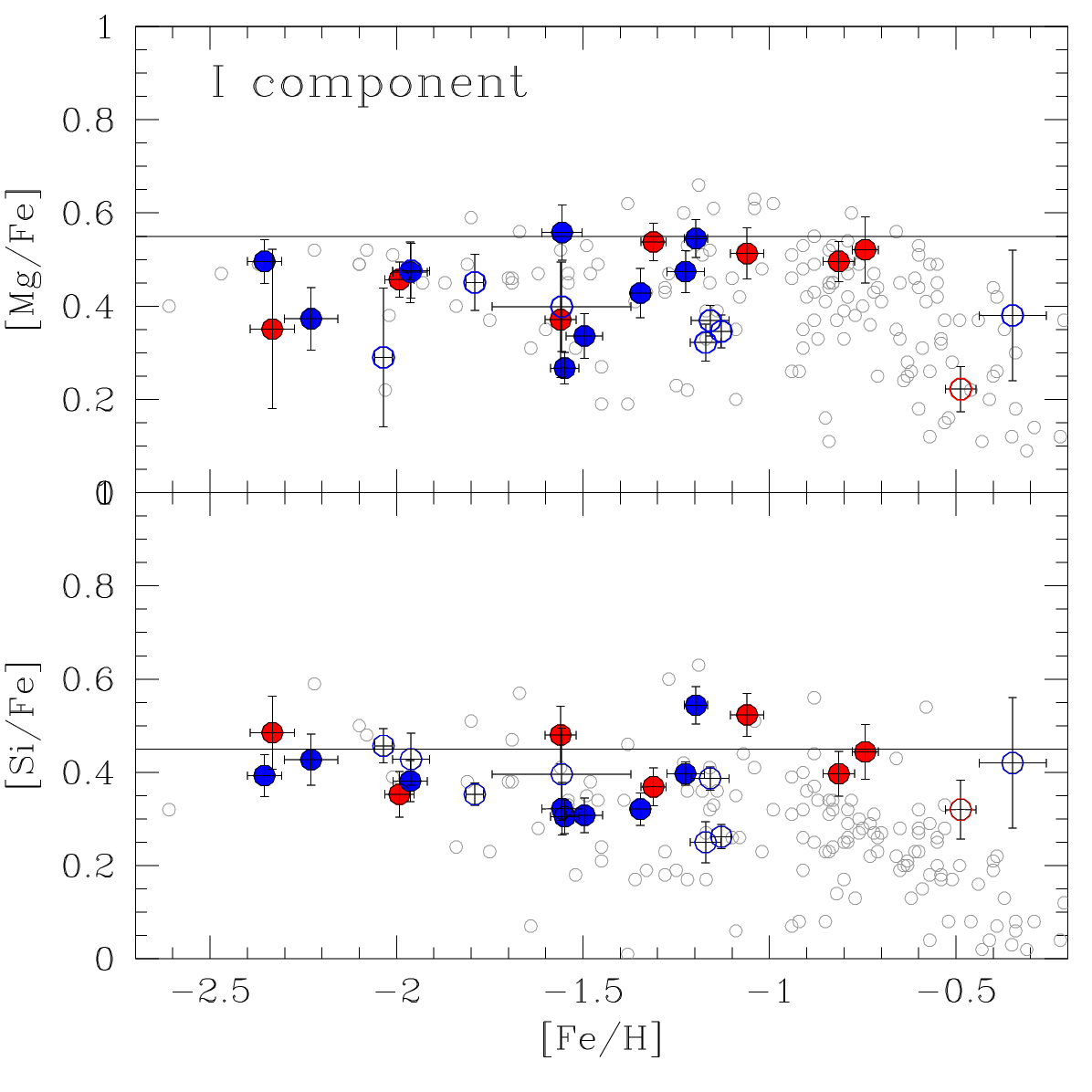}\includegraphics[scale=0.30]{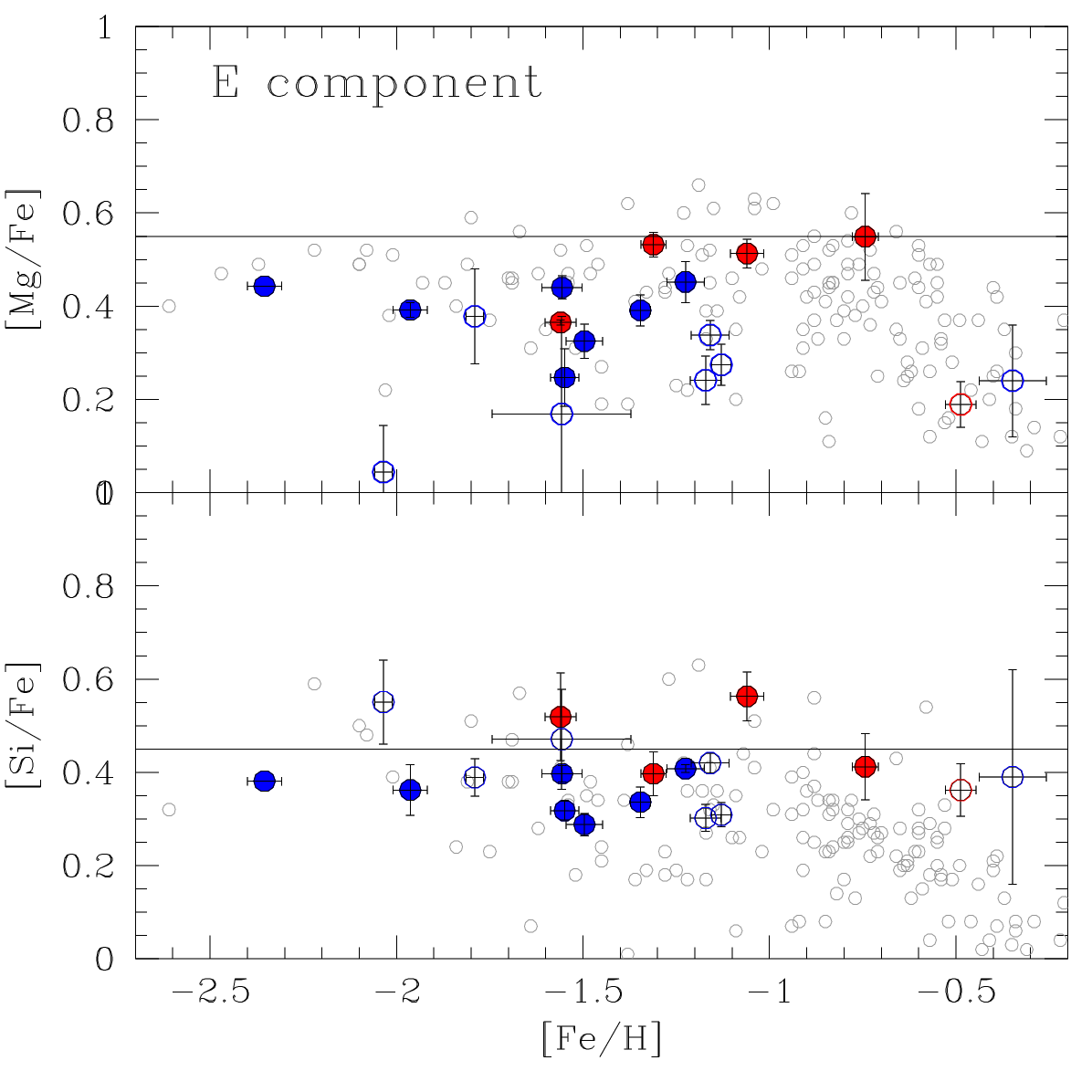}
\caption{Average abundance ratios [Mg/Fe] and [Si/Fe] for GCs in our extended
sample, using only the unpolluted P fraction (left panel), the polluted
component with intermediate I and extreme E composition (middle and right panel,
respectively) in each GC. The horizontal lines trace by eye the upper
envelope of abundances in the left panel and are reproduced in the other two
panels.}
\label{f:mediesolo}
\end{figure*}

Despite the generally small variations in these species, dwarfed e.g. by the
changes in O, Na, and Al, a first indication is coming even looking at the
average [element/Fe] ratios, as in Fig.~\ref{f:mediesolo}. In the three panels,
the mean values [Mg/Fe] and [Si/Fe] are re-computed again for all GCs using only
stars of the primordial P component (left panel), and the stars of the polluted
populations with intermediate I and extreme E composition (middle and right
panels, respectively). The E component, however, is not present in all
GCs. When using as reference the line traced on the upper
envelope distributions of Mg and Si in the left panel, it is easy to see that Mg
is decreasing and Si slightly increasing proceeding toward stars with more and
more altered composition, that is along the sequence P-I-E.

\begin{figure*}[t]
\centering
\includegraphics[scale=0.30]{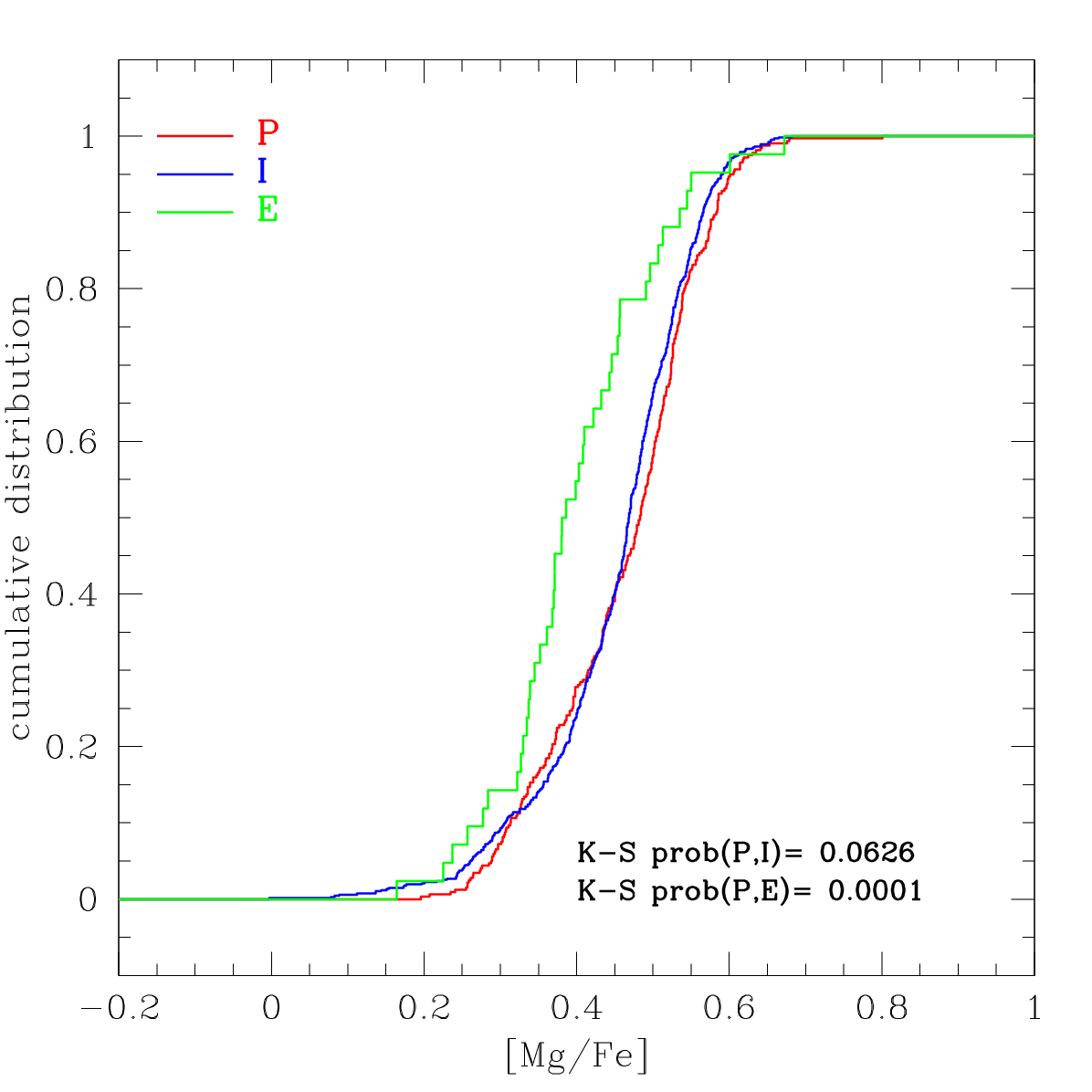}\includegraphics[scale=0.30]{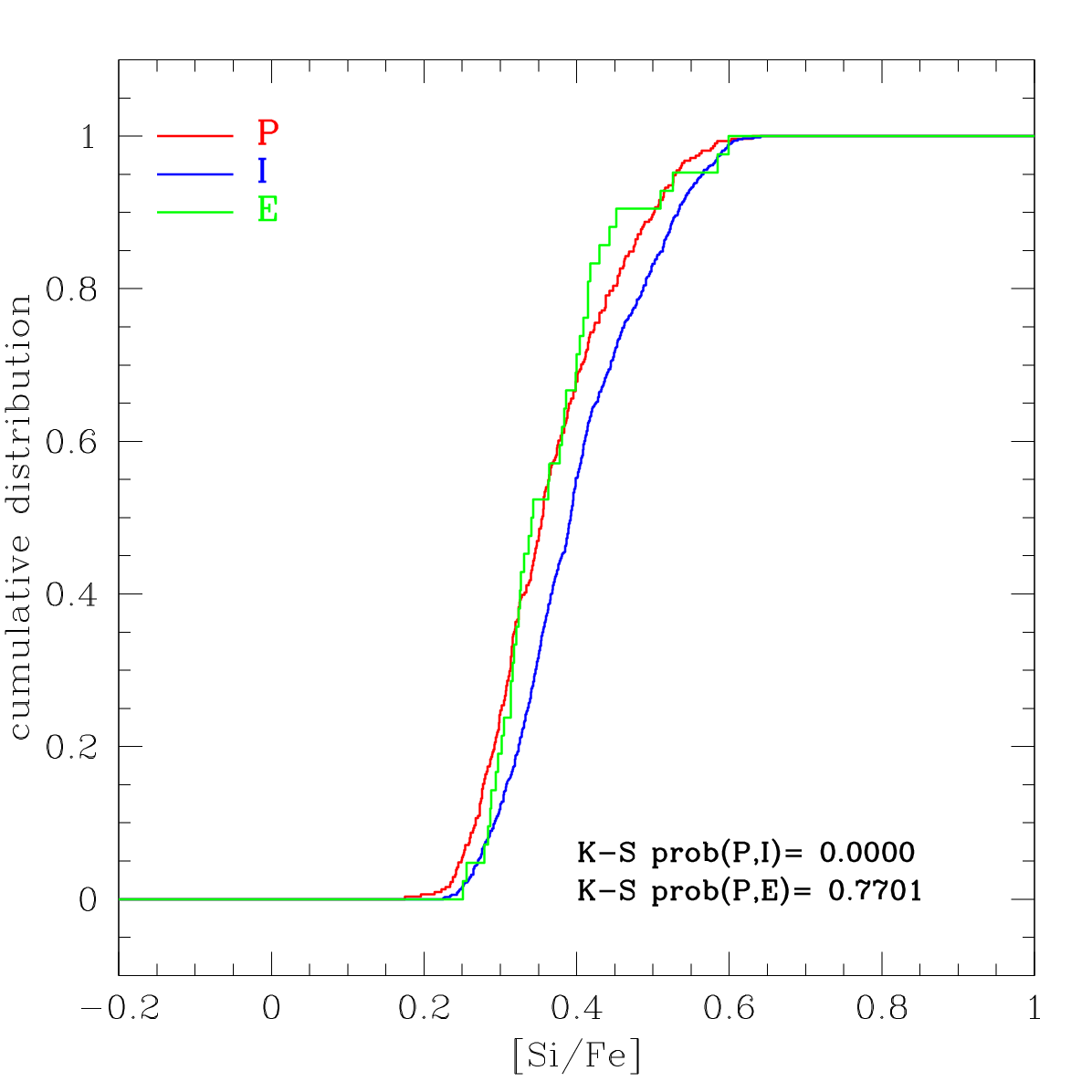}\includegraphics[scale=0.30]{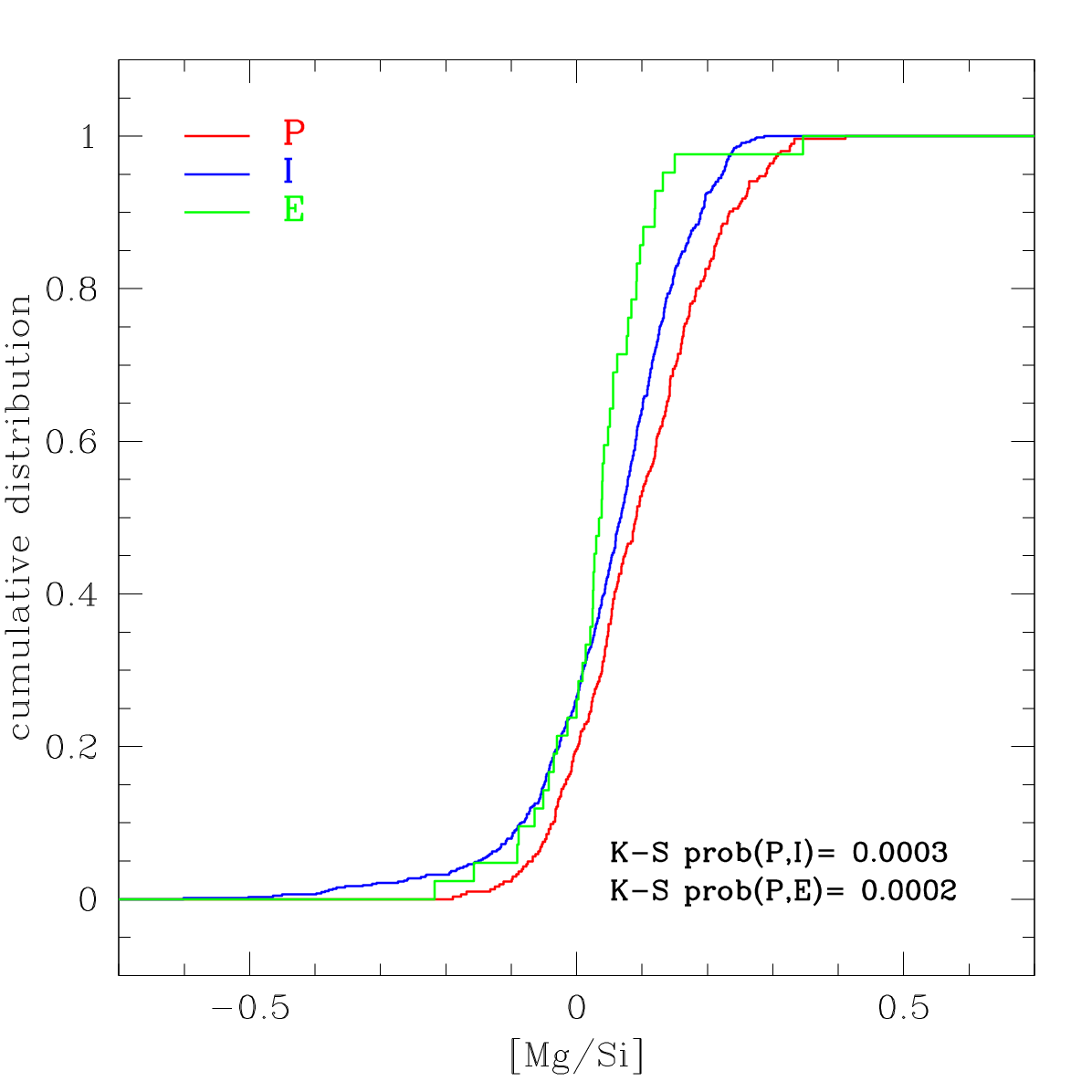}
\caption{Results of the Kolmogorov-Smirnov test on the cumulative distributions
of the [Mg/Fe], [Si/Fe], and [Mg/Si] of the P, I, and E components in the 16
program GCs. In each panel the probability of the K-S test for the comparison
P-I and P-E is listed.}
\label{f:ksMGNA}
\end{figure*}

A more quantitative statistics is represented in Fig.~\ref{f:ksMGNA}. We applied
a K-S test to the cumulative distributions of the [Mg/Fe], [Si/Fe], and
[Mg/Si] ratios in the P, I, and E components in the 16 program GCs. For
each ratio it was possible to safely reject the null hypothesis that the
samples are extracted from the same parent population, at least concerning
the polluted and the unpolluted components, since variations in Mg and Si are
not equally pronounced in all GCs.

\begin{figure*}[t]
\centering
\includegraphics[scale=0.15]{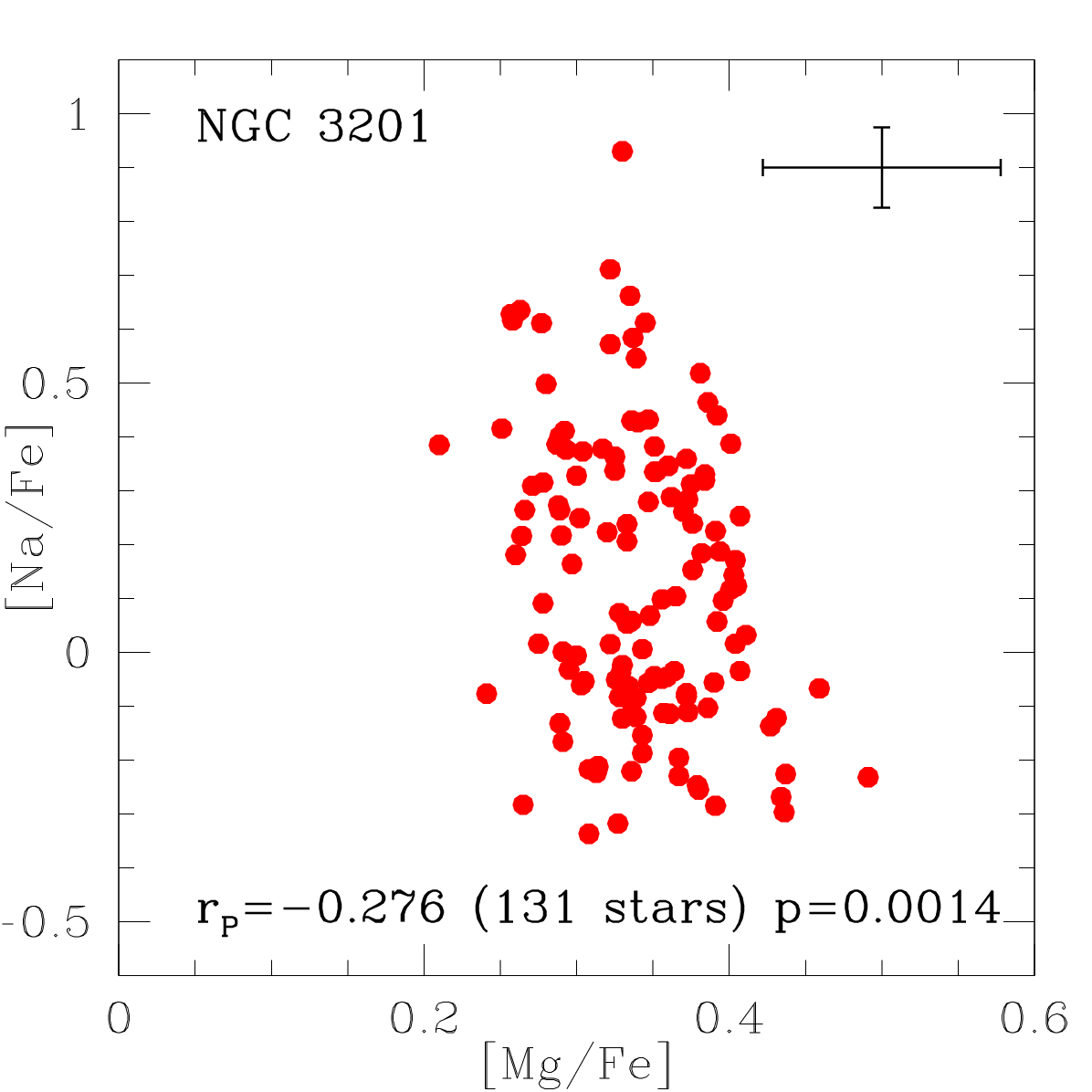}\includegraphics[scale=0.15]{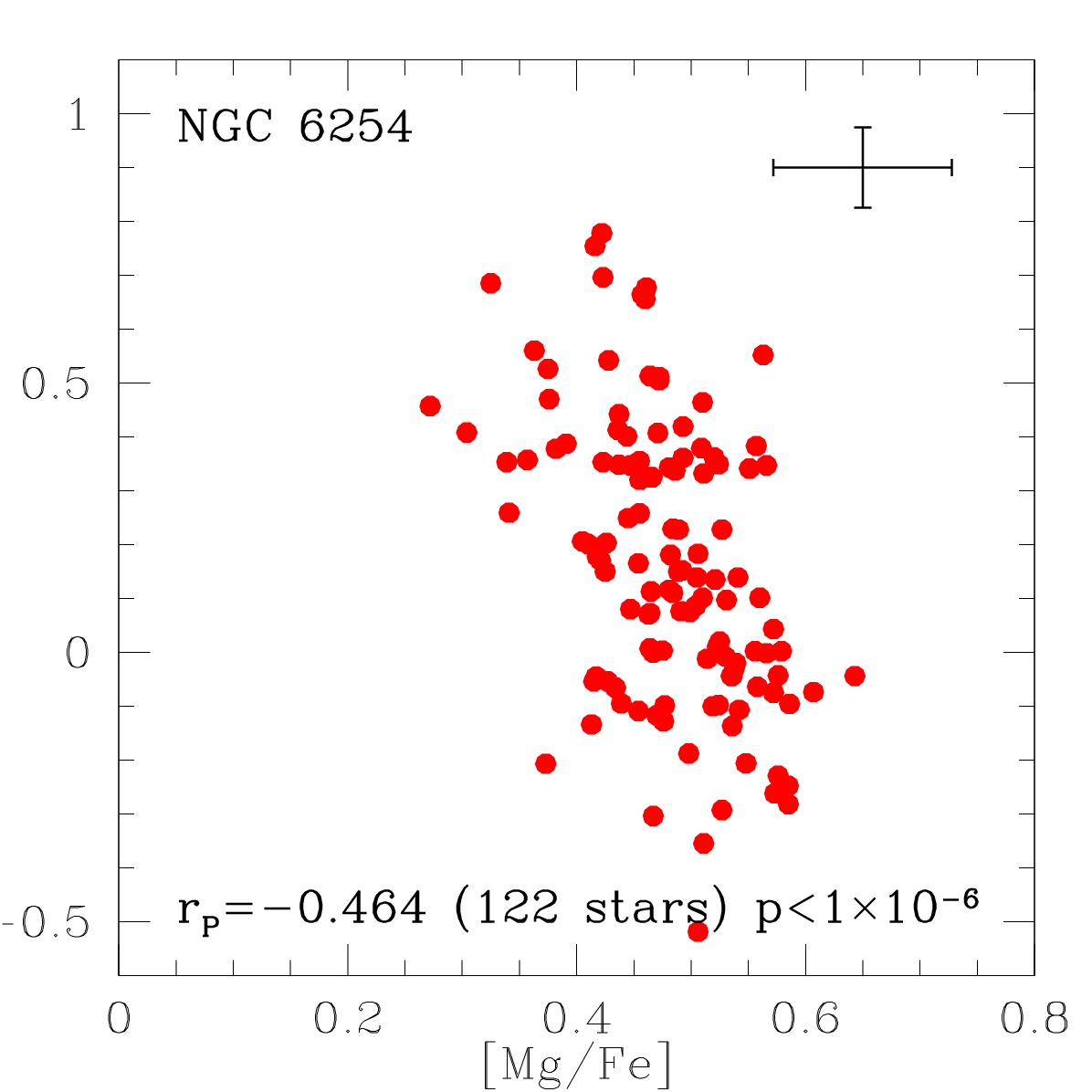}\includegraphics[scale=0.15]{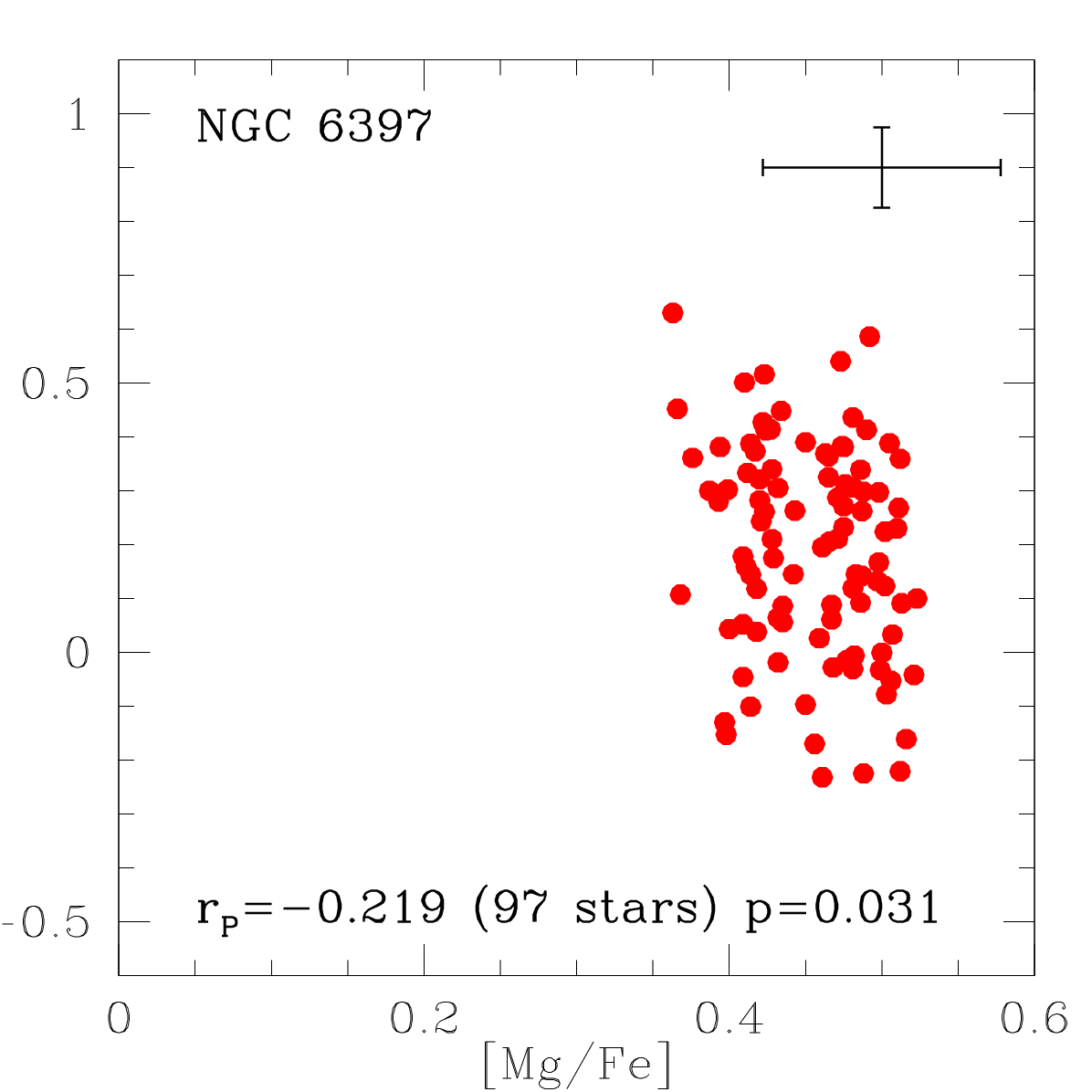}\includegraphics[scale=0.15]{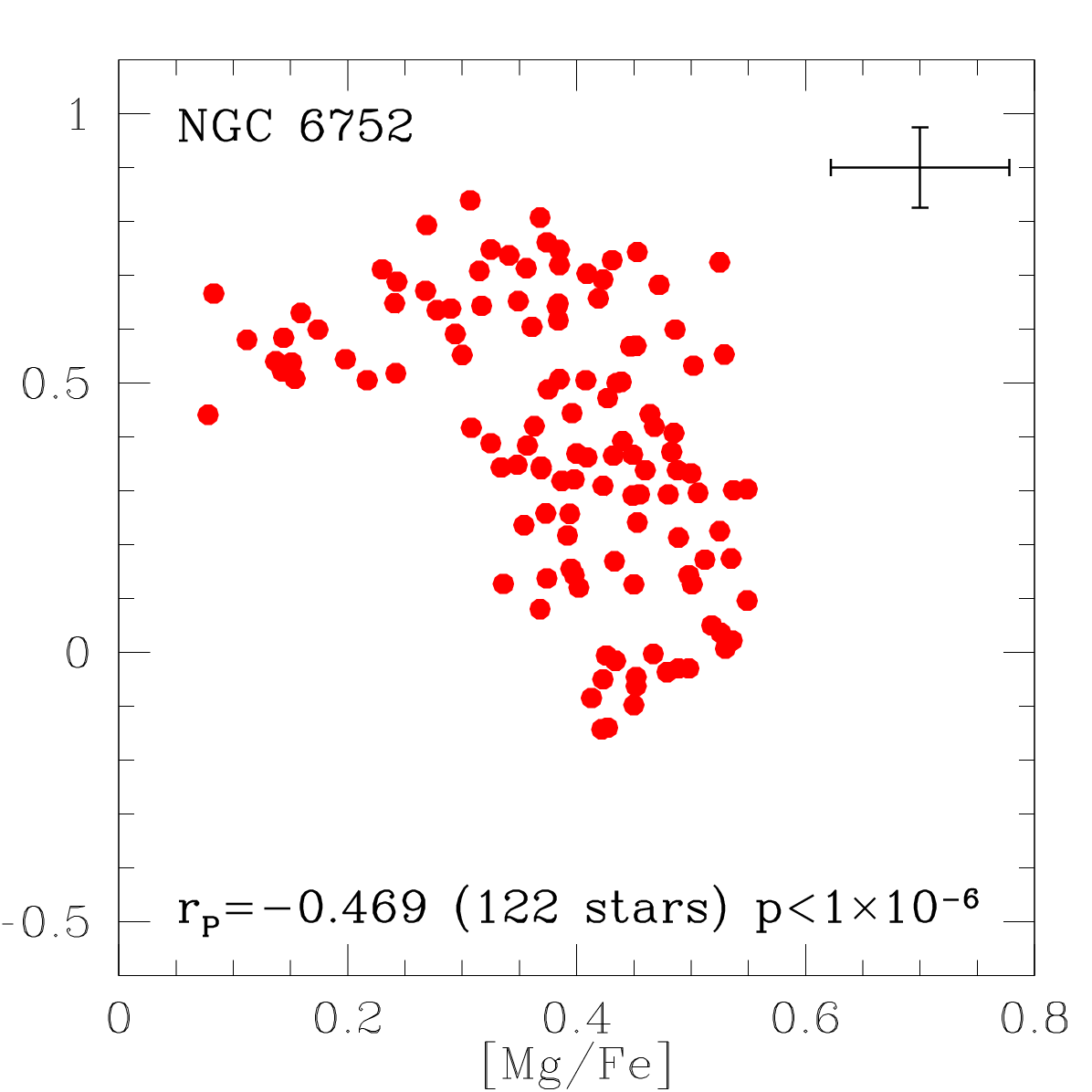}\includegraphics[scale=0.15]{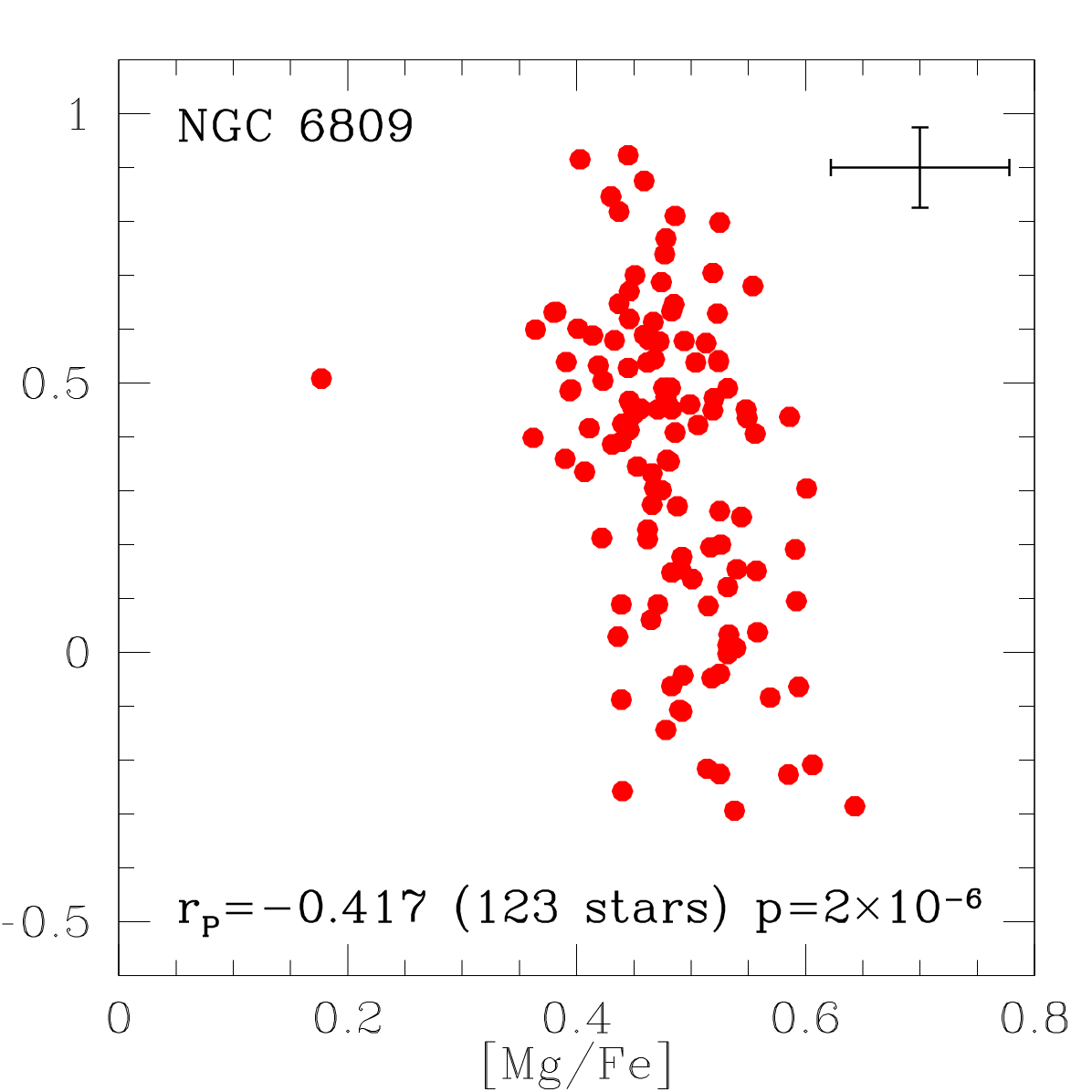}\includegraphics[scale=0.15]{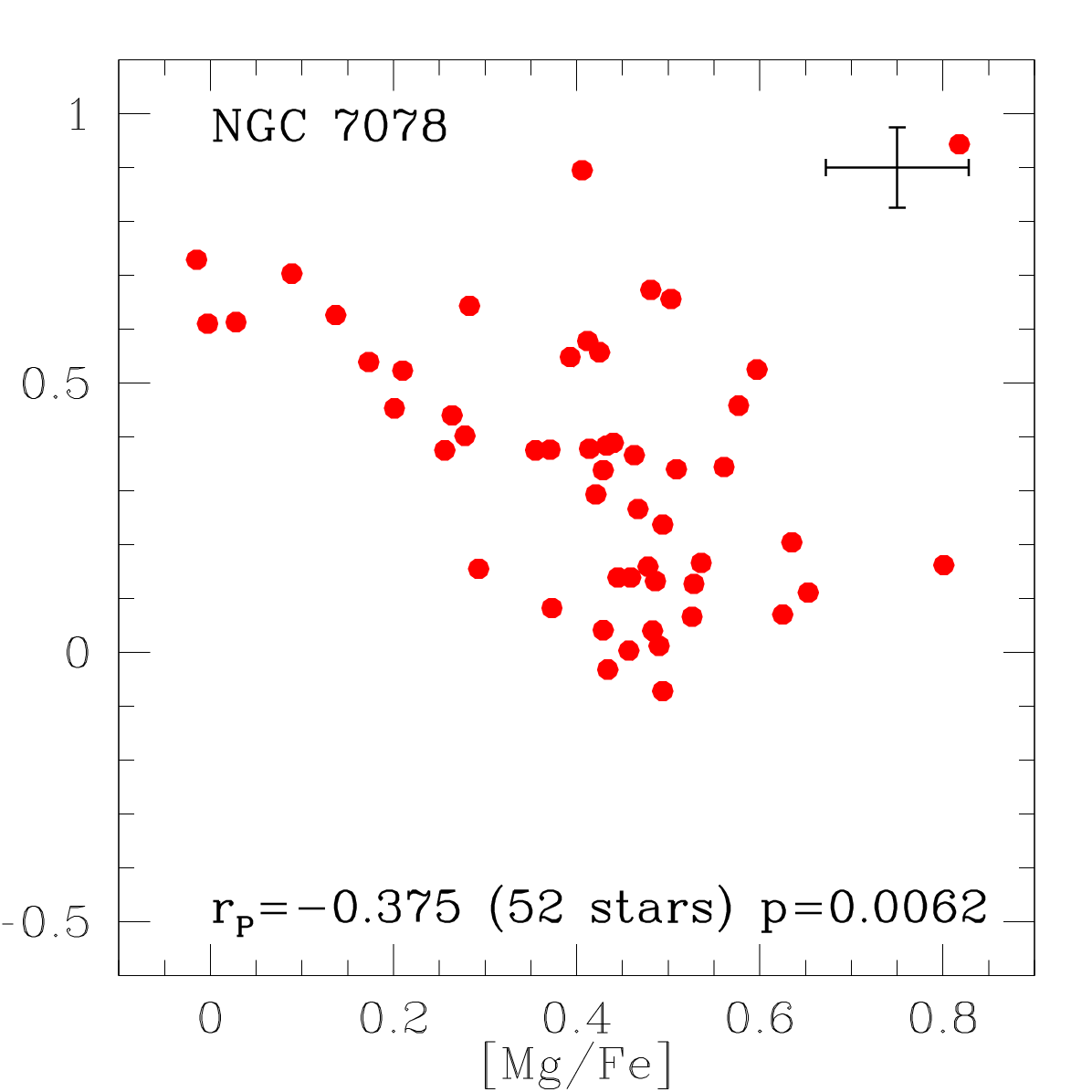}
\caption{The Na-Mg anti-correlation in the GCs of our program sample where the
linear regression between Na and Mg is found to be statistically significant
(p-value $<$ 0.05). The Pearson's correlation coefficient and probability are 
listed in each panel together with a typical error bar.}
\label{f:mgna}
\end{figure*}

We also know that Na and O are modified in practically all GCs (e.g. Carretta et
al. 2009a,b,2010a for extensive surveys), and the changes in [Mg/Na] and [Si/Na]
are probably driven by the large variations in the Na content among the P,I,and
E components. For a more detailed screening it is possible to see in what GCs Mg
is more involved in MPs by evaluating the anti-correlation with Na or the Mg-O
correlation. In Fig.~\ref{f:mgna} we plot the results for the six GCs where a
linear regression is statistically significant (p-value$<$0.05), according to the Pearson's
correlation coefficient and the two-tailed probability that the correlation is
real, both listed in each panel. The Mg-Na anti-correlation is matched rather
well by the complementary Mg-O correlation\footnote{Apart from NGC~3201, where
the correlation, with p=0.076, is formally not statistically significant.}, but
the first relation represents a better indication because Na is available for
many more stars than O.

We confirm the discovery made in Carretta et al. (2009b) that the largest star to
star variations in the Mg abundance are found among the MPs of metal-poor and
massive GCs. The conclusion in the present study rests on a sample more
than a factor 6 larger with respect to the sample used in Carretta et al. (2009b).

\begin{figure*}[t]
\centering
\includegraphics[scale=0.15]{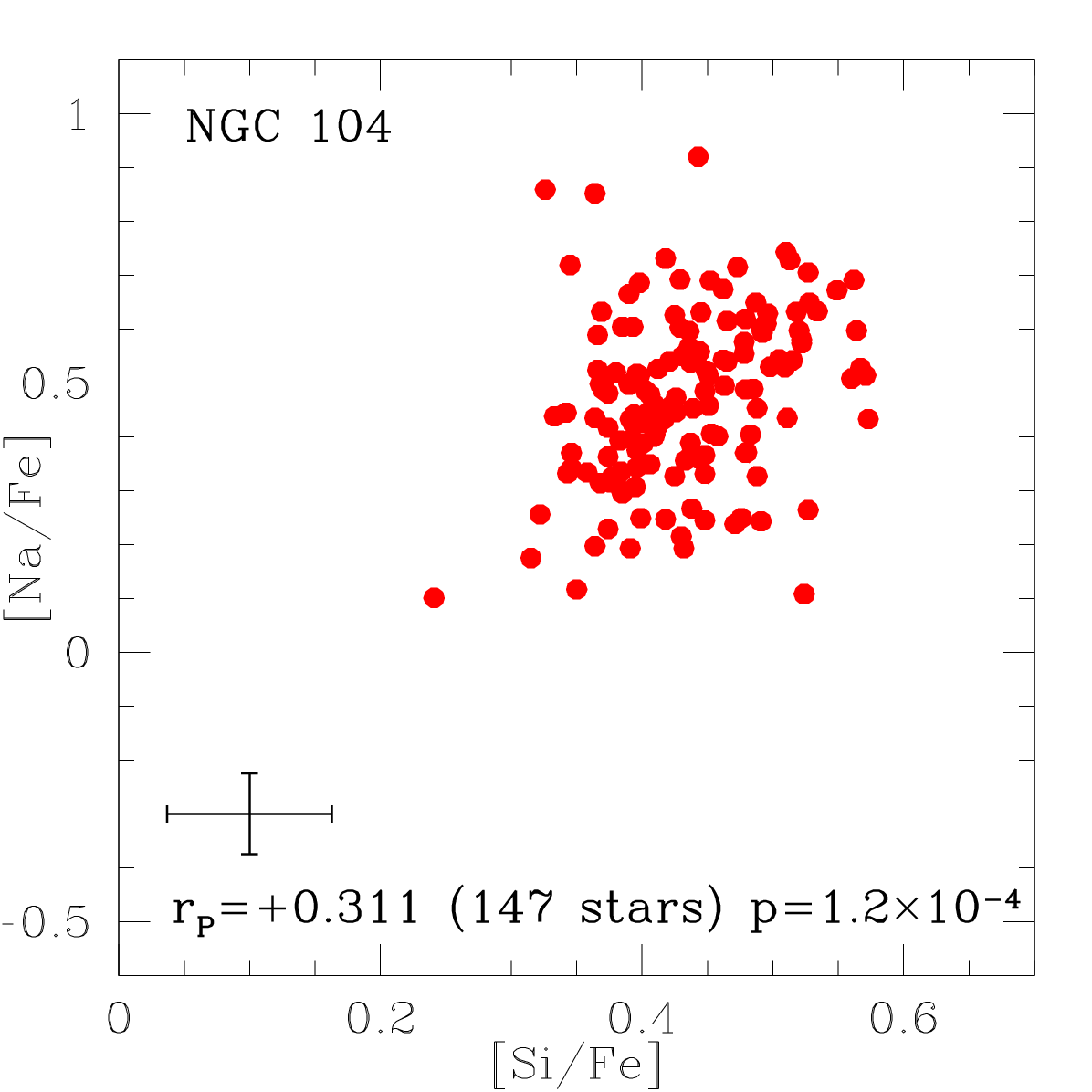}\includegraphics[scale=0.15]{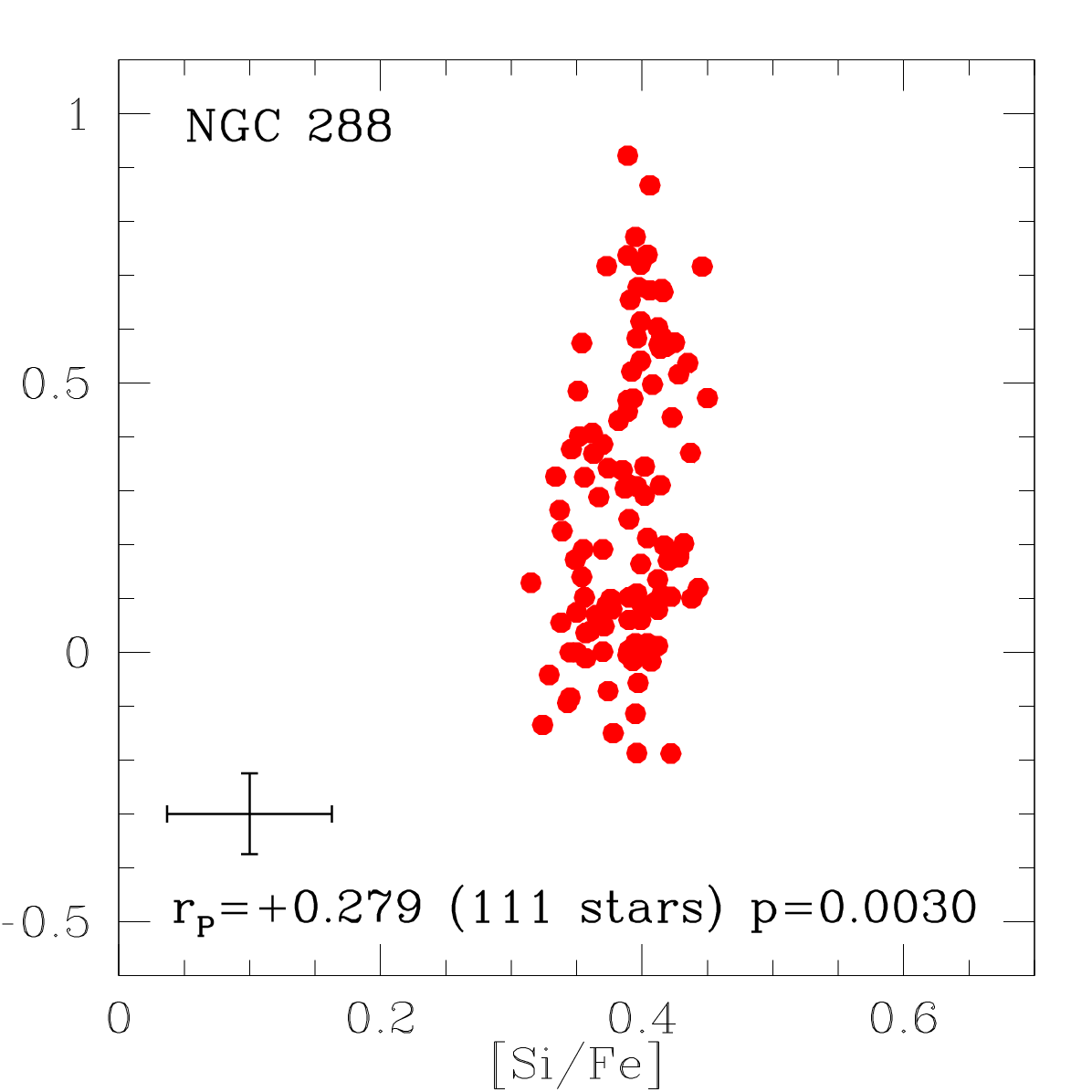}\includegraphics[scale=0.15]{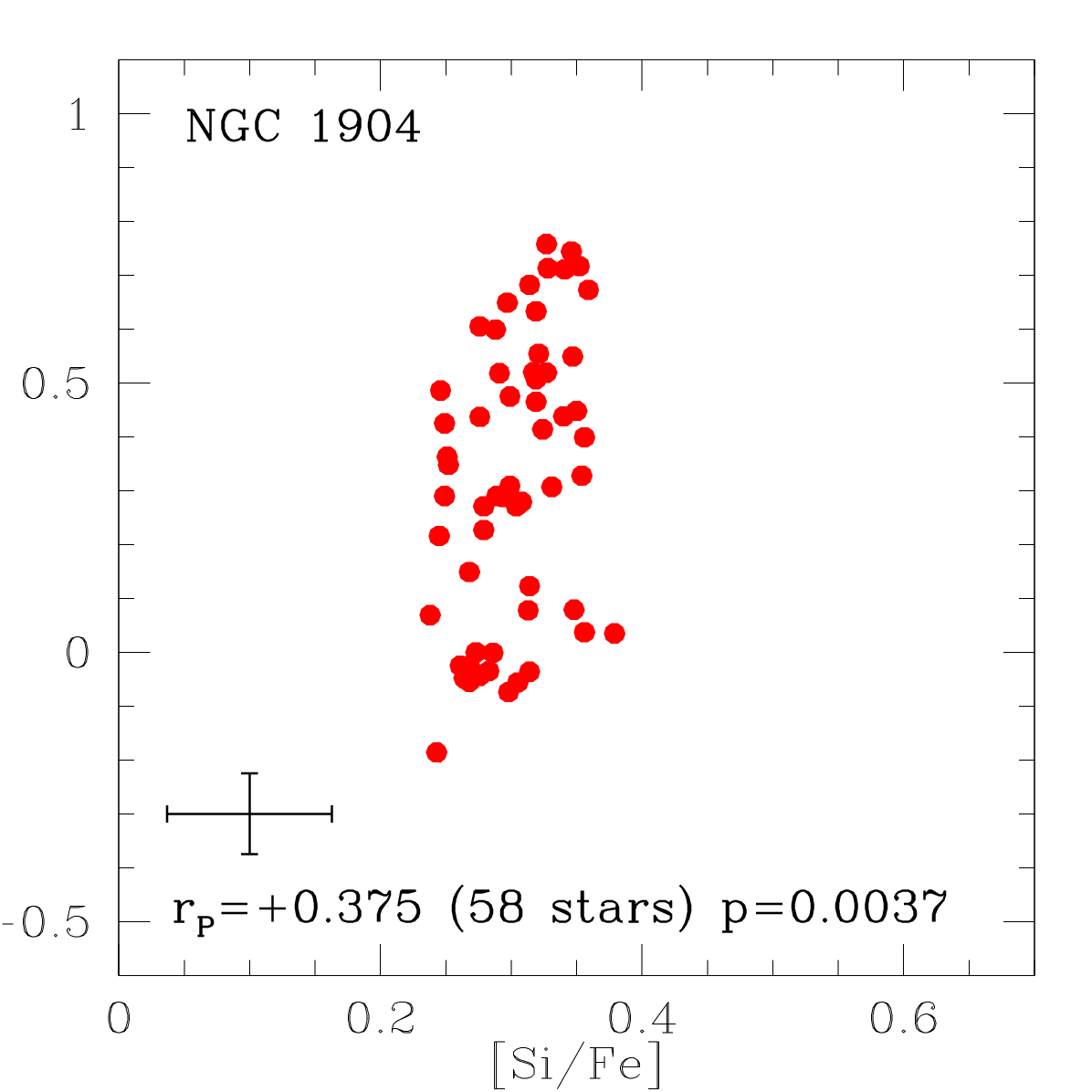}\includegraphics[scale=0.15]{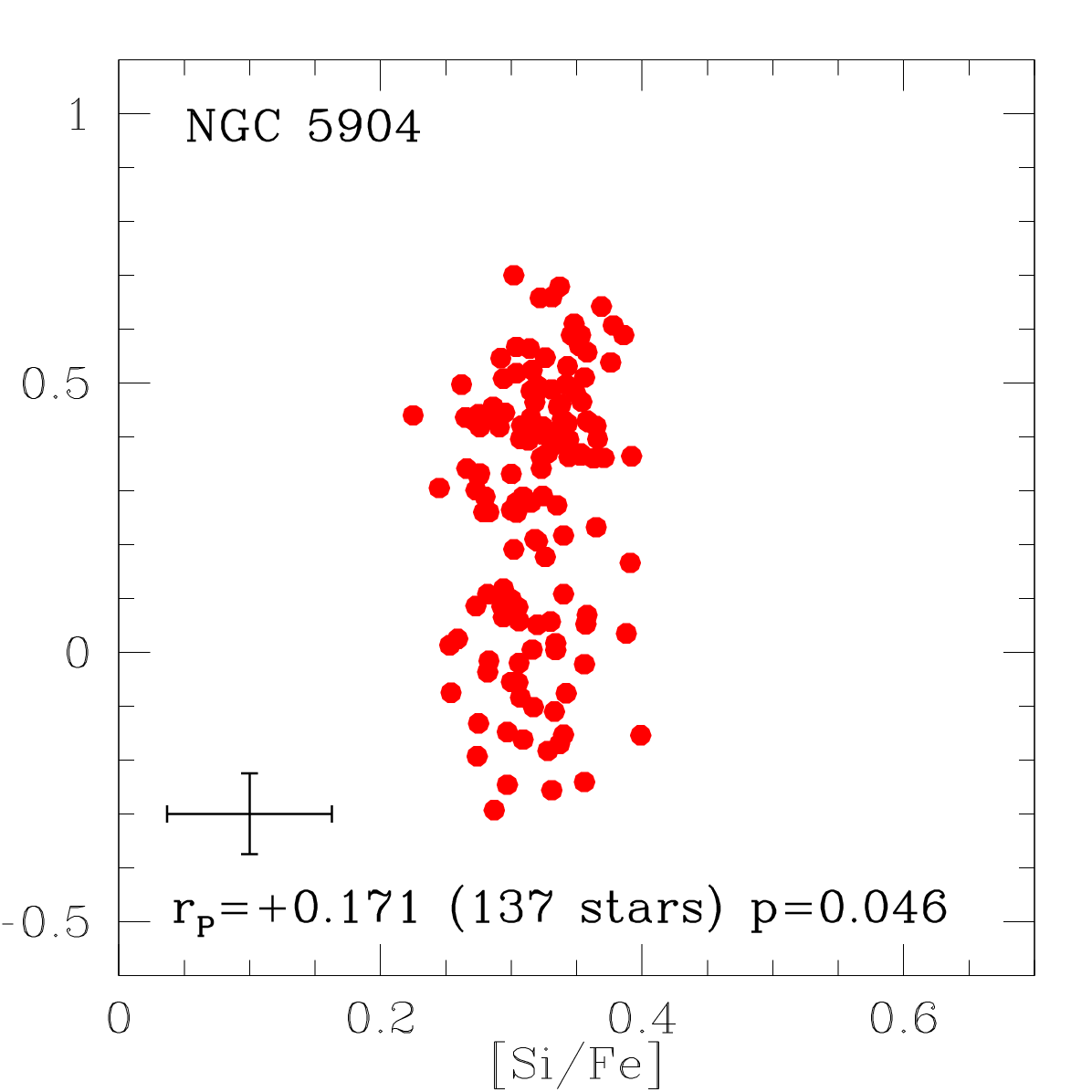}\includegraphics[scale=0.15]{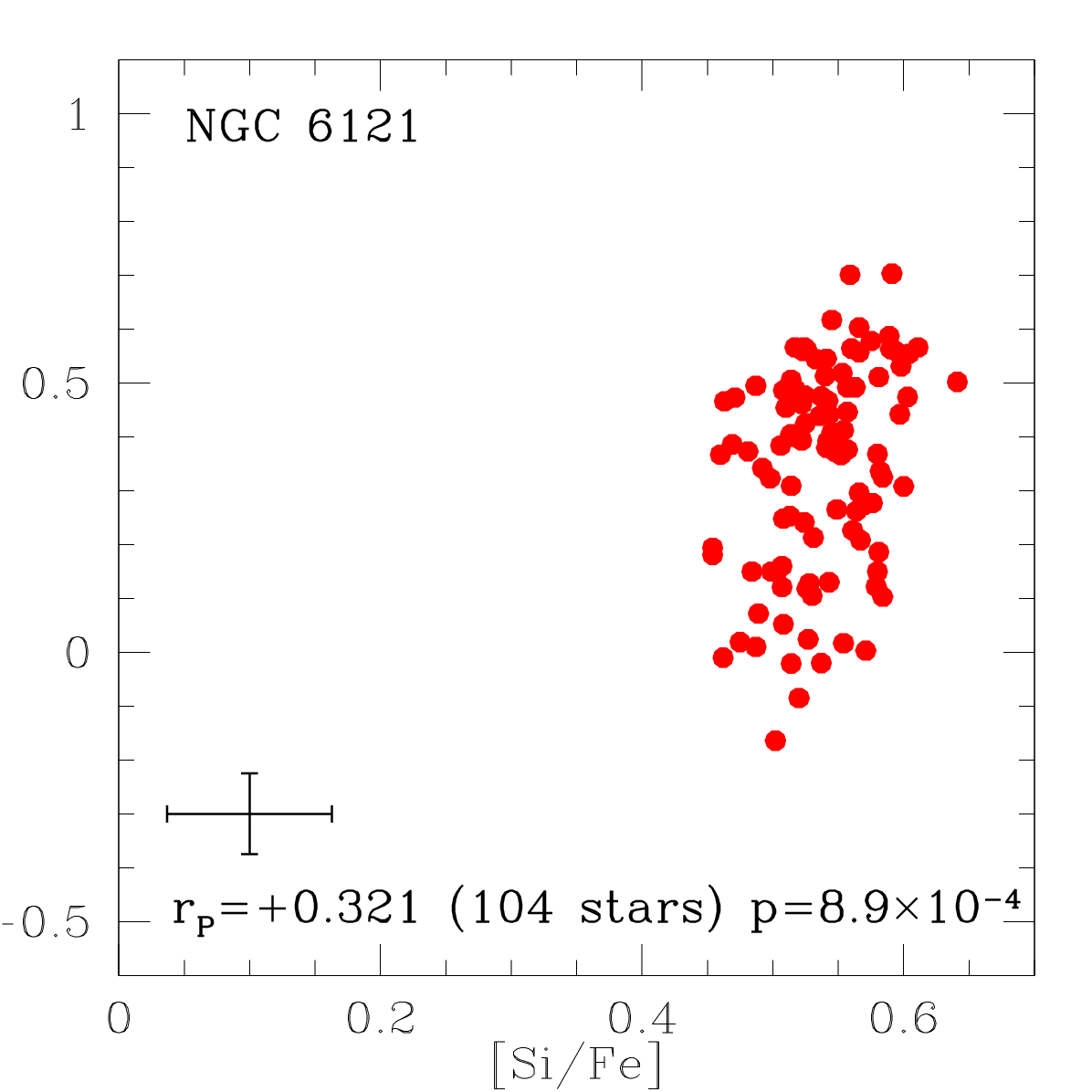}
\includegraphics[scale=0.15]{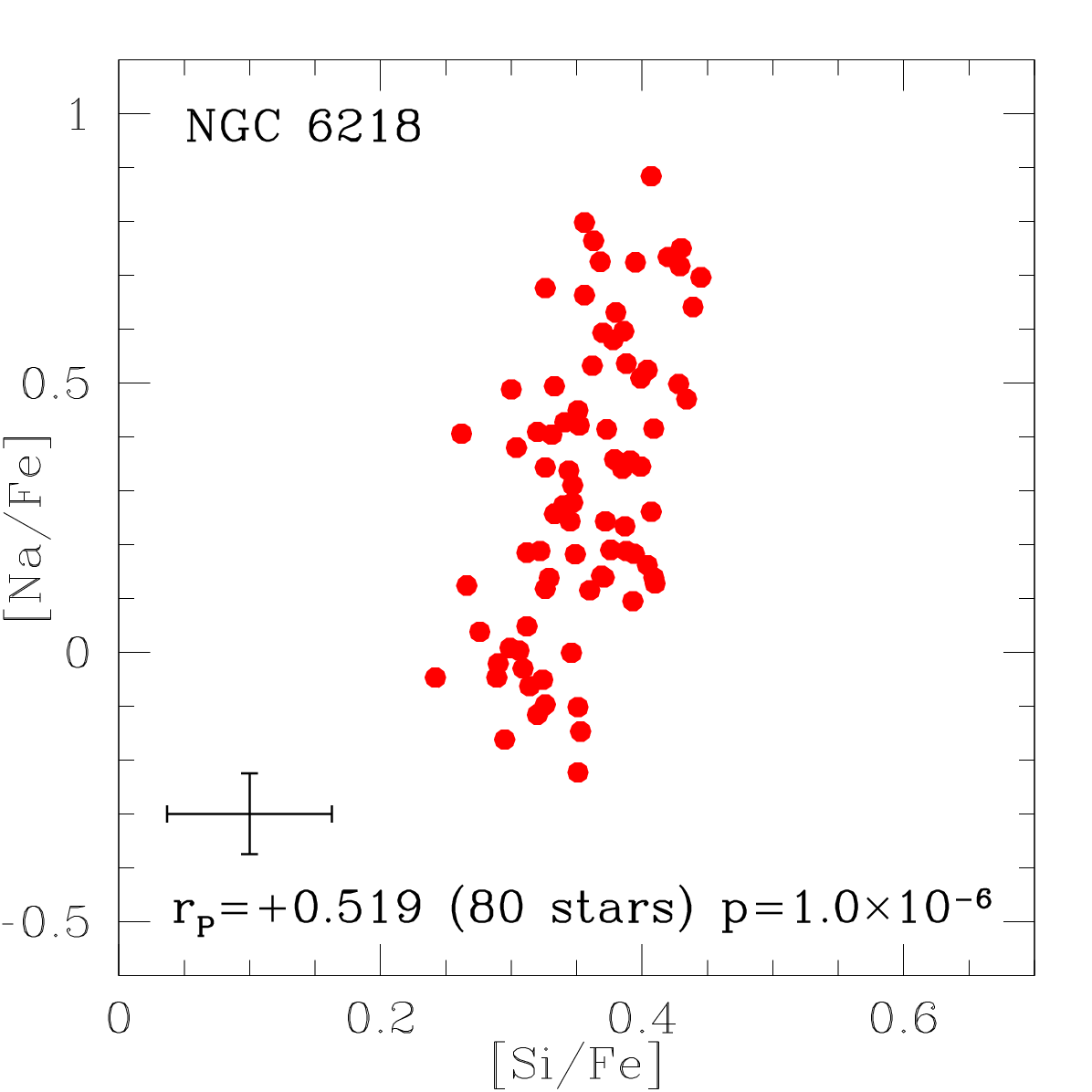}\includegraphics[scale=0.15]{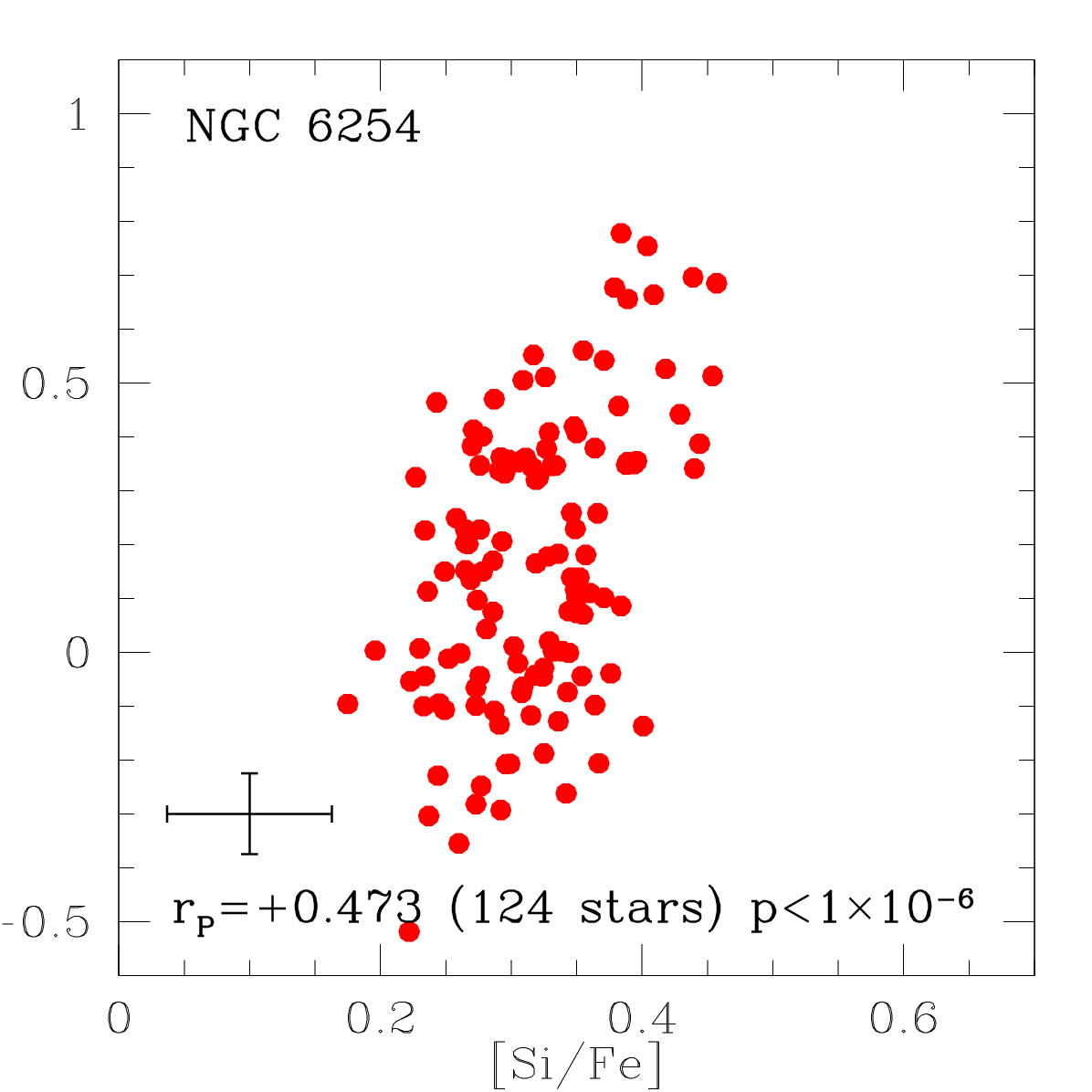}\includegraphics[scale=0.15]{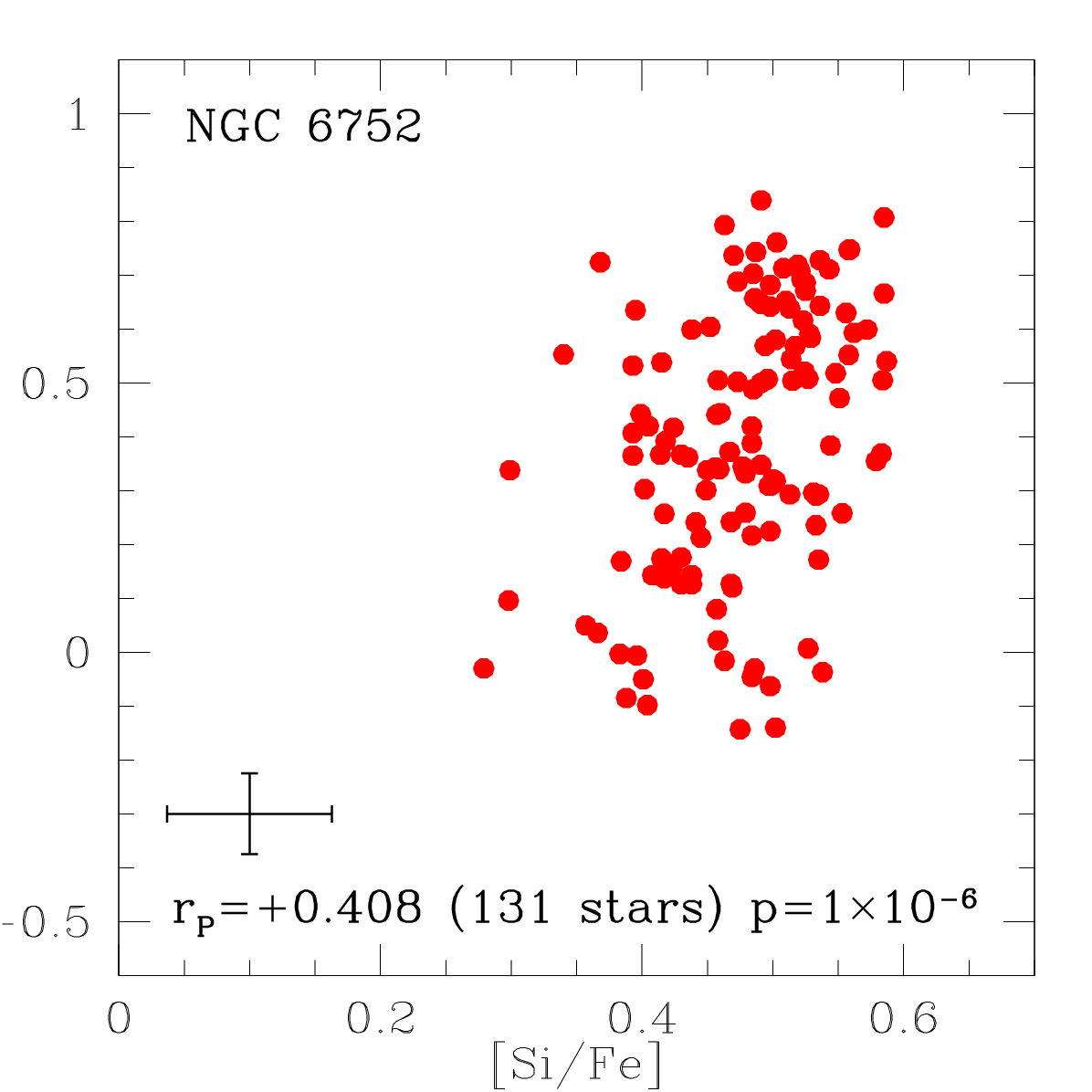}\includegraphics[scale=0.15]{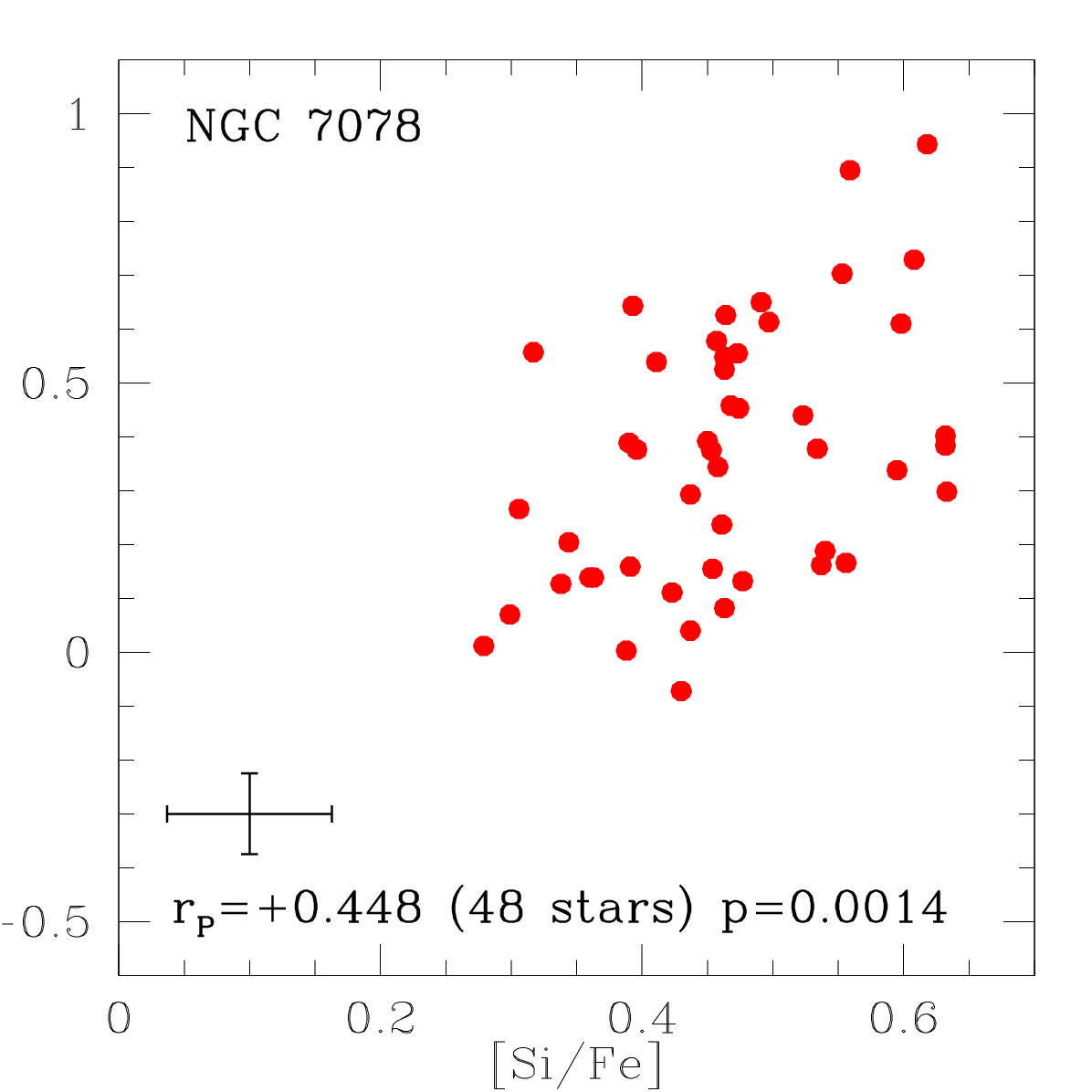}\includegraphics[scale=0.15]{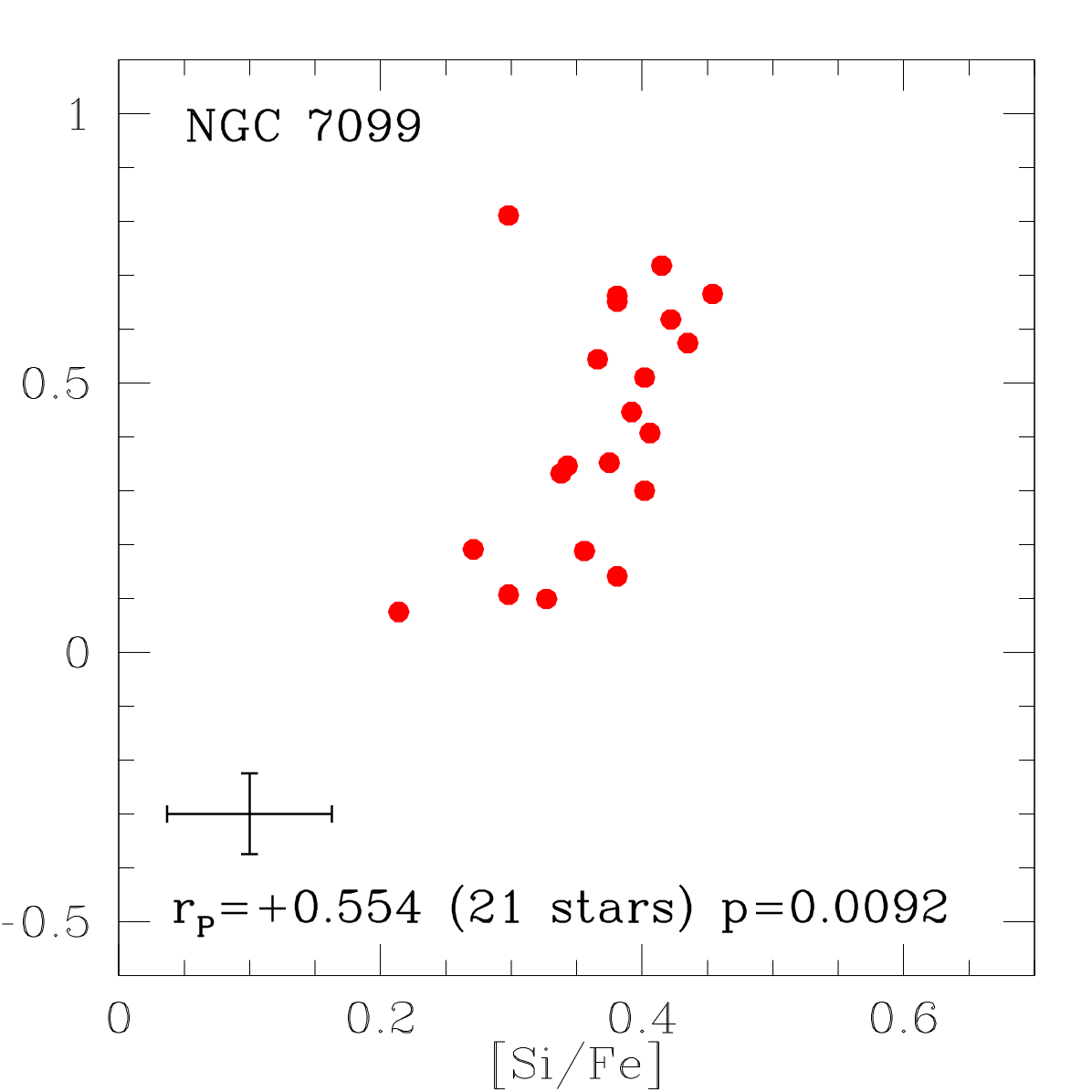}
\caption{The Si-Na correlation in the GCs of our program sample where the linear
regression is found to be statistically significant. A typical error bar is
shown.}
\label{f:sinaV}
\end{figure*}

The distributions of the [Mg/Si] ratios are much closer to each other than those
including the specie with the largest variation (Na), as expected from the
involved mechanism that is a simple leakage from the main MgAl cycle (e.g. Yong
et al. 2005, Carretta et al.2009b). In our data, 10 out of 16  GCs show a
statistically significant correlation Si-Na (Fig.~\ref{f:sinaV}). For the
other 6 GCs, the probabilities for the linear regression are
p=0.71,0.08,0.15,0.07,0.07, and 0.66 (for NGC~3201, NGC~4590, NGC~6171,
NGC~6397, NGC~6809, and NGC~6838, respectively). This evidence suggests that the
leakage from Mg on Si is not a relevant phenomenon in these GCs.

\subsection{The very high temperature regime: Ca}

\begin{figure*}[t]
\centering
\includegraphics[scale=0.30]{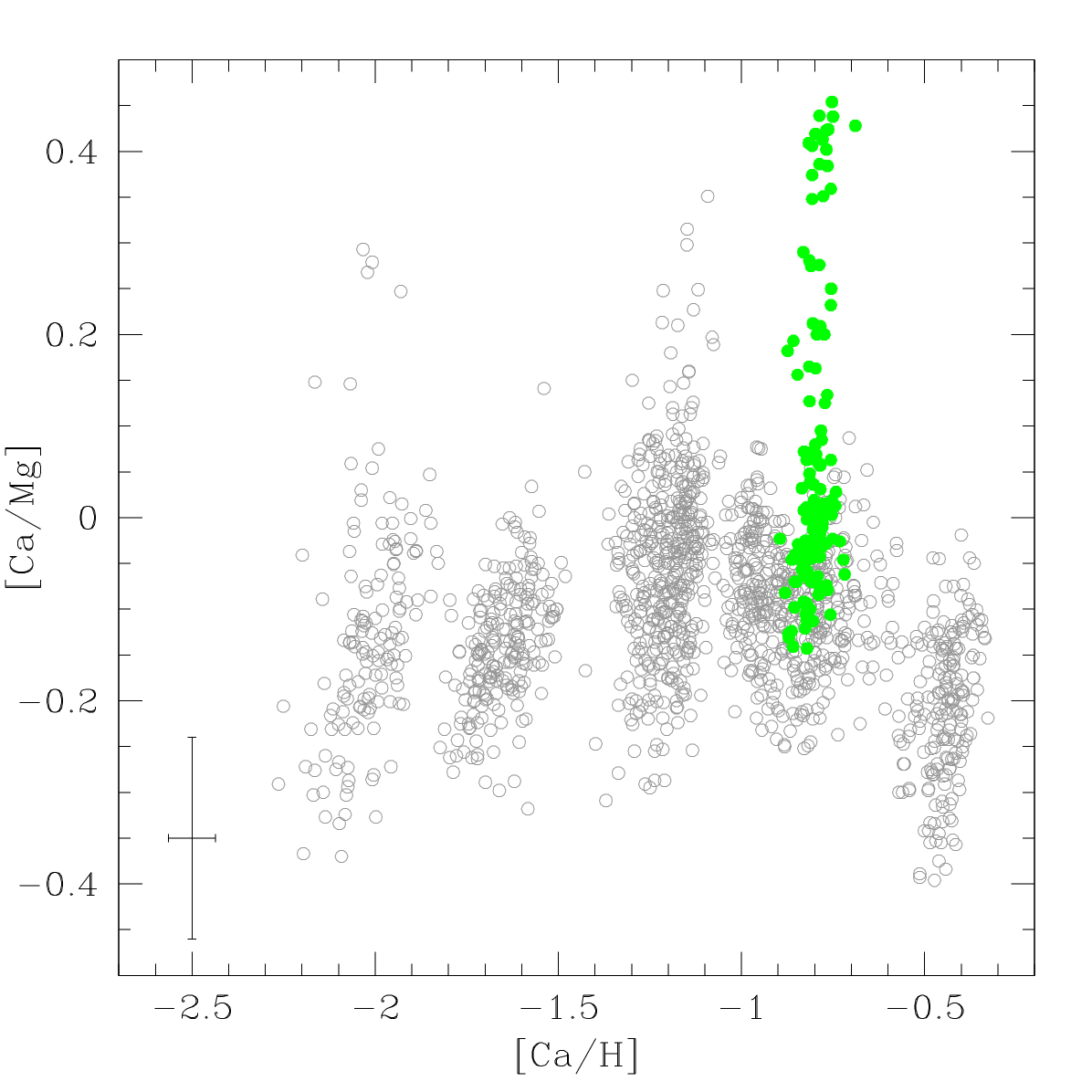}\includegraphics[scale=0.30]{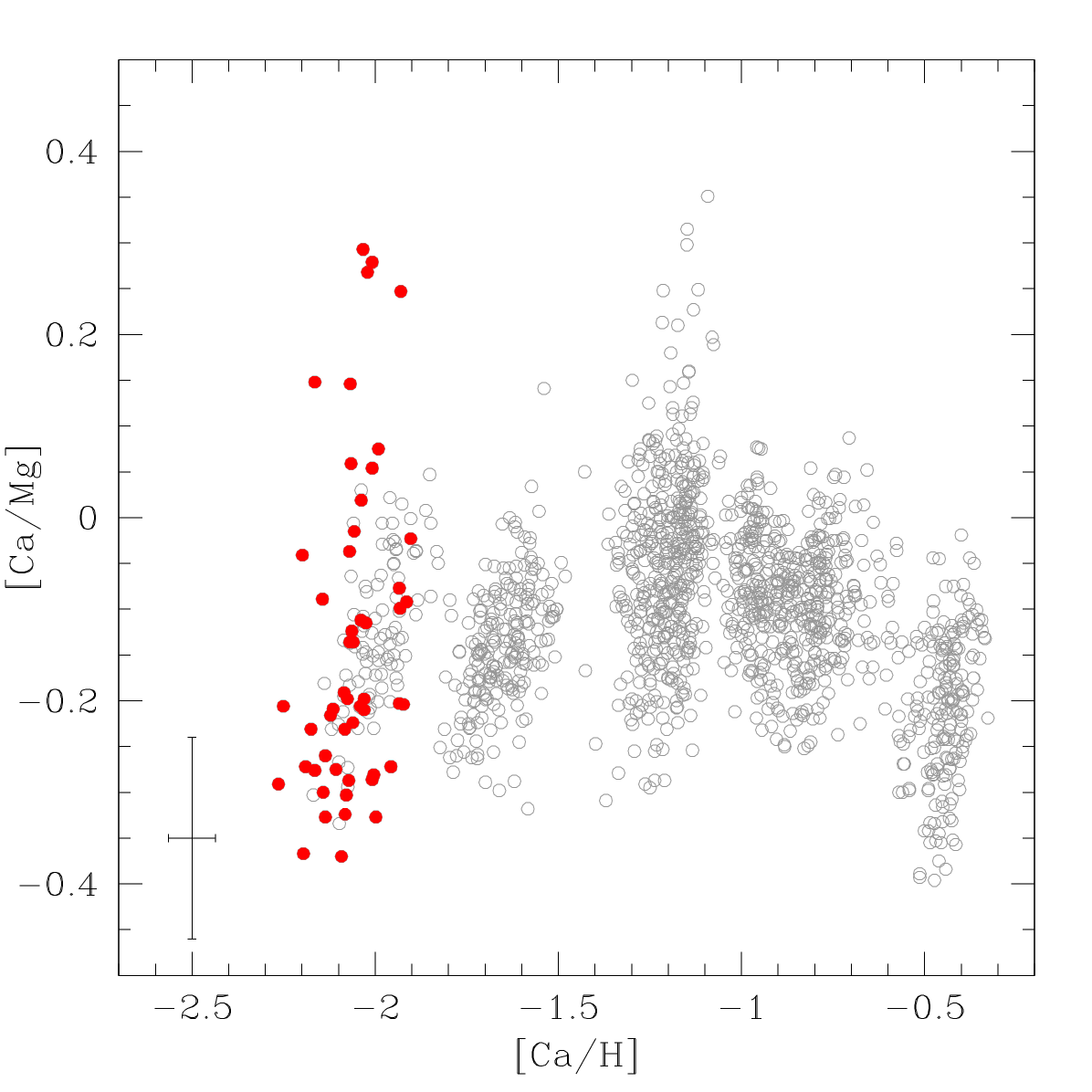}\includegraphics[scale=0.30]{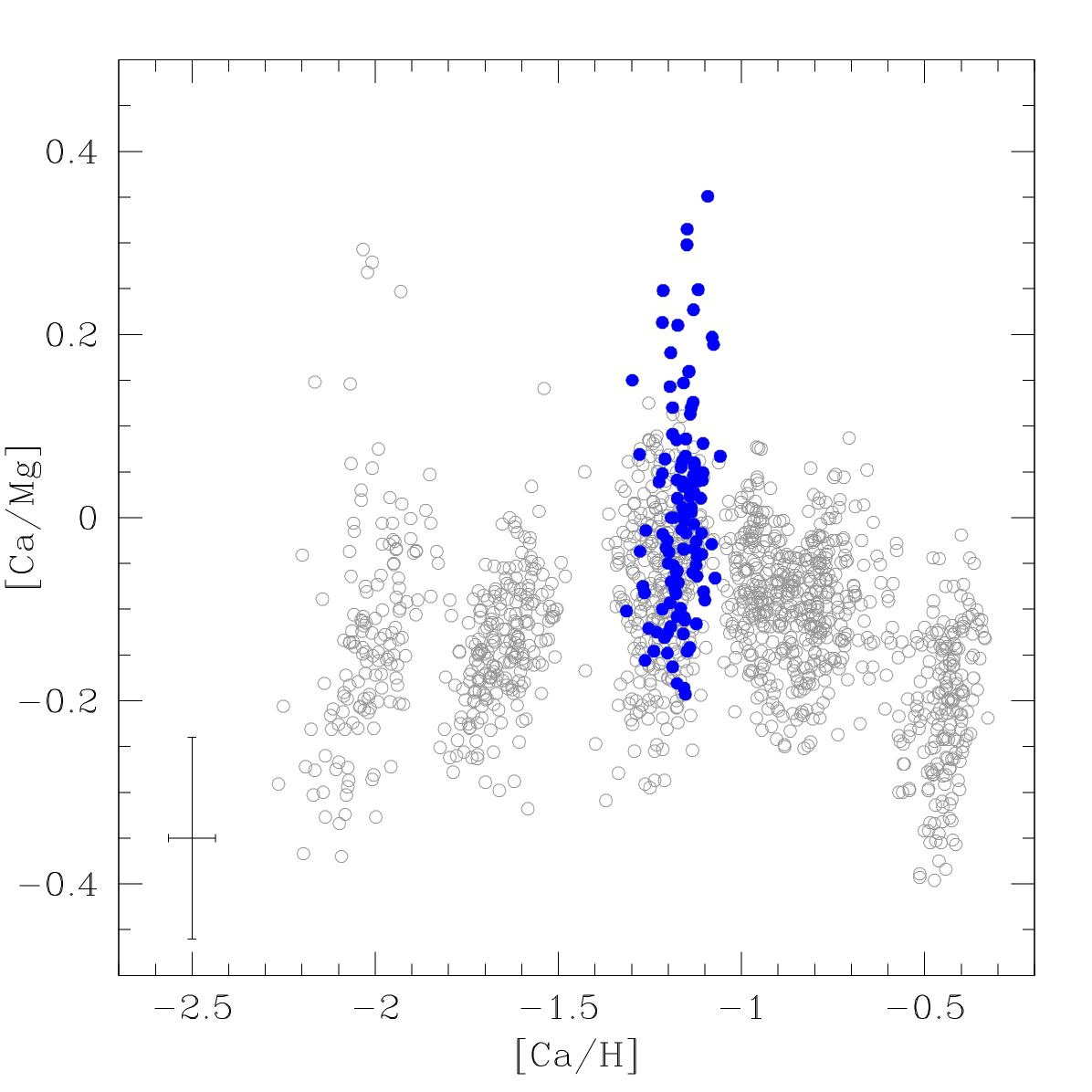}
\caption{Diagnostic plot [Ca/Mg] as a function of [Ca/H] for the stars of 16 GCs
in the present study (empty grey points). In the left panel, green filled circles are stars in
NGC~2808 from Carretta (2015), in the middle and right panels, red and blue
filled circles are stars in NGC~7078 and NGC~6752, respectively, from the
present work.}
\label{f:camg}
\end{figure*}

The regime of MPs with chemistry generated by proton-capture reactions at very
high temperature can be studied with the aid of the diagnostic plot introduced
in Carretta et al. (2013c), measuring possible excesses of Ca linked to Mg
depletions. In this regime, the normal production of Al from Mg destruction may
be bypassed favouring the production of heavier species such as K, Ca, and even
Sc from proton-captures on Ar nuclei (Ventura et al. 2012). The Ca and Mg
abundances measured for a large number of stars in the present work provide an
useful statistics to investigate the relevance of these very high temperature
nuclear cycles involved in polluters of the first stellar generation in GCs 
(see also Carretta and Bragaglia 2021 and references therein).  In 
Fig.~\ref{f:camg} we plot the abundance ratio [Ca/Mg] as a function of the
[Ca/H] ratio for all stars in the 16 program GCs.

In the left panel of Fig.~\ref{f:camg} we superimposed the stars in NGC~2808
(green filled circles) from Carretta (2015). In the MPs of this GC correlations
and anti-correlations with Mg are observed up to the heaviest proton-capture species,
like K (Mucciarelli et al. 2015), Ca and Sc  (Carretta 2015). The large
extension of the [Ca/Mg] ratios is explained with large depletions in Mg
accompanied by moderate excesses in the Ca content.

In the middle and right panels our strictly homogeneous analysis allows to
highlight the same behaviour also for stars in NGC~6752 and NGC~7078, although
to a lesser extent. These are the only two GCs in our present sample where a
significant excess in the [Ca/Mg] ratio was observed. A depletion in Mg was
assessed already in previous sections. To check whether the depletion in Mg is
also matched by a slight enhancement in the Ca abundance we resort to the
procedure used in Carretta and Bragaglia (2021). In their systematic survey to
find excess of Ca and Sc in GC stars they used as reference the same sample of
unpolluted field stars from Gratton et al. (2003) used in the present work.
The differential procedure compares in the plane [Ca/H]-[Mg/H] GC stars with
the abundance pattern of field stars that are expected to incorporate only the
yields from supernova nucleosynthesis.

\begin{figure*}[t]
\centering
\includegraphics[scale=0.40]{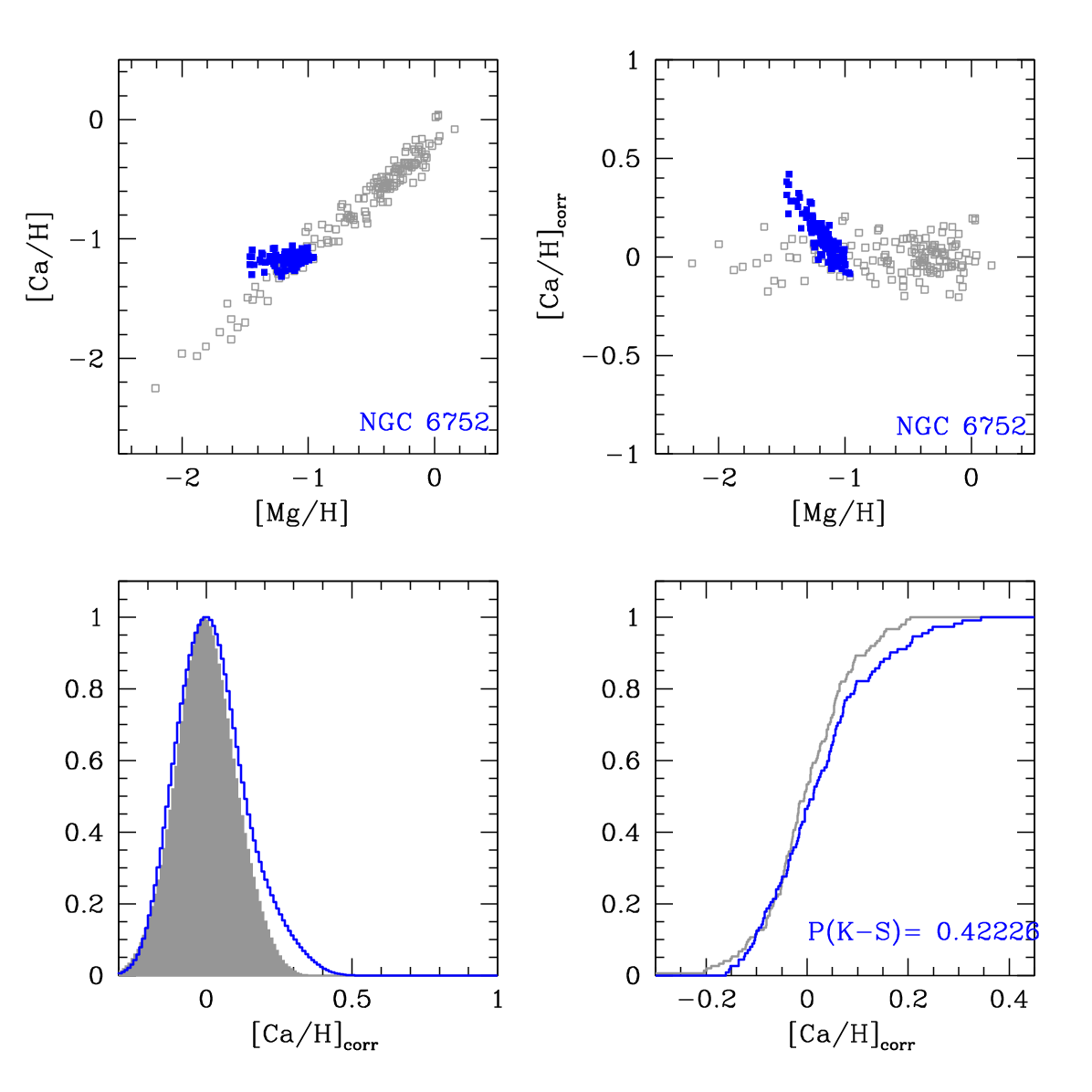}\includegraphics[scale=0.40]{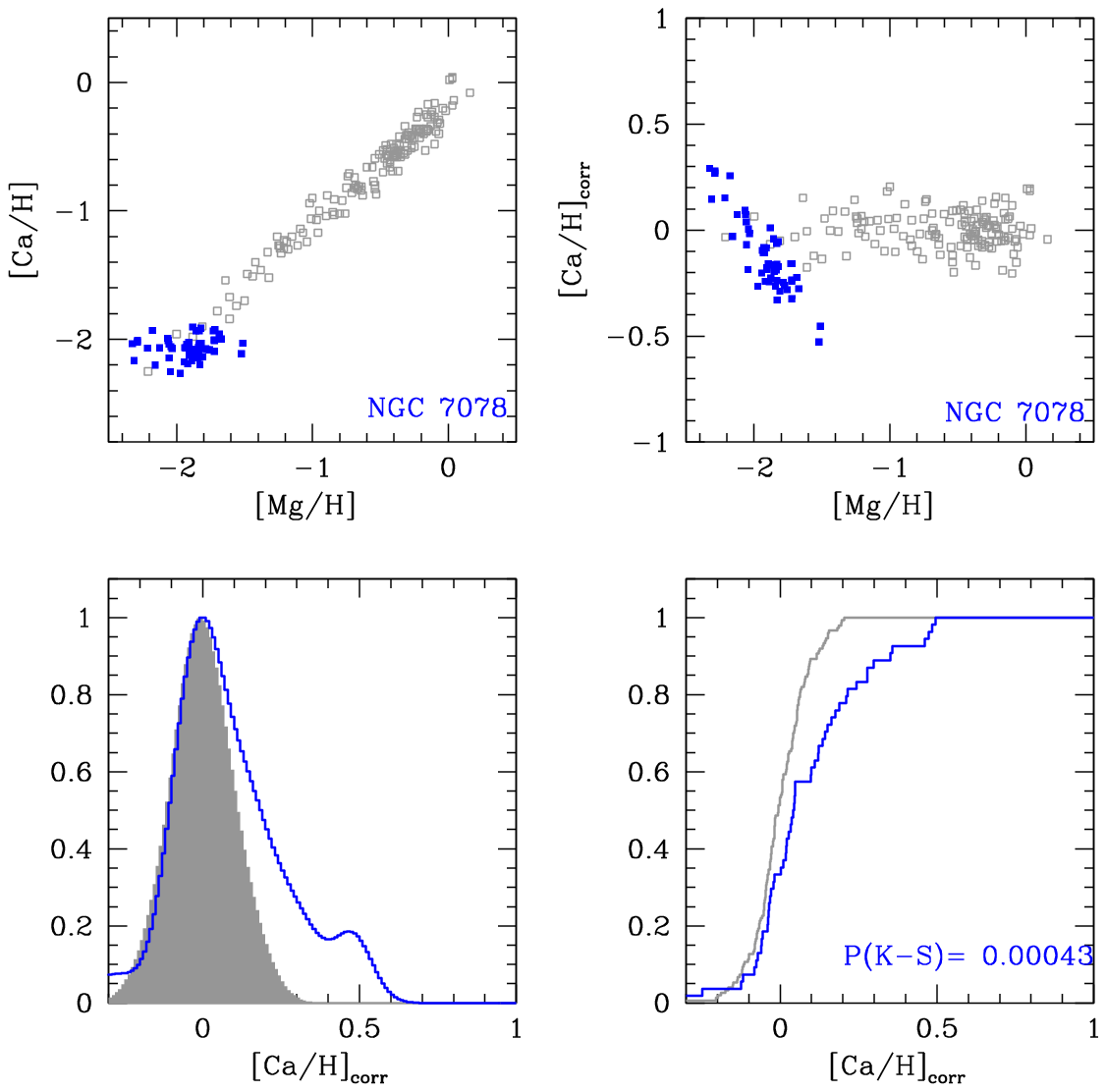}
\caption{Detection of Ca excesses in NGC~6752 (left panels) and NGC~7078 (right
panels). In the upper left panels GC stars (blue squares) are superimposed to
field stars (with typical error of 0.08 dex), and all abundances are linearised in the upper right panels (see text).
Generalised histograms of the [Ca/H] values for field (grey area) and GC stars
(blue area) are in the lower left panels and the cumulative distributions are
shown in the lower right panels, together with the probability of the two-tailed
K-S test.}
\label{f:censo}
\end{figure*}

We repeated the steps used by Carretta and Bragaglia (2021) on all the GCs of
their silver sample, that is the homogeneously analysed, but limited samples
of stars observed with the dedicated UVES-FLAMES fibres in Carretta et al.
(2009b). In the present analysis we significantly increased the samples by
exploiting also the homogeneous analysis of the GIRAFFE spectra. The results
found to be statistically significant are illustrated in Fig.~\ref{f:censo}. Stars
in NGC~6752 and NGC~7078 (blue squares) are superimposed to field stars
from Gratton et al. (2003: empty grey squares) in the [Ca/H]-[Mg/H] plane (upper
left panels). We then subtract the linear fit traced to field stars to both
samples to obtain linearised [Ca/H]$_{corr}$ values, for an easier evaluation
of excesses (upper right panels). The generalised histograms in the lower left
panels, aligned using the mode of the distributions, allow to estimate the excess of [Ca/H] of GC stars with respect to the
reference sample of unpolluted field stars. Finally, a K-S test is applied to
the cumulative distributions of field and GC stars in the lower right panels. 
The two-tailed probability is listed. Despite the evidence in Fig.~\ref{f:camg}
and the excess (although small) shown in Fig.~\ref{f:censo} the value for
NGC~6752 does not allow to safely reject the null hypothesis that the two
samples are extracted  from the same parent population. The test is instead even
formally significant in the case of NGC~7078, indicating an excess of Ca together
with a Mg depletion and signalling the outcome of proton-capture reactions at the
very high temperature regime.

\section{Discussion and summary}

In the present work we derived homogeneous abundances of Mg, Si, Ca, and Ti for a
large sample of RGB stars in 16 GCs covering most of the metallicity range of
Milky Way GCs. Thanks to the identical methodology used in the analysis, the
sample can be safely coupled to previous studies of individual GCs by our group,
assembling one of the largest sample of giants (more than 2600) in GCs with 
abundances from high resolution optical spectra. The extremely homogeneous
analysis minimises possible systematics. 

\begin{figure}
\centering
\includegraphics[bb=20 151 585 485, clip, scale=0.40]{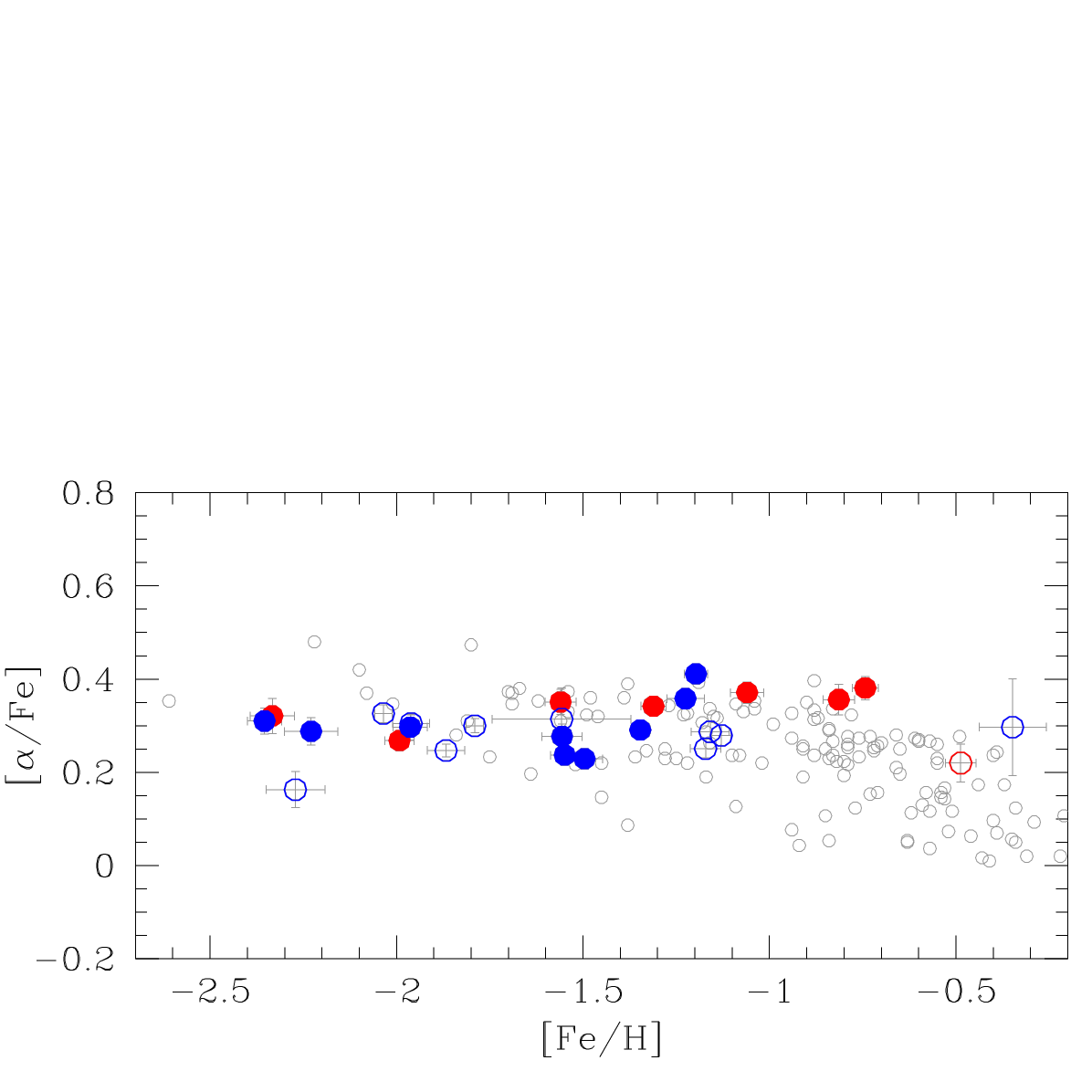}
\caption{Ratios [$\alpha$/Fe] (average of [Si/Fe], [Ca/Fe], [Ti/Fe]) as a
function of metallicity. Grey circles are field stars from Gratton et al.
(2003). Blue and red filled circles are accreted and in situ GCs in our extended
sample.}
\label{f:alphaGC}
\end{figure}

In our data the abundance of $\alpha-$elements is found to be
approximately constant in all studied GCs (Fig.~\ref{f:mediepfield}). This is
more evident in Fig.~\ref{f:alphaGC} where we show the mean $<[\alpha/$Fe]$>$
ratio computed as the average of [Si/Fe], [Ca/Fe], and [Ti/Fe], the three
elements which are the least affected by the abundance variations related to
MPs. The ratio  $<[\alpha/$Fe]$>$ is flat as a function of [Fe/H] up to the
metallicity of the GCs in the MW bulge (NGC~6388 and NGC~6441). The signature of
type Ia SNe (lower $<[\alpha/$Fe]$>$) is not apparent in all GCs examined
here, whereas the classical downturn at [Fe/H]=-1 dex is seen among field
stars, and in the [Si/Fe] ratios from APOGEE.

Thus, the constancy of $\alpha-$elements in GCs born in a variety of
environments implies that the effects of SN Ia are never seen in individual GCs,
apart from a few notable cases like $\omega$ Cen (see Gratton et al. 2004). 
Observationally this is supported by the small dispersions in [Fe/H] in almost all
GCs (see e.g. Carretta and Bragaglia 2025, Latour et al. 2025).
The main theories for the origin of MPs in GCs suggest that the formation of 
polluted stars happens on very short timescales (few million years) 
for enrichment from massive stars (e.g.Gieles et al. 2025), reaching up to ~80-100 Myr for the AGB scenario as
this is the timescale for the first SNe Ia to stop further generations of stars
forming within the cluster (D'Ercole et al. 2008).
The near homogeneity of heavy elements in GCs suggests that these objects were
already gas free when SN exploded or lost the gas following the explosions. 
Analytical models computed to reproduce the observed metallicity spreads
advocate that star formation ended in GCs after 3-4 Myr (Wirth et al. 2022),
hence most of the enrichment should take place within a few Myr, well before 
the explosions of the first thermonuclear SNe. 

The scatter in the mean values in Fig.~\ref{f:mediepfield} is compatible with a
stochastic enrichment from a few core-collapse SNe, as first noted by Carney
(1996). This is the main reason why the origin of in situ and accreted GCs
cannot be recognised from the abundance of $\alpha-$elements alone.

In more than half of the GC sample we found star to star variations in Si 
correlated to changes in the Na abundance. This evidence, corroborated by
robust statistical tests, indicates that in these GCs the stars enriching
the gas that formed the polluted population reached temperatures high enough for
a significant leakage out of the MgAl chain on $^{28}$Si.
According to models based on rates from NACRE (Arnould et al. 1999), this implies
to exceed a threshold temperature of about 65 MK in these still unidentified
sources of proton-capture reactions. From our present data, this seems to be a
rather common occurrence.

Less frequent is the presence of large changes in the Mg abundance established
by core-collapse SNe. Significant variations in Mg are only found in GCs that 
are metal-poor, massive or both (like NGC~7078, where a few stars show 
[Mg/Fe]$<$0.0), confirming the early findings by Carretta et al. (2009b), and 
later additional evidence (e.g. M\'esz\'aros et al. 2020), on a more 
statistically robust sample.
Anyway, the higher temperatures involved in the MgAl chain with respect to those
required to produce the widespread Na-O anti-correlation and the much larger
primordial abundances of Mg and Si in GC stars imply that MPs have a small
impact on the average abundances of $\alpha-$elements in GCs.

\begin{figure}
\centering
\includegraphics[scale=0.40]{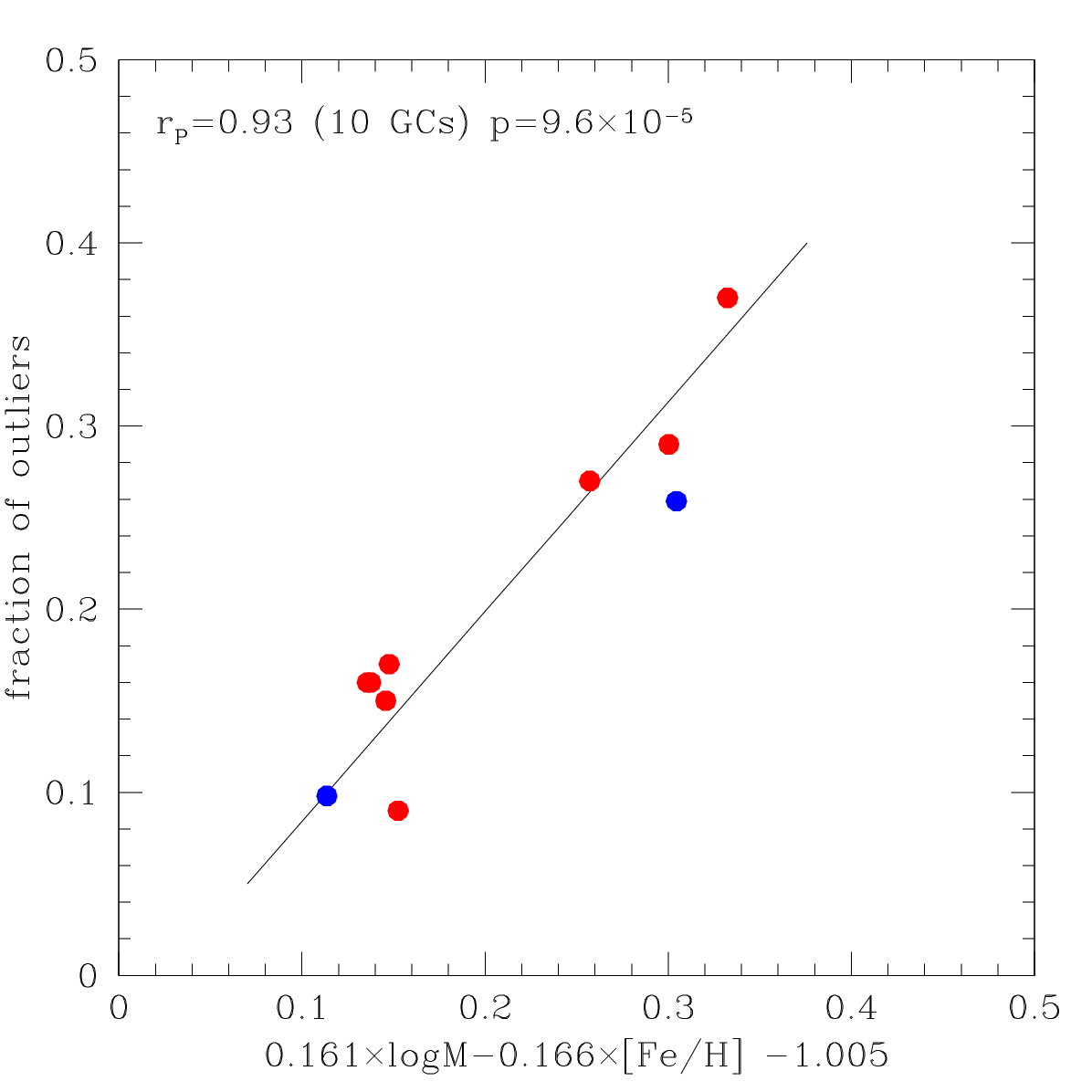}
\caption{Fraction of stars with excesses of Ca in 8 GCs from the census in 
Carretta and Bragaglia (2021; red points) as a function of a linear combination
of GC mass (Baumgardt et al. 2018) and metallicity (Harris 2010). Blue points are for NGC~7078 and NGC~6752 from the
present work. A linear fit is given by the solid line, with the Pearson's
correlation coefficient and the two-tailed probability listed in the panel.} 
\label{f:bivarcenso}
\end{figure}

Finally, we found the signature of the action of proton-capture reactions at
very high temperature in two out of 16 GCs. Following the approach detailed in 
Carretta and Bragaglia (2021) we detected excesses of Ca with respect to field 
stars of similar metallicity in NGC~6752 and NGC~7078. These excesses
possibly trace the production of Ca activated on Ar nuclei by proton-captures
in a particularly energetic regime (Ventura et al. 2012). When testing the
null hypothesis that the sample populations of GC stars and unpolluted
field stars are drawn from the same parent distribution a K-S test unambiguously
show that the result for NGC~7078 has a high significance. The case for NGC~6752
is formally not statistically significant. However, the amount of the excess can be 
quantified by computing the number of outliers in the GC linearised
distribution. Following Carretta and Bragaglia (2021) the outliers are defined
as those whose corrected ratios [Ca/H]$_{corr}$ exceed the 3$\sigma$ range with
respect to the average for field stars. The measured fractions of outliers 
(0.098 for NGC~6752 and 0.259 for NGC~7078) fit very well the relation with a
linear combination of cluster mass and metallicity found in Carretta and
Bragaglia (2021). In Fig.~\ref{f:bivarcenso} we plot in red the eight GCs where a
significant excess of Ca was found by Carretta and Bragaglia (2021). By adding
the GCs from the present work (blue points) we repeated the fit finding a
correlation with very high significance as indicated by the two-tailed
probability listed in the figure, with a net improvement from the 
p-value ($3.0\times 10^{-4}$) found in Carretta and Bragaglia (2021).

The dependence on global cluster mass and metallicity is the same combination
affecting other characteristics of MPs in GCs like the amount of Al produced in
the polluters (Carretta et al. 2009b) or the minimum abundance of O observed in
GCs [O/Fe]$_{min}$ (Carretta et al. 2009a). This bivariate dependence seems to
be thus effective in driving the behaviour of MPs (Na-O and Mg-Al 
anti-correlations, Ca excesses) over several regimes distinct by the involved
temperature range in the putative polluters.

The era of artisan surveys of abundances is ending. Upcoming large-scale
spectroscopic surveys such as 4MOST and WEAVE will deliver high-quality data for
million of stars and a large fraction of Galactic stellar clusters. Our precise,
large, and very homogeneous sample of giant stars in GCs, all certified members
by radial velocity and abundance, may be a useful benchmark for these
``industrial" surveys, contributing to their calibration and validation.

\begin{acknowledgements}
This research has made large use of the SIMBAD database (in particular  Vizier),
operated at CDS, Strasbourg, France, of the NASA's Astrophysical Data System,
and TOPCAT (Taylor 2005). I especially thanks Angela Bragaglia for several
valuable suggestions. I also aknowledge funding from Bando Astrofisica
Fondamentale INAF 2023 (PI A. Vallenari) and from Prin INAF 2019 (PI S.
Lucatello).
\end{acknowledgements}

\FloatBarrier

\begin{appendix}
\onecolumn
\section{Abundances as a function of effective temperature}

We show plots of all the derived abundances of Mg, Si, Ca, and Ti as a function
of the effective temperature in all our 16 program GCs.

%\FloatBarrier
\begin{figure*}[h]
\centering
\includegraphics[scale=0.23]{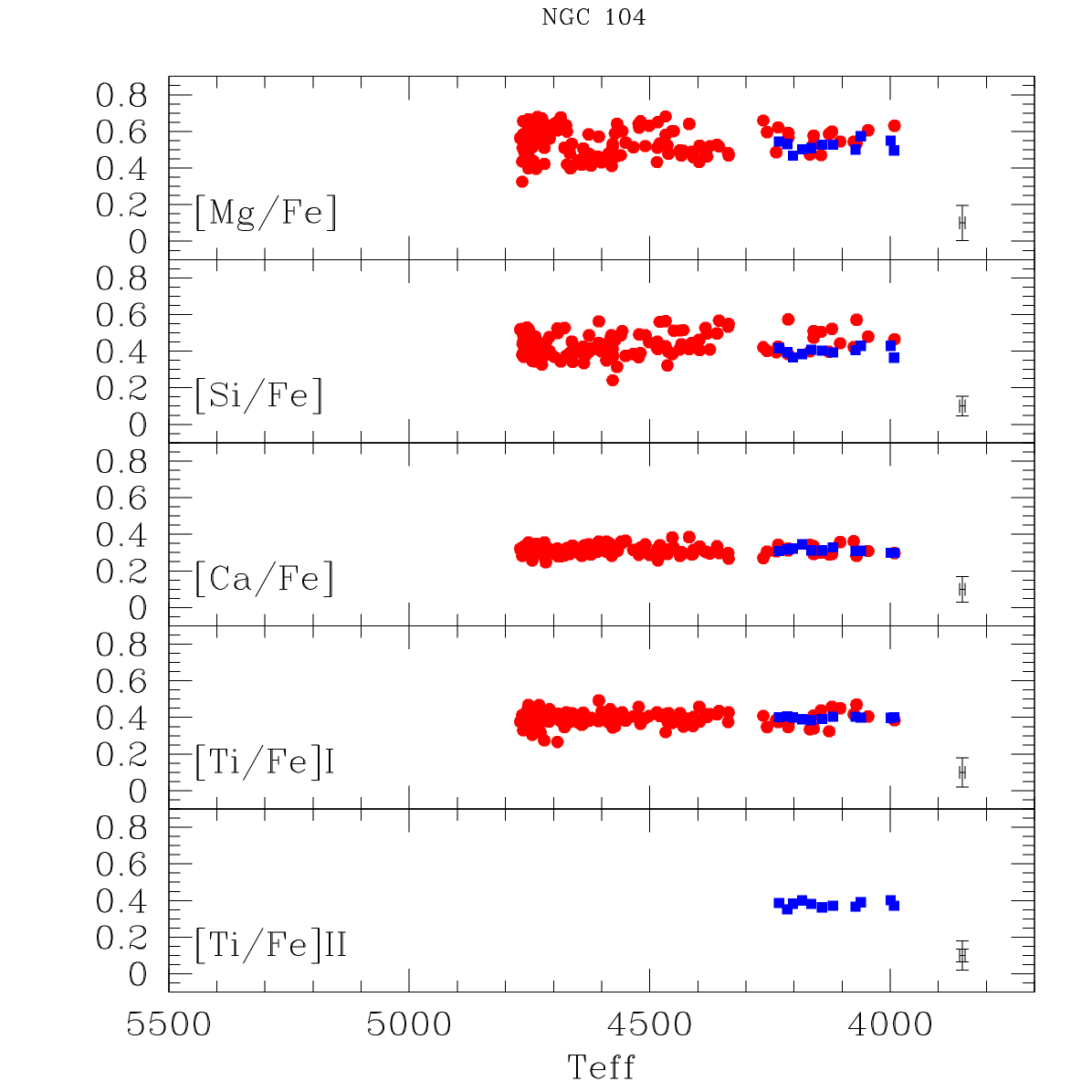}\includegraphics[scale=0.23]{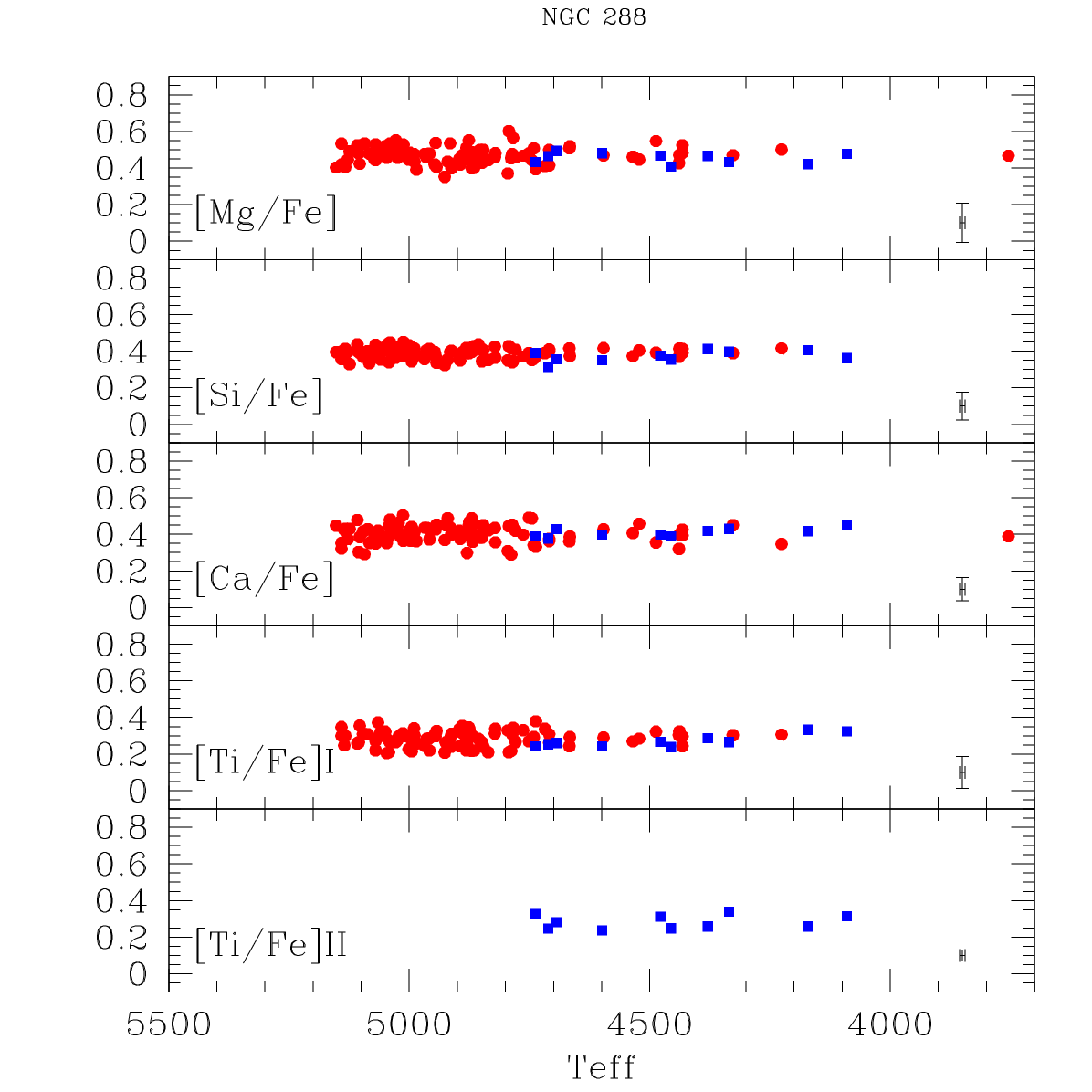}\includegraphics[scale=0.23]{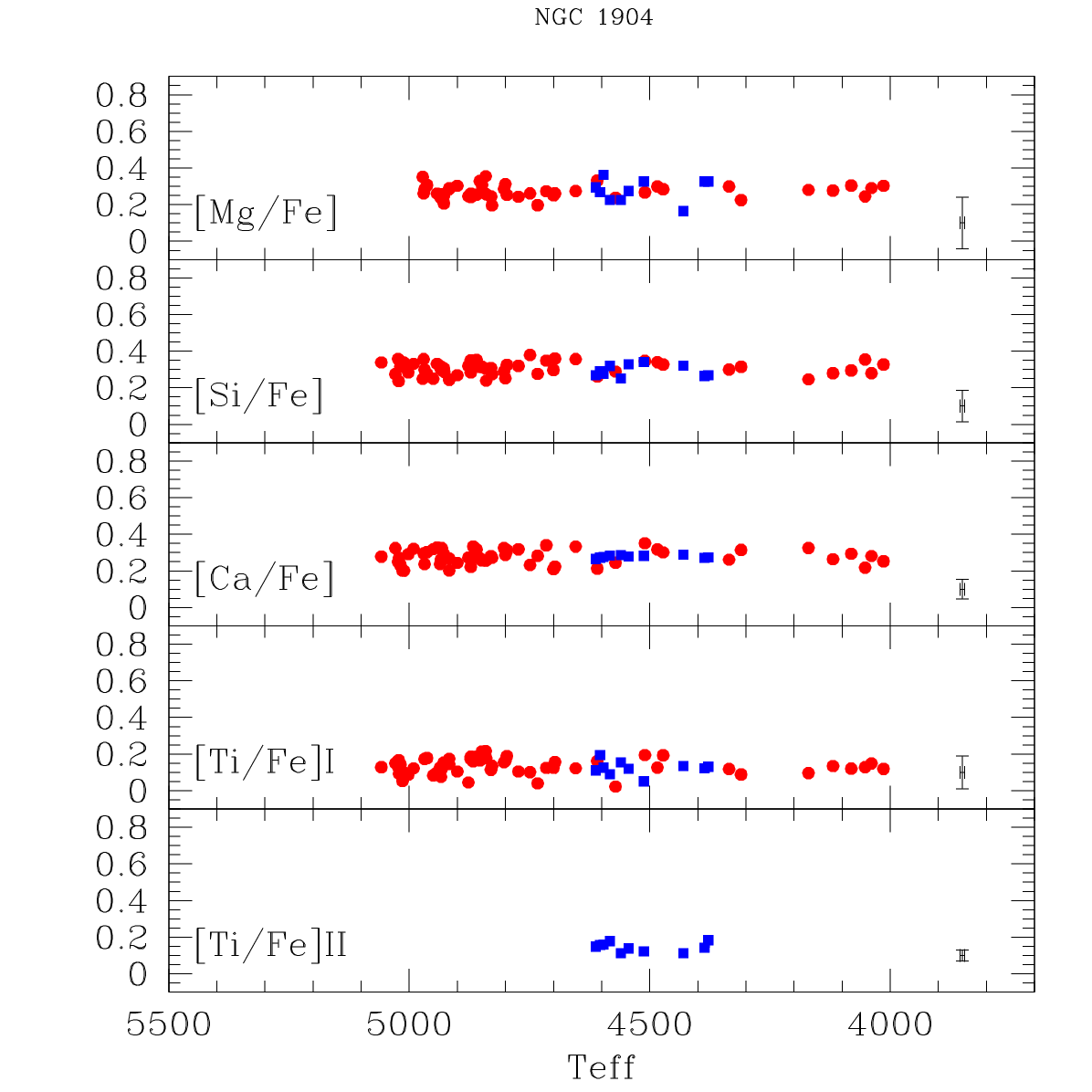}\includegraphics[scale=0.23]{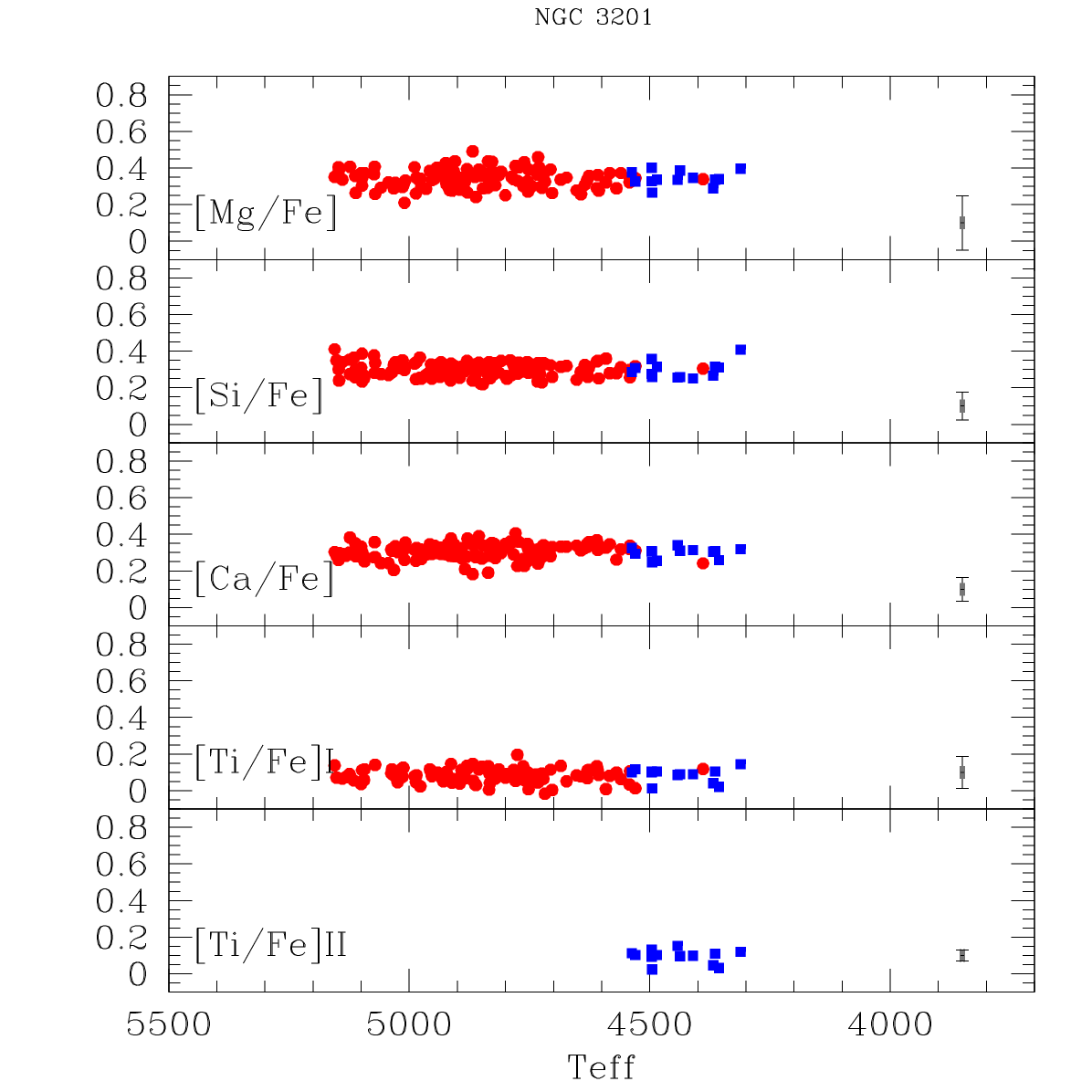}
\includegraphics[scale=0.23]{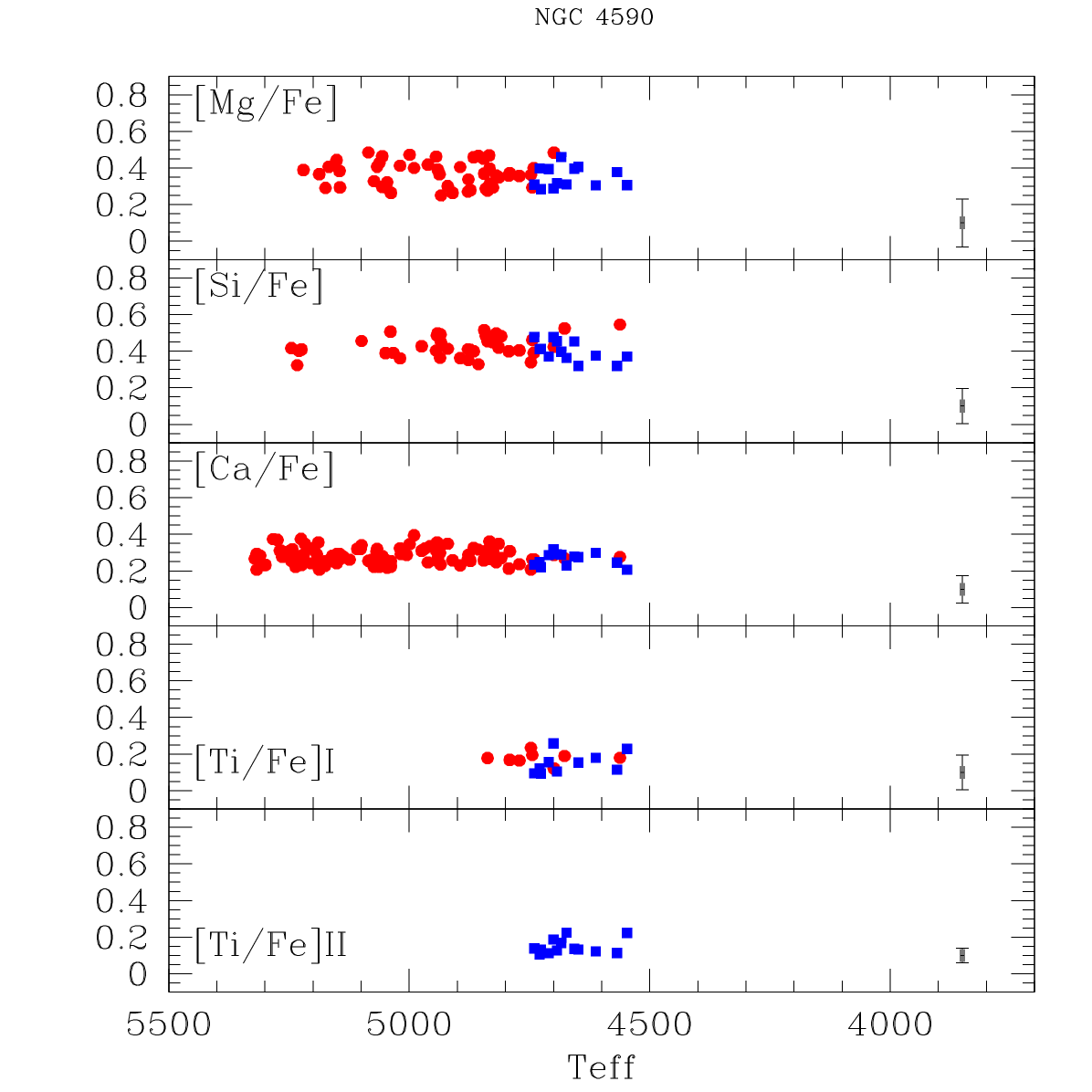}\includegraphics[scale=0.23]{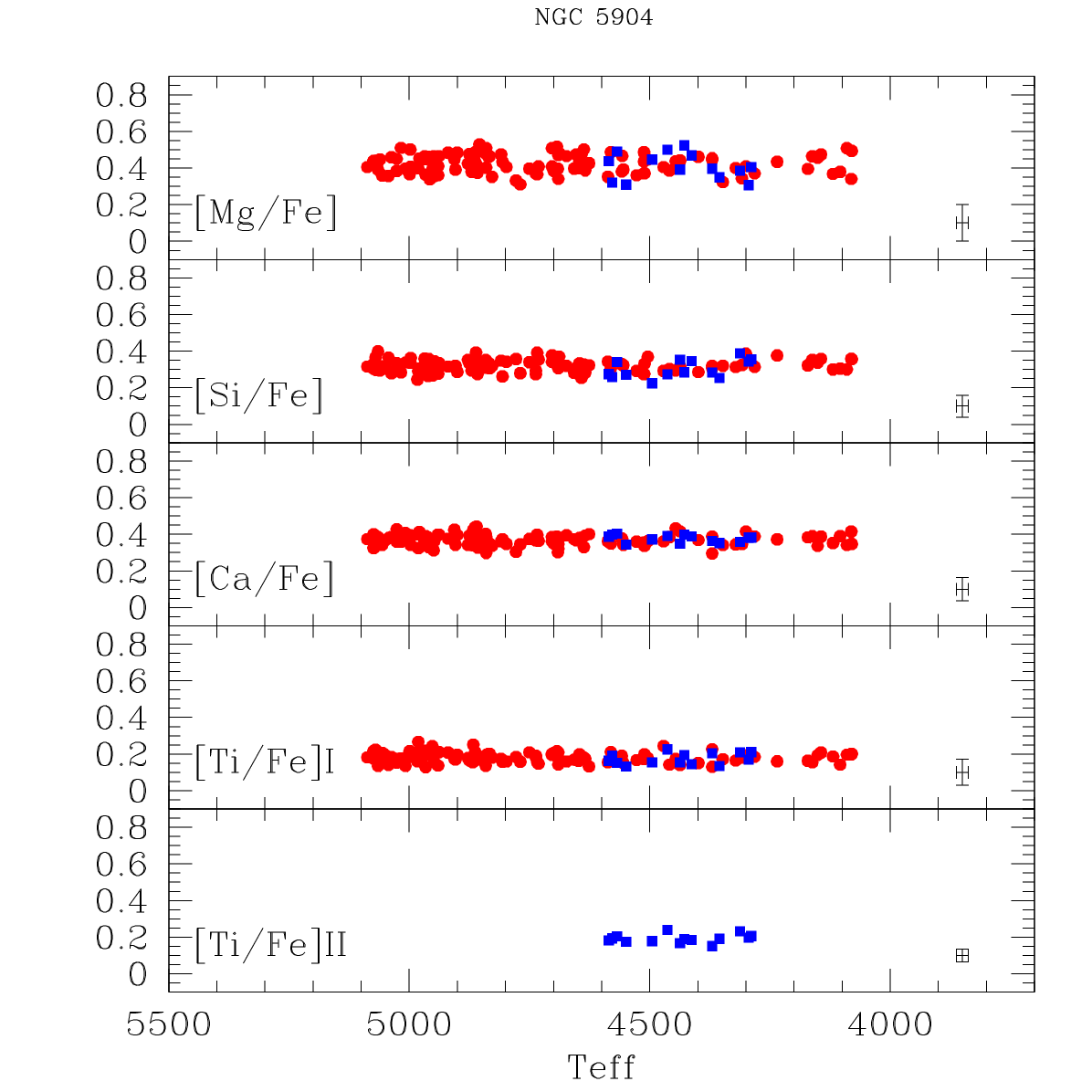}\includegraphics[scale=0.23]{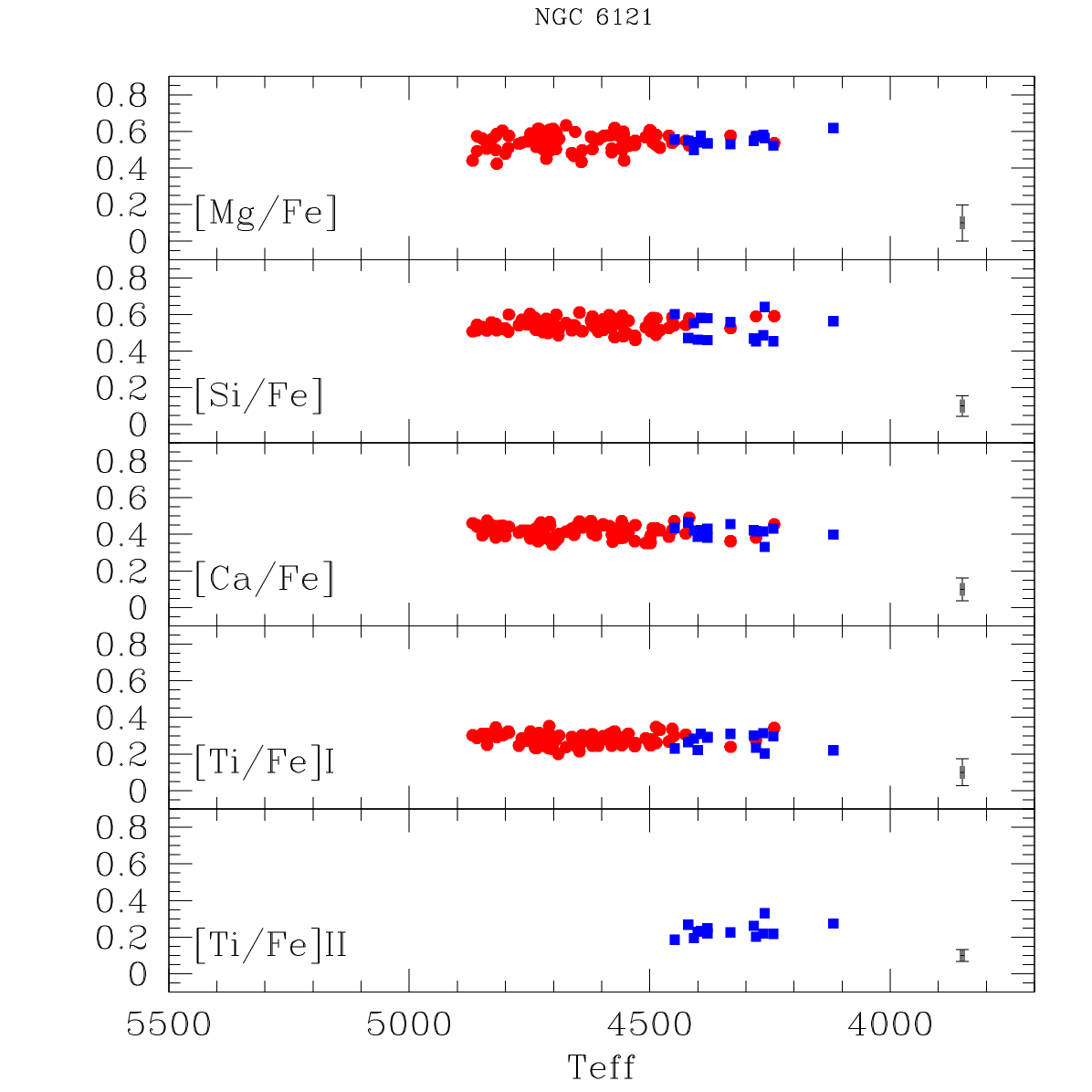}\includegraphics[scale=0.23]{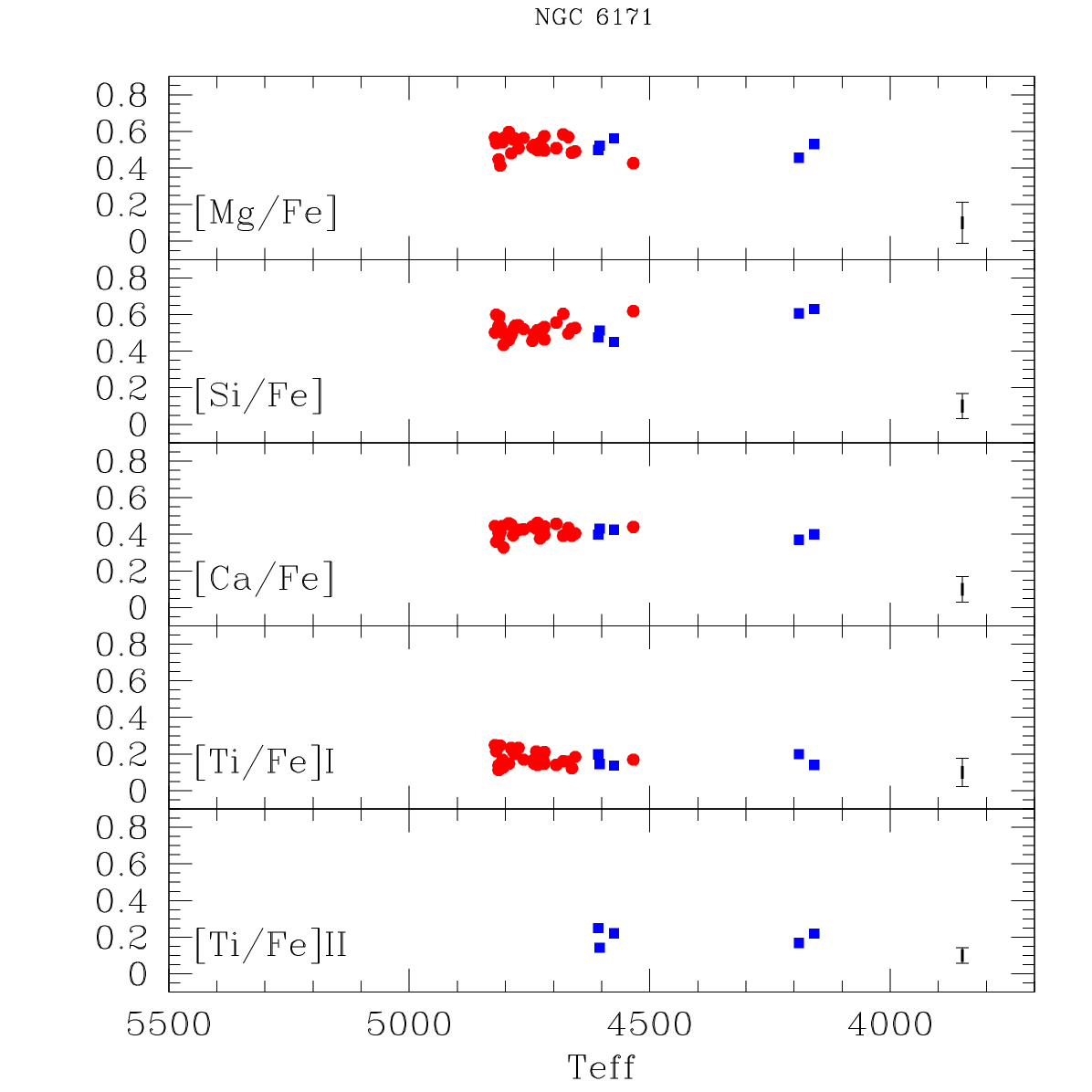}
\includegraphics[scale=0.23]{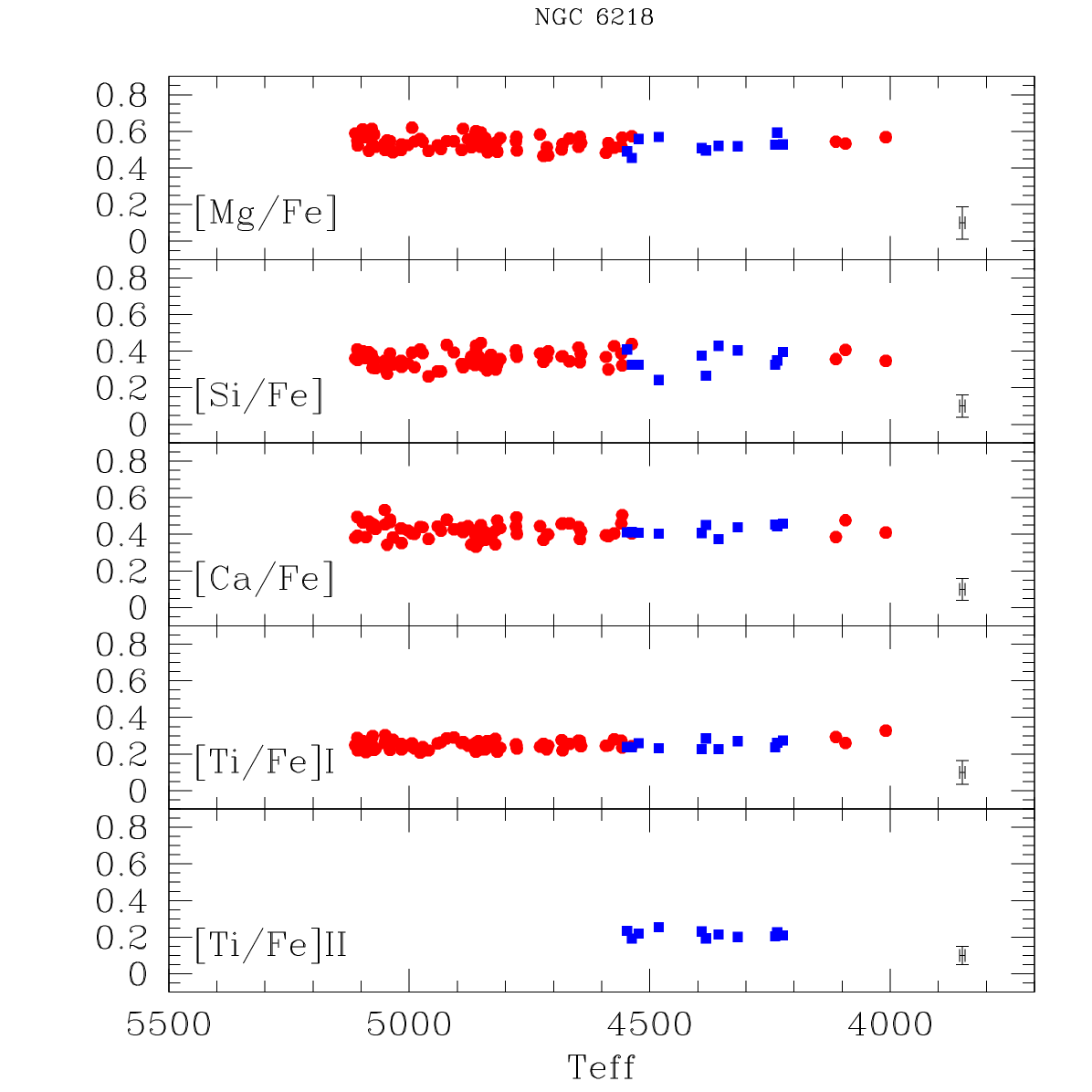}\includegraphics[scale=0.23]{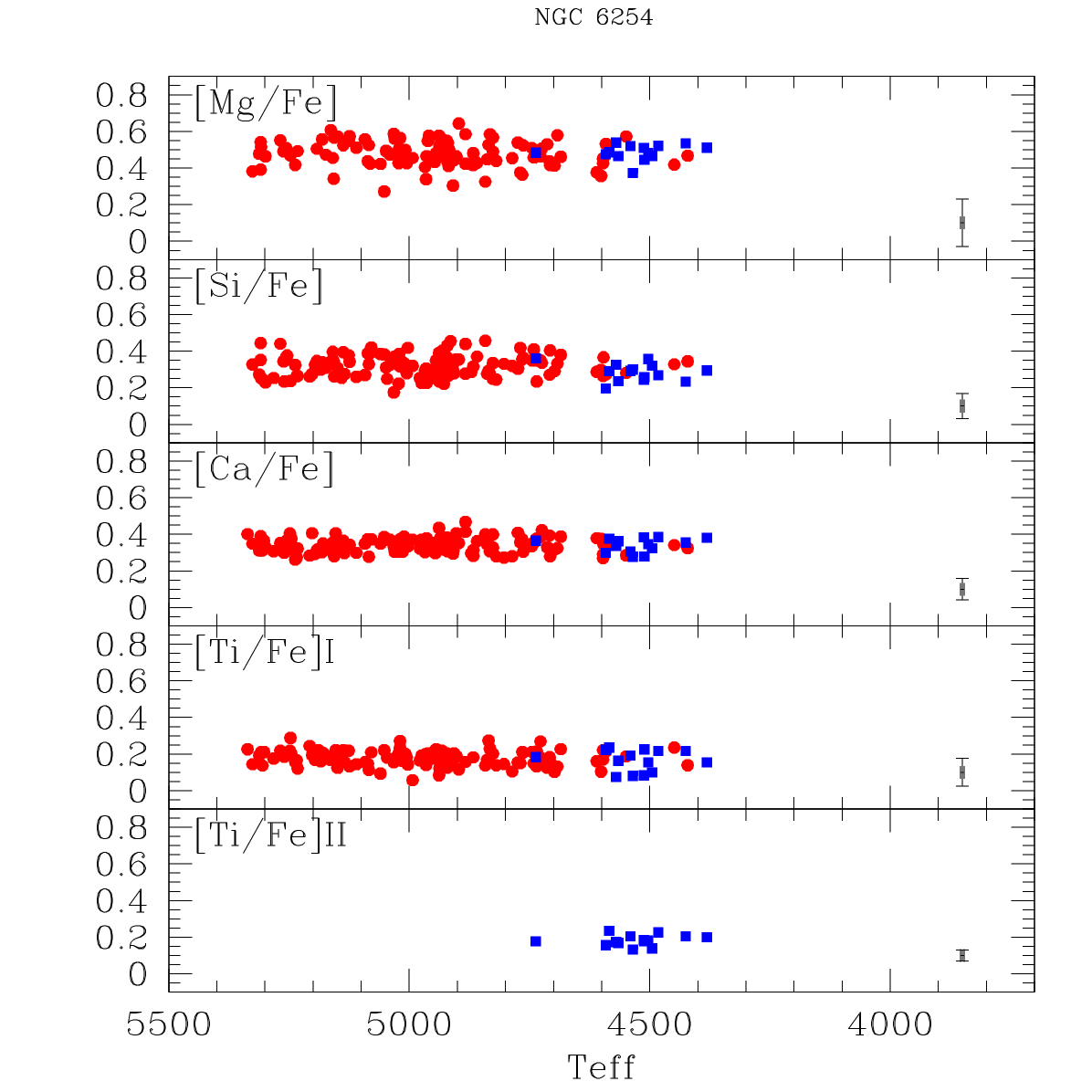}\includegraphics[scale=0.23]{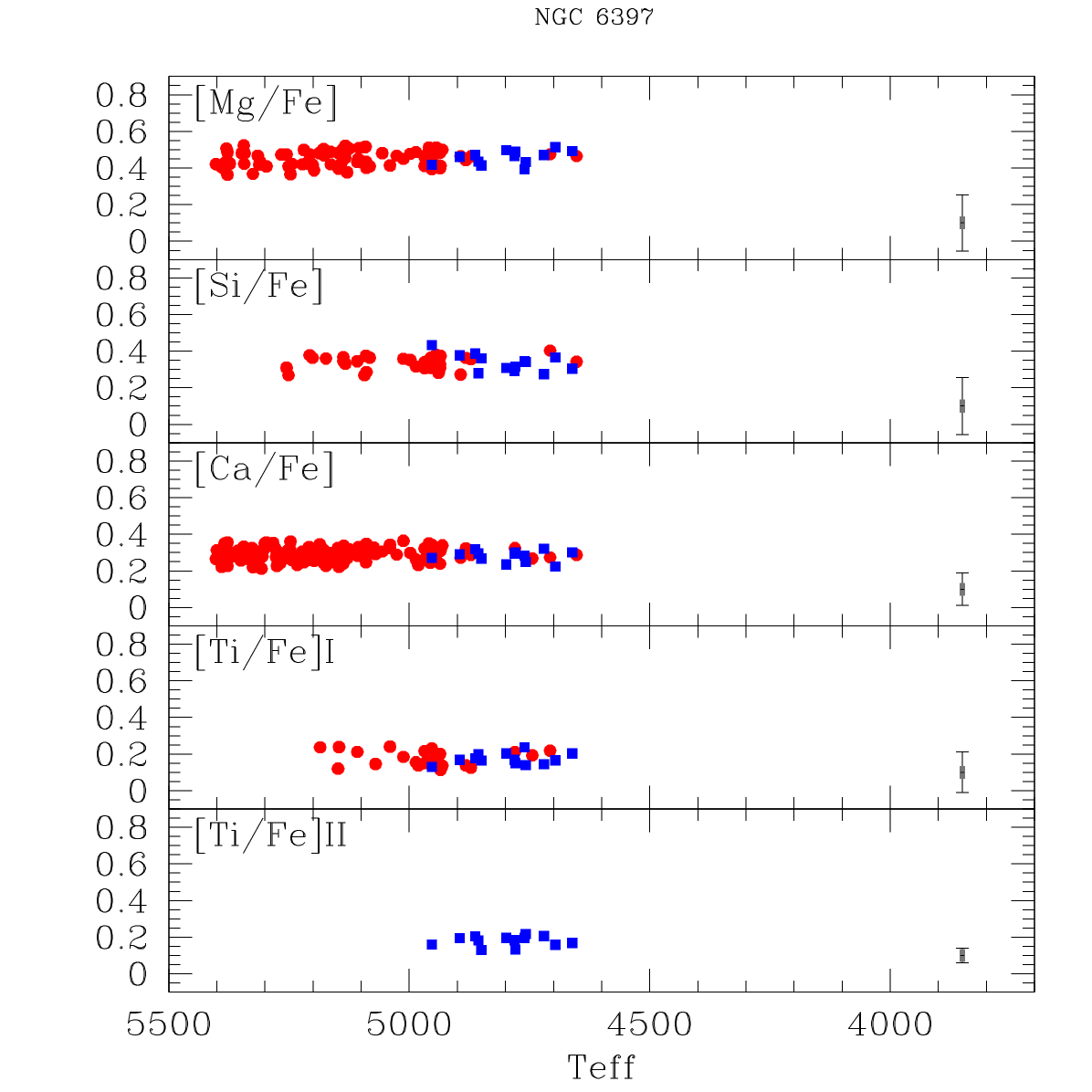}\includegraphics[scale=0.23]{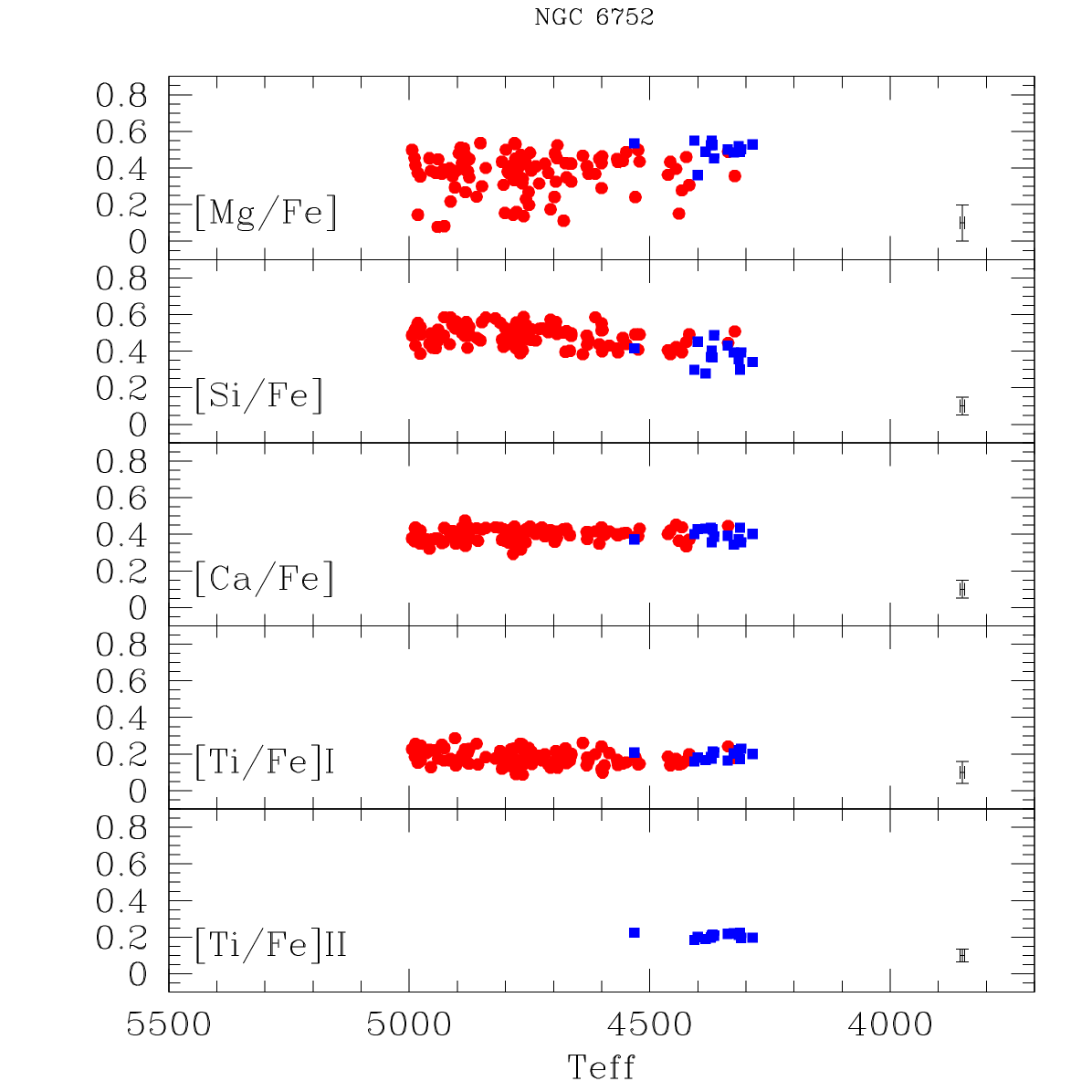}
\includegraphics[scale=0.23]{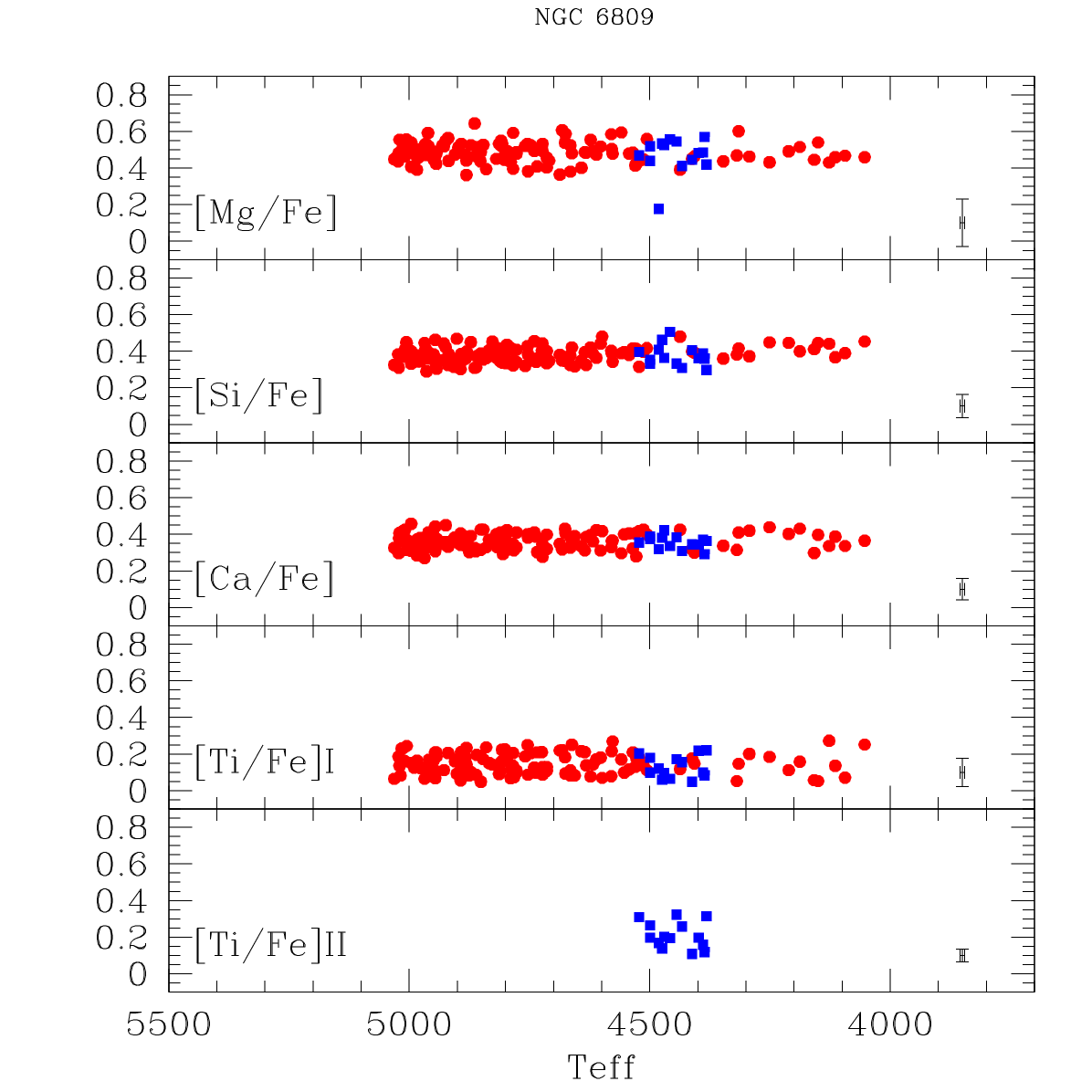}\includegraphics[scale=0.23]{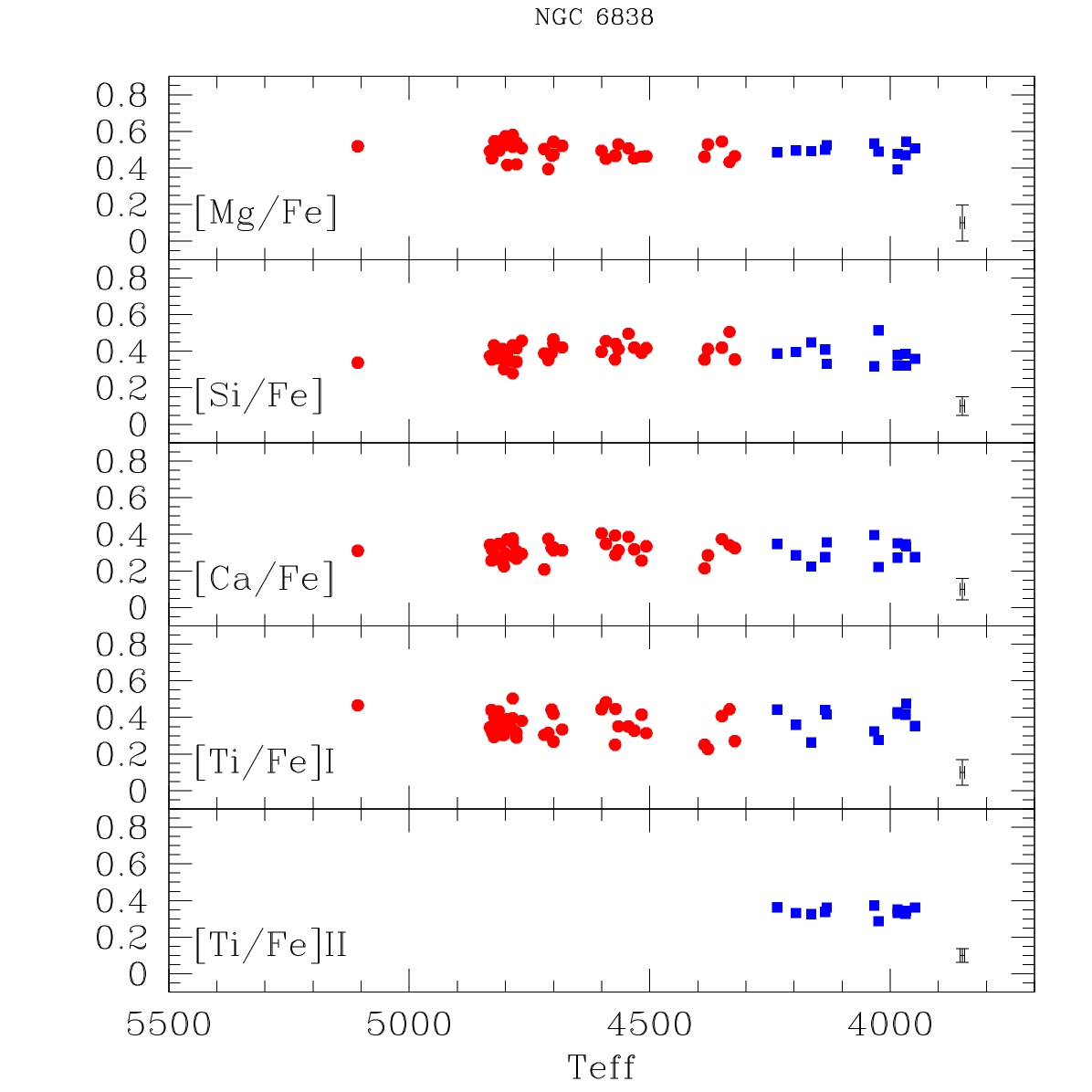}\includegraphics[scale=0.23]{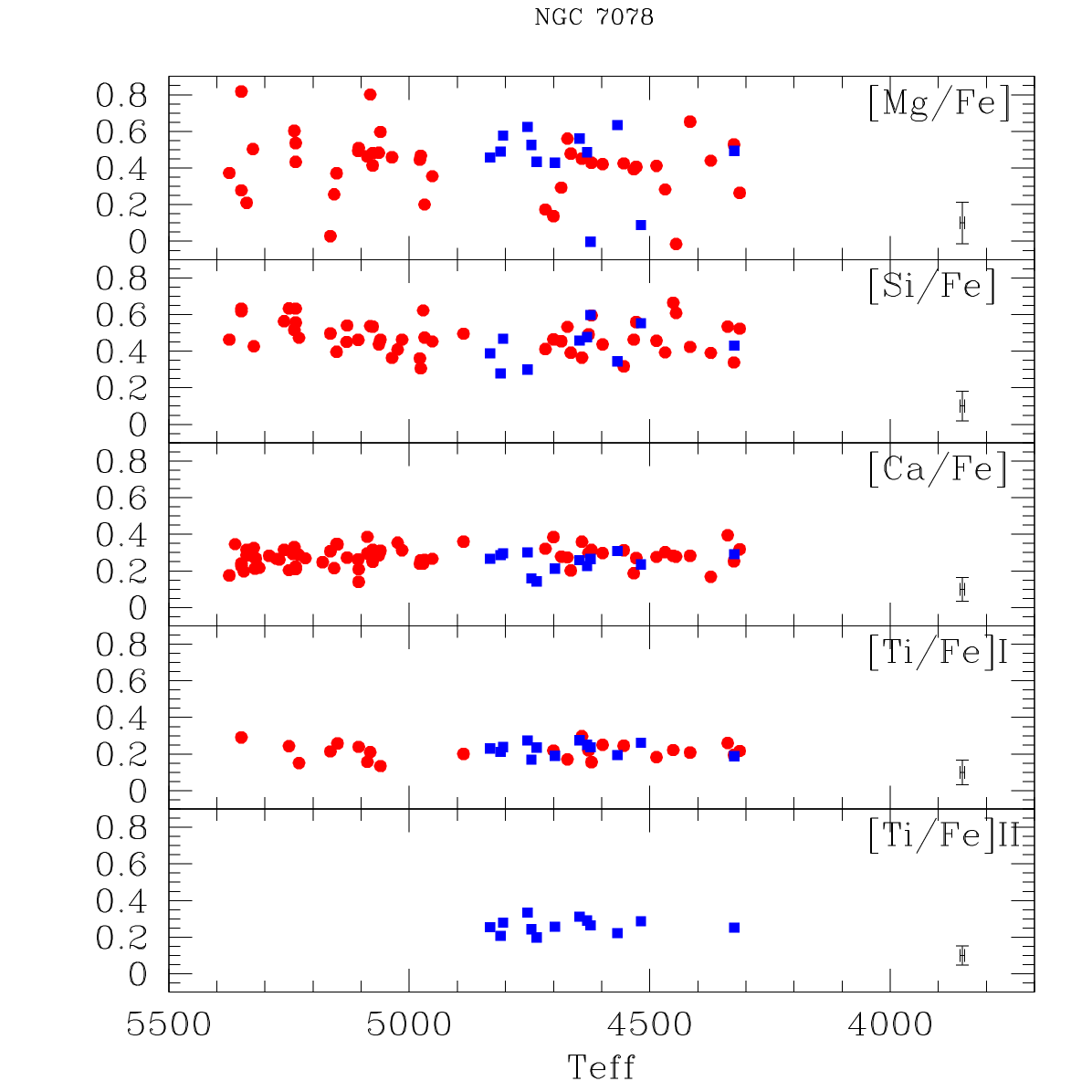}\includegraphics[scale=0.23]{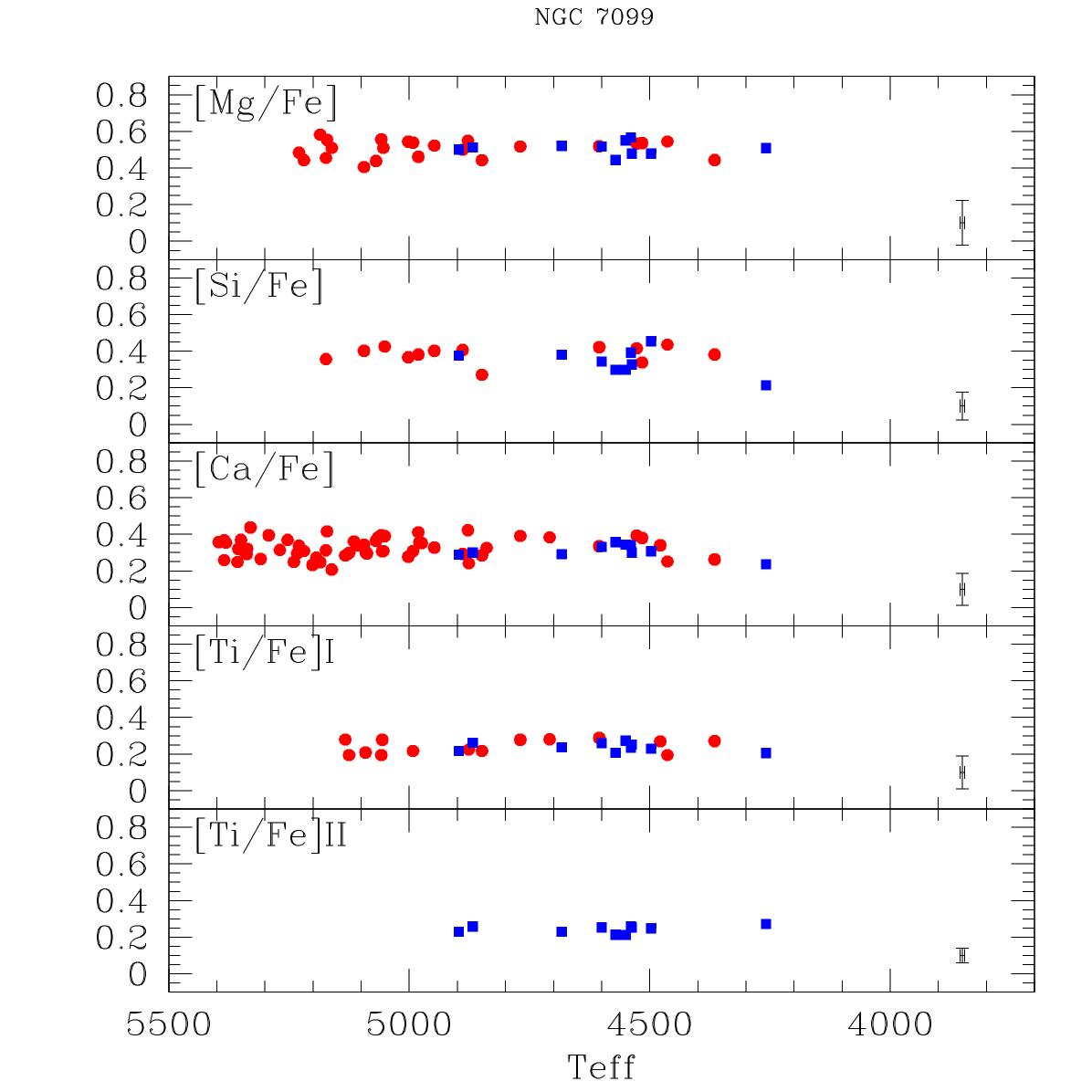}
\caption{Abundance ratios [Mg/Fe],[Si/Fe],
[Ca/Fe],[Ti/Fe]~{\sc i},and [Ti/Fe]~{\sc ii} from this work
as a function of temperature. Blue squares are for UVES stars and red circles for 
GIRAFFE stars. Error bars from Table~\ref{t:errabuTOT} are those referred to
GIRAFFE (except for [Ti/Fe]~{\sc ii}).}
\label{f:alphateff}
\end{figure*}

\FloatBarrier
\newpage
\section{Tables of sensitivity and internal errors}

\begin{table*}[h]
\tiny
\centering
\setlength{\tabcolsep}{1.0mm}
\caption{Sensitivities to errors in the atmospheric parameters (UVES)}
\begin{tabular}{llllllcrrrrr}
\hline
        & \multicolumn{5}{c}{$\Delta$T$_{\rm eff}=50$~K} & &\multicolumn{5}{c}{$\Delta V_t = +0.1$~km/s} \\
	\cline{2-6} \cline{8-12}\\

cluster &$\Delta$[Mg/Fe]&$\Delta$[Si/Fe]&$\Delta$[Ca/Fe]&$\Delta$[Ti/Fe]I& $\Delta$[Ti/Fe]II &  &$\Delta$[Mg/Fe]&$\Delta$[Si/Fe]&$\Delta$[Ca/Fe]&$\Delta$[Ti/Fe]I& $\Delta$[Ti/Fe]II  \\
NGC 104 & $-$0.023& $-$0.071&   +0.032& +0.054&   +0.027 &  &  +0.010&   +0.017& $-$0.029& $-$0.037 & $-$0.035\\
NGC 288 & $-$0.025& $-$0.054&   +0.009& +0.029&   +0.016 &  &  +0.014&   +0.020& $-$0.017&   +0.007 & $-$0.019\\
NGC~1904& $-$0.016& $-$0.045&   +0.002& +0.024&   +0.018 &  &  +0.017&   +0.022& $-$0.003&   +0.019 & $-$0.018\\
NGC~3201& $-$0.017& $-$0.052&   +0.005& +0.026&   +0.015 &  &  +0.010&   +0.017& $-$0.014&   +0.011 & $-$0.014\\
NGC~4590& $-$0.014& $-$0.025& $-$0.005& +0.020&   +0.018 &  &$-$0.052& $-$0.051& $-$0.062& $-$0.051 & $-$0.004\\
NGC~5904& $-$0.018& $-$0.055&   +0.008& +0.029&   +0.020 &  &  +0.014&   +0.021& $-$0.017&   +0.009 & $-$0.020\\
NGC~6121& $-$0.018& $-$0.062&   +0.017& +0.040& $-$0.034 &  &  +0.013&   +0.019& $-$0.024& $-$0.003 & $-$0.011\\
NGC~6171& $-$0.019& $-$0.060&   +0.018& +0.045&   +0.025 &  &  +0.009&   +0.013& $-$0.022& $-$0.021 & $-$0.024\\
NGC~6218& $-$0.014& $-$0.058&   +0.011& +0.044&   +0.012 &  &$-$0.002&   +0.011& $-$0.028& $-$0.023 & $-$0.022\\
NGC~6254& $-$0.013& $-$0.040&   +0.007& +0.041&   +0.011 &  &  +0.006&   +0.014& $-$0.012& $-$0.007 & $-$0.015\\
NGC~6397& $-$0.008& $-$0.022& $-$0.005& +0.032&   +0.019 &  &$-$0.011&   +0.007& $-$0.005& $-$0.006 & $-$0.010\\
NGC~6752& $-$0.017& $-$0.058&   +0.005& +0.037&   +0.015 &  &  +0.004&   +0.020& $-$0.019& $-$0.012 & $-$0.017\\
NGC~6809& $-$0.014& $-$0.043&   +0.000& +0.039&   +0.014 &  &  +0.005&   +0.011& $-$0.001& $-$0.006 & $-$0.017\\
NGC~6838& $-$0.045& $-$0.091&   +0.010& +0.043&   +0.016 &  &  +0.008&   +0.017& $-$0.023& $-$0.044 & $-$0.030\\
NGC~7078& $-$0.015& $-$0.027& $-$0.012& +0.026&   +0.023 &  &$-$0.008&   +0.006& $-$0.003& $-$0.007 & $-$0.011\\
NGC~7099& $-$0.010& $-$0.024& $-$0.006& +0.036&   +0.000 &  &$-$0.005&   +0.006& $-$0.003& $-$0.005 & $-$0.009\\
\hline
        & \multicolumn{5}{c}{$\Delta \log g = +0.2$~dex} &\multicolumn{1}{c}{} &\multicolumn{5}{c}{$\Delta$ [A/H] = +0.1~dex} \\
	\cline{2-6} \cline{8-12}\\ 
	 
cluster &$\Delta$[Mg/Fe]&$\Delta$[Si/Fe]&$\Delta$[Ca/Fe]&$\Delta$[Ti/Fe]I& $\Delta$[Ti/Fe]II &  &$\Delta$[Mg/Fe]&$\Delta$[Si/Fe]&$\Delta$[Ca/Fe]&$\Delta$[Ti/Fe]I& $\Delta$[Ti/Fe]II  \\	
NGC 104 & $-$0.005& +0.035& $-$0.040& $-$0.014& $-$0.026 &  &   +0.002& $-$0.006& $-$0.011& $-$0.014 & $-$0.006\\
NGC 288 &   +0.003& +0.037& $-$0.011& $-$0.002& $-$0.012 &  &   +0.000&   +0.015& $-$0.008& $-$0.007 &   +0.002\\
NGC~1904& $-$0.001& +0.031& $-$0.010& $-$0.004& $-$0.010 &  & $-$0.002&   +0.011& $-$0.007& $-$0.005 &   +0.000\\
NGC~3201&   +0.000& +0.035& $-$0.012& $-$0.005& $-$0.010 &  &   +0.000&   +0.015& $-$0.006& $-$0.005 &   +0.001\\
NGC~4590& $-$0.004& +0.011& $-$0.008& $-$0.011& $-$0.006 &  & $-$0.005& $-$0.002& $-$0.006& $-$0.007 &   +0.004\\
NGC~5904& $-$0.001& +0.036& $-$0.017& $-$0.006& $-$0.011 &  &   +0.000&   +0.016& $-$0.007& $-$0.007 &   +0.002\\
NGC~6121&   +0.002& +0.043& $-$0.019&   +0.000&   +0.093 &  &   +0.000&   +0.017& $-$0.009& $-$0.012 &   +0.025\\
NGC~6171&   +0.000& +0.041& $-$0.030& $-$0.014& $-$0.015 &  &   +0.000&   +0.015& $-$0.006& $-$0.018 &   +0.000\\
NGC~6254& $-$0.016& +0.037& $-$0.013& $-$0.009& $-$0.010 &  &   +0.003&   +0.016& $-$0.009& $-$0.010 &   +0.004\\
NGC~6254& $-$0.002& +0.029& $-$0.011& $-$0.011& $-$0.005 &  & $-$0.003&   +0.011& $-$0.008& $-$0.012 &   +0.006\\
NGC~6397& $-$0.018& +0.009& $-$0.009& $-$0.015& $-$0.004 &  & $-$0.010& $-$0.003& $-$0.009& $-$0.012 &   +0.004\\
NGC~6752& $-$0.019& +0.034& $-$0.015& $-$0.014& $-$0.014 &  &   +0.000&   +0.015& $-$0.009& $-$0.012 & $-$0.004\\
NGC~6809& $-$0.008& +0.024& $-$0.015& $-$0.018& $-$0.010 &  & $-$0.004&   +0.010& $-$0.008& $-$0.013 &   +0.004\\
NGC~6838&   +0.011& +0.049& $-$0.033& $-$0.009& $-$0.022 &  &   +0.004&   +0.015& $-$0.009& $-$0.012 & $-$0.003\\
NGC~7078& $-$0.011& +0.012& $-$0.006& $-$0.015& $-$0.009 &  & $-$0.009& $-$0.003& $-$0.008& $-$0.012 &   +0.002\\
NGC~7099& $-$0.017& +0.000& $-$0.015& $-$0.022& $-$0.010 &  & $-$0.011& $-$0.004& $-$0.011& $-$0.016 &   +0.003\\
\hline
\end{tabular}
\label{t:sensAuves}
\end{table*}

\begin{table*}[h]
\tiny
\centering
\setlength{\tabcolsep}{1.0mm}
\caption{Sensitivities to errors in the atmospheric parameters (GIRAFFE)}
\begin{tabular}{lllllcrrrr}
\hline
        & \multicolumn{4}{c}{$\Delta$T$_{\rm eff}=50$~K} & &\multicolumn{4}{c}{$\Delta V_t = +0.1$~km/s} \\
	\cline{2-5} \cline{7-10}\\

cluster &$\Delta$[Mg/Fe]&$\Delta$[Si/Fe]&$\Delta$[Ca/Fe]&$\Delta$[Ti/Fe]I &  &$\Delta$[Mg/Fe]&$\Delta$[Si/Fe]&$\Delta$[Ca/Fe]&$\Delta$[Ti/Fe]I  \\
NGC 104 & $-$0.006& $-$0.047&   +0.027&   +0.045 &  &+0.012&   +0.023& $-$0.016& $-$0.012 \\
NGC 288 & $-$0.020& $-$0.040& $-$0.001&   +0.013 &  &+0.016& $-$0.019& $-$0.005&   +0.012 \\
NGC~1904& $-$0.021& $-$0.045& $-$0.004&   +0.013 &  &+0.018&   +0.024& $-$0.002&   +0.017 \\
NGC~3201& $-$0.021& $-$0.044& $-$0.006&   +0.013 &  &+0.015&   +0.019& $-$0.002&   +0.013 \\
NGC~4590& $-$0.020& $-$0.029& $-$0.012&   +0.017 &  &+0.009&   +0.010& $-$0.003&   +0.009 \\
NGC~5904& $-$0.020& $-$0.046&   +0.001&   +0.017 &  &+0.017&   +0.024& $-$0.007&   +0.013 \\
NGC~6121& $-$0.014& $-$0.048&   +0.012& $-$0.049 &  &+0.016&   +0.024& $-$0.009&   +0.024 \\
NGC~6171& $-$0.013& $-$0.047&   +0.013&   +0.027 &  &+0.013&   +0.019& $-$0.007&   +0.005 \\
NGC~6218& $-$0.020& $-$0.042& $-$0.001&   +0.016 &  &+0.008&   +0.015& $-$0.011&   +0.007 \\
NGC~6254& $-$0.019& $-$0.035& $-$0.007&   +0.009 &  &+0.010&   +0.016& $-$0.004&   +0.012 \\
NGC~6397& $-$0.019& $-$0.024& $-$0.012&   +0.011 &  &+0.006&   +0.008& $-$0.004&   +0.007 \\
NGC~6752& $-$0.023& $-$0.046& $-$0.011&   +0.009 &  &+0.016&   +0.021&   +0.006&   +0.018 \\
NGC~6809& $-$0.025& $-$0.043& $-$0.011&   +0.008 &  &+0.010&   +0.013& $-$0.003&   +0.011 \\
NGC~6838& $-$0.021& $-$0.057&   +0.006&   +0.021 &  &+0.012&   +0.022& $-$0.013& $-$0.001 \\
NGC~7078& $-$0.020& $-$0.029& $-$0.013&   +0.012 &  &+0.006&   +0.007& $-$0.004&   +0.006 \\
NGC~7099& $-$0.018& $-$0.026& $-$0.011&   +0.014 &  &+0.006&   +0.008& $-$0.005&   +0.007 \\
\hline
        & \multicolumn{4}{c}{$\Delta \log g = +0.2$~dex} &\multicolumn{1}{c}{} &\multicolumn{4}{c}{$\Delta$ [A/H] = +0.1~dex} \\
	\cline{2-5} \cline{7-10}\\ 
	 
cluster &$\Delta$[Mg/Fe]&$\Delta$[Si/Fe]&$\Delta$[Ca/Fe]&$\Delta$[Ti/Fe]I&  &$\Delta$[Mg/Fe]&$\Delta$[Si/Fe]&$\Delta$[Ca/Fe]&$\Delta$[Ti/Fe]I  \\	
NGC 104 & $-$0.023& +0.022& $-$0.055& $-$0.023 &  & $-$0.005&   +0.010& $-$0.009& $-$0.018 \\
NGC 288 & $-$0.002& +0.028& $-$0.022& $-$0.001 &  & $-$0.001&   +0.009& $-$0.002& $-$0.004 \\
NGC~1904&   +0.001& +0.011& $-$0.014& $-$0.001 &  &   +0.001&   +0.027& $-$0.003& $-$0.002 \\
NGC~3201& $-$0.001& +0.028& $-$0.016& $-$0.002 &  &   +0.001&   +0.011& $-$0.001& $-$0.001 \\
NGC~4590&   +0.002& +0.014& $-$0.005& $-$0.009 &  & $-$0.001&   +0.000& $-$0.001& $-$0.006 \\
NGC~5904& $-$0.004& +0.027& $-$0.021& $-$0.005 &  & $-$0.001&   +0.011& $-$0.004& $-$0.005 \\
NGC~6121& $-$0.005& +0.035& $-$0.030&   +0.034 &  & $-$0.003&   +0.011& $-$0.005&   +0.011 \\
NGC~6171& $-$0.008& +0.034& $-$0.047& $-$0.005 &  & $-$0.004&   +0.011&   +0.000& $-$0.012 \\
NGC~6254& $-$0.005& +0.025& $-$0.018& $-$0.005 &  & $-$0.001&   +0.009& $-$0.005& $-$0.005 \\
NGC~6254& $-$0.001& +0.002& $-$0.010& $-$0.001 &  &   +0.001&   +0.007& $-$0.001&   +0.000 \\
NGC~6397&   +0.001& +0.011& $-$0.004& $-$0.005 &  & $-$0.002& $-$0.001& $-$0.003& $-$0.004 \\
NGC~6752& $-$0.004& +0.024& $-$0.012& $-$0.006 &  & $-$0.001&   +0.008& $-$0.002& $-$0.003 \\
NGC~6809&   +0.000& +0.021& $-$0.009& $-$0.005 &  &   +0.001&   +0.008& $-$0.002& $-$0.002 \\
NGC~6838& $-$0.014& +0.030& $-$0.042& $-$0.012 &  & $-$0.007&   +0.008& $-$0.010& $-$0.016 \\
NGC~7078&   +0.001& +0.013& $-$0.002& $-$0.005 &  & $-$0.003& $-$0.001& $-$0.003& $-$0.006 \\
NGC~7099& $-$0.001& +0.010& $-$0.005& $-$0.007 &  & $-$0.003& $-$0.001& $-$0.002& $-$0.005 \\
\hline
\end{tabular}
\label{t:sensAgira}
\end{table*}

\begin{table*}[h]
\tiny
\centering
\setlength{\tabcolsep}{1.0mm}
\caption{Errors in abundances due to errors in atmospheric
parameters and in the EWs.}
\begin{tabular}{lllllrlc|lllllrlll}
\hline
& \multicolumn{7}{c}{UVES} & &\multicolumn{7}{c}{GIRAFFE} &\\
  
                    & T$_{\rm eff}$& $\log g$ & [A/H]   & $v_t$  &nr & EW  &  	  all   &      & T$_{\rm eff}$& $\log g$ & [A/H]   & $v_t$  &nr & EW  &  	  all &\\
\hline
$[$Mg/Fe$]$         &     +0.003   &$-$0.001  &  +0.002 &  +0.006&  3   & 0.053  &  	0.054 &&   $-$0.001   &$-$0.002  &$-$0.005 &  +0.012&  2   & 0.095  &	   0.096 &NGC 104\\
$[$Si/Fe$]$         &   $-$0.009   &$-$0.004  &$-$0.006 &  +0.010&  8   & 0.033  &  	0.035 &&   $-$0.006   &  +0.002  &  +0.010 &  +0.023&  8   & 0.048  &	   0.054 &\\
$[$Ca/Fe$]$         &     +0.004   &$-$0.004  &$-$0.011 &$-$0.017& 15   & 0.024  &  	0.031 &&     +0.003   &$-$0.006  &$-$0.009 &$-$0.016&  4   & 0.068  &	   0.070 &\\
$[$Ti/Fe$]${\sc i}  &     +0.006   &$-$0.001  &$-$0.014 &$-$0.016&  8   & 0.033  &  	0.038 &&     +0.005   &$-$0.002  &$-$0.018 &$-$0.012&  3   & 0.078  &	   0.079 &\\
$[$Ti/Fe$]${\sc ii} &     +0.003   &$-$0.003  &$-$0.006 &$-$0.021& 11   & 0.028  &  	0.035 &&\\
\hline
$[$Mg/Fe$]$         &   $-$0.003   &  +0.001  &  +0.000 &  +0.011&  3   & 0.050  &  	0.052 &&   $-$0.002   &  +0.000  &  +0.000 &  +0.016&  2   & 0.098  &	   0.107 &NGC 288\\
$[$Si/Fe$]$         &   $-$0.006   &  +0.008  &  +0.008 &  +0.016&  7   & 0.033  &  	0.039 &&   $-$0.005   &  +0.006  &  +0.004 &$-$0.019&  7   & 0.052  &	   0.075 &\\
$[$Ca/Fe$]$         &     +0.001   &$-$0.002  &$-$0.004 &$-$0.014& 16   & 0.022  &  	0.026 &&     +0.000   &$-$0.005  &$-$0.001 &$-$0.005&  5   & 0.062  &	   0.063 &\\
$[$Ti/Fe$]${\sc i}  &     +0.003   &  +0.000  &$-$0.004 &  +0.006&  9   & 0.029  &  	0.030 &&     +0.002   &  +0.000  &$-$0.002 &  +0.012&  3   & 0.080  &	   0.087 &\\
$[$Ti/Fe$]${\sc ii} &     +0.002   &$-$0.002  &  +0.001 &$-$0.015& 12   & 0.025  &  	0.030 &&\\
\hline
$[$Mg/Fe$]$         &   $-$0.002   &  +0.000  &$-$0.001 &  +0.010&  1   & 0.085  &  	0.086 &&   $-$0.002   &  +0.000  &  +0.000 &  +0.050&  1   & 0.131  &	   0.140 &NGC 1904\\
$[$Si/Fe$]$         &   $-$0.005   &  +0.006  &  +0.004 &  +0.013&  6   & 0.035  &  	0.038 &&   $-$0.005   &  +0.002  &  +0.010 &$-$0.067&  6   & 0.053  &	   0.087 &\\
$[$Ca/Fe$]$         &     +0.000   &$-$0.002  &$-$0.002 &$-$0.002& 16   & 0.021  &  	0.022 &&     +0.000   &$-$0.003  &$-$0.001 &$-$0.006&  6   & 0.053  &	   0.054 &\\
$[$Ti/Fe$]${\sc i}  &     +0.002   &$-$0.001  &$-$0.002 &  +0.011&  8   & 0.030  &  	0.032 &&     +0.001   &  +0.000  &$-$0.001 &  +0.048&  3   & 0.076  &	   0.089 &\\
$[$Ti/Fe$]${\sc ii} &     +0.002   &$-$0.002  &  +0.000 &$-$0.011& 10   & 0.027  &  	0.029 &&\\
\hline
$[$Mg/Fe$]$         &   $-$0.001   &  +0.000  &  +0.000 &  +0.005&  2   & 0.064  &  	0.064 &&   $-$0.002   &  +0.000  &  +0.000 &  +0.029&  1   & 0.146  &	   0.149 &NGC 3201\\
$[$Si/Fe$]$         &   $-$0.004   &  +0.007  &  +0.010 &  +0.009&  8   & 0.032  &  	0.035 &&   $-$0.004   &  +0.006  &  +0.005 &  +0.036&  5   & 0.065  &	   0.075 &\\
$[$Ca/Fe$]$         &     +0.000   &$-$0.002  &$-$0.004 &$-$0.007& 18   & 0.021  &  	0.023 &&     +0.000   &$-$0.003  &  +0.000 &$-$0.004&  5   & 0.065  &	   0.065 &\\
$[$Ti/Fe$]${\sc i}  &     +0.002   &$-$0.001  &$-$0.003 &  +0.006&  9   & 0.030  &  	0.031 &&     +0.001   &  +0.000  &  +0.000 &  +0.025&  3   & 0.084  &	   0.088 &\\
$[$Ti/Fe$]${\sc ii} &     +0.001   &$-$0.002  &  +0.001 &$-$0.007& 10   & 0.028  &  	0.029 &&\\
\hline
$[$Mg/Fe$]$         &   $-$0.001   &$-$0.001  &$-$0.002 &$-$0.198&  1   & 0.094  &  	0.219 &&   $-$0.002   &  +0.000  &$-$0.001 &  +0.025&  1   & 0.129  &	   0.131 &NGC 4590\\
$[$Si/Fe$]$         &   $-$0.002   &  +0.002  &$-$0.001 &$-$0.195&  2   & 0.066  &  	0.205 &&   $-$0.002   &  +0.003  &  +0.000 &  +0.028&  2   & 0.091  &	   0.095 &\\
$[$Ca/Fe$]$         &     +0.000   &$-$0.002  &$-$0.003 &$-$0.236& 13   & 0.026  &  	0.237 &&   $-$0.001   &$-$0.001  &$-$0.001 &$-$0.008&  3   & 0.074  &	   0.075 &\\
$[$Ti/Fe$]${\sc i}  &     +0.002   &$-$0.002  &$-$0.003 &$-$0.194&  3   & 0.054  &  	0.201 &&     +0.001   &$-$0.002  &$-$0.004 &  +0.025&  2   & 0.091  &	   0.095 &\\
$[$Ti/Fe$]${\sc ii} &     +0.001   &$-$0.001  &  +0.002 &$-$0.015&  7   & 0.036  &  	0.039 &&\\
\hline
$[$Mg/Fe$]$         &   $-$0.004   &  +0.000  &  +0.000 &  +0.007&  2   & 0.069  &  	0.070 &&   $-$0.005   &$-$0.001  &  +0.000 &  +0.019&  2   & 0.098  &	   0.100 &NGC 5904\\
$[$Si/Fe$]$         &   $-$0.013   &  +0.007  &  +0.008 &  +0.011&  8   & 0.035  &  	0.040 &&   $-$0.011   &  +0.006  &  +0.003 &  +0.026&  7   & 0.053  &	   0.060 &\\
$[$Ca/Fe$]$         &     +0.002   &$-$0.003  &$-$0.004 &$-$0.009& 17   & 0.024  &  	0.026 &&     +0.000   &$-$0.004  &$-$0.001 &$-$0.008&  5   & 0.062  &	   0.063 &\\
$[$Ti/Fe$]${\sc i}  &     +0.007   &$-$0.001  &$-$0.004 &  +0.005& 10   & 0.031  &  	0.032 &&     +0.004   &$-$0.001  &$-$0.001 &  +0.014&  4   & 0.070  &	   0.071 &\\
$[$Ti/Fe$]${\sc ii} &     +0.005   &$-$0.002  &  +0.001 &$-$0.010&  9   & 0.033  &  	0.035 &&\\
\hline
$[$Mg/Fe$]$         &   $-$0.001   &  +0.000  &  +0.000 &  +0.005&  3   & 0.057  &  	0.057 &&   $-$0.001   &$-$0.001  &$-$0.001 &  +0.019&  2   & 0.096  &	   0.098 &NGC 6121\\
$[$Si/Fe$]$         &   $-$0.005   &  +0.009  &  +0.008 &  +0.008&  9   & 0.033  &  	0.036 &&   $-$0.004   &  +0.007  &  +0.003 &  +0.029&  8   & 0.048  &	   0.057 &\\
$[$Ca/Fe$]$         &     +0.001   &$-$0.004  &$-$0.004 &$-$0.010& 18   & 0.023  &  	0.026 &&     +0.001   &$-$0.006  &$-$0.001 &$-$0.011&  5   & 0.061  &	   0.062 &\\
$[$Ti/Fe$]${\sc i}  &     +0.003   &  +0.000  &$-$0.006 &$-$0.001&  9   & 0.033  &  	0.034 &&   $-$0.004   &  +0.007  &  +0.003 &  +0.029&  4   & 0.068  &	   0.074 &\\
$[$Ti/Fe$]${\sc ii} &     +0.002   &$-$0.003  &  +0.000 &$-$0.010& 11   & 0.030  &  	0.032 &&\\
\hline
$[$Mg/Fe$]$         &   $-$0.002   &  +0.000  &  +0.000 &  +0.006&  2   & 0.081  &  	0.082 &&   $-$0.001   &$-$0.002  &$-$0.002 &  +0.027&  2   & 0.109  &	   0.112 &NGC 6171\\
$[$Si/Fe$]$         &   $-$0.006   &  +0.008  &  +0.010 &  +0.009&  7   & 0.043  &  	0.047 &&   $-$0.002   &  +0.007  &  +0.005 &  +0.040&  8   & 0.054  &	   0.068 &\\
$[$Ca/Fe$]$         &     +0.002   &$-$0.006  &$-$0.004 &$-$0.015& 18   & 0.027  &  	0.032 &&     +0.001   &$-$0.010  &  +0.000 &$-$0.015&  5   & 0.069  &	   0.070 &\\
$[$Ti/Fe$]${\sc i}  &     +0.005   &$-$0.003  &$-$0.012 &$-$0.015& 32   & 0.020  &  	0.028 &&     +0.001   &$-$0.001  &$-$0.005 &  +0.011&  4   & 0.077  &	   0.078 &\\
$[$Ti/Fe$]${\sc ii} &     +0.003   &$-$0.003  &  +0.000 &$-$0.017&  9   & 0.038  &  	0.042 &&\\
\hline
$[$Mg/Fe$]$         &   $-$0.002   &$-$0.002  &  +0.001 &$-$0.003&  4   & 0.062  &  	0.062 &&   $-$0.001   &$-$0.002  &$-$0.002 &  +0.018&  2   & 0.087  &	   0.089 &NGC 6218\\
$[$Si/Fe$]$         &   $-$0.007   &  +0.004  &  +0.005 &  +0.017&  9   & 0.041  &  	0.045 &&   $-$0.006   &  +0.002  &  +0.003 &  +0.035&  6   & 0.050  &	   0.061 &\\
$[$Ca/Fe$]$         &     +0.001   &$-$0.001  &$-$0.003 &$-$0.042& 21   & 0.027  &  	0.050 &&     +0.003   &$-$0.006  &$-$0.003 &$-$0.024&  5   & 0.055  &	   0.060 &\\
$[$Ti/Fe$]${\sc i}  &     +0.005   &$-$0.001  &$-$0.003 &$-$0.035& 27   & 0.024  &  	0.042 &&     +0.005   &$-$0.002  &$-$0.005 &$-$0.018&  4   & 0.062  &	   0.065 &\\
$[$Ti/Fe$]${\sc ii} &     +0.001   &$-$0.001  &  +0.001 &$-$0.033& 10   & 0.039  &  	0.051 &&\\
\hline
$[$Mg/Fe$]$         &   $-$0.001   &  +0.000  &$-$0.002 &  +0.005&  2   & 0.063  &  	0.063 &&   $-$0.002   &  +0.000  &  +0.001 &  +0.013&  1   & 0.129  &	   0.130 &NGC 6254\\
$[$Si/Fe$]$         &   $-$0.003   &  +0.006  &  +0.006 &  +0.013&  6   & 0.036  &  	0.040 &&   $-$0.003   &  +0.000  &  +0.004 &  +0.021&  4   & 0.065  &	   0.068 &\\
$[$Ca/Fe$]$         &     +0.001   &$-$0.002  &$-$0.005 &$-$0.011& 17   & 0.022  &  	0.025 &&   $-$0.001   &$-$0.002  &$-$0.001 &$-$0.005&  5   & 0.058  &	   0.058 &\\
$[$Ti/Fe$]${\sc i}  &     +0.003   &$-$0.002  &$-$0.007 &$-$0.006& 23   & 0.019  &  	0.021 &&     +0.001   &  +0.000  &  +0.000 &  +0.016&  3   & 0.074  &	   0.076 &\\
$[$Ti/Fe$]${\sc ii} &     +0.001   &$-$0.001  &  +0.004 &$-$0.014& 10   & 0.028  &  	0.031 &&\\
\hline
$[$Mg/Fe$]$         &   $-$0.001   &$-$0.004  &$-$0.004 &$-$0.009&  2   & 0.073  &  	0.074 &&   $-$0.002   &  +0.000  &$-$0.001 &  +0.020&  1   & 0.153  &	   0.154 &NGC 6397\\
$[$Si/Fe$]$         &   $-$0.002   &  +0.002  &$-$0.001 &  +0.006&  3   & 0.059  &  	0.060 &&   $-$0.002   &  +0.002  &  +0.000 &  +0.027&  1   & 0.153  &	   0.155 &\\
$[$Ca/Fe$]$         &     +0.000   &$-$0.002  &$-$0.004 &$-$0.004& 15   & 0.027  &  	0.027 &&   $-$0.001   &$-$0.001  &$-$0.001 &$-$0.014&  3   & 0.088  &	   0.089 &\\
$[$Ti/Fe$]${\sc i}  &     +0.003   &$-$0.003  &$-$0.005 &$-$0.005& 13   & 0.029  &  	0.030 &&     +0.001   &$-$0.001  &$-$0.002 &  +0.024&  2   & 0.108  &	   0.111 &\\
$[$Ti/Fe$]${\sc ii} &     +0.002   &$-$0.001  &  +0.002 &$-$0.008&  7   & 0.039  &  	0.040 &&\\
\hline
$[$Mg/Fe$]$         &   $-$0.002   &$-$0.004  &  +0.000 &  +0.001&  3   & 0.068  &  	0.068 &&   $-$0.002   &  +0.000  &  +0.000 &  +0.021&  1   & 0.096  &	   0.098 &NGC 6752\\
$[$Si/Fe$]$         &   $-$0.006   &  +0.007  &  +0.008 &  +0.006&  9   & 0.039  &  	0.041 &&   $-$0.005   &  +0.002  &  +0.002 &  +0.027&  6   & 0.039  &	   0.048 &\\
$[$Ca/Fe$]$         &     +0.001   &$-$0.003  &$-$0.005 &$-$0.006& 20   & 0.026  &  	0.027 &&   $-$0.001   &$-$0.001  &$-$0.001 &  +0.008&  4   & 0.048  &	   0.049 &\\
$[$Ti/Fe$]${\sc i}  &     +0.004   &$-$0.003  &$-$0.006 &$-$0.004& 29   & 0.022  &  	0.023 &&     +0.001   &$-$0.001  &$-$0.001 &  +0.023&  3   & 0.055  &	   0.060 &\\
$[$Ti/Fe$]${\sc ii} &     +0.002   &$-$0.003  &$-$0.002 &$-$0.005& 12   & 0.034  &  	0.034 &&\\
\hline
$[$Mg/Fe$]$         &   $-$0.001   &$-$0.002  &$-$0.003 &  +0.007&  1   & 0.084  &  	0.084 &&   $-$0.003   &  +0.000  &  +0.000 &  +0.020&  1   & 0.128  &	   0.130 &NGC 6809\\
$[$Si/Fe$]$         &   $-$0.004   &  +0.005  &  +0.006 &  +0.014&  4   & 0.042  &  	0.045 &&   $-$0.004   &  +0.004  &  +0.004 &  +0.026&  5   & 0.057  &	   0.063 &\\
$[$Ca/Fe$]$         &     +0.000   &$-$0.003  &$-$0.005 &$-$0.001& 16   & 0.021  &  	0.022 &&   $-$0.001   &$-$0.002  &$-$0.001 &$-$0.006&  5   & 0.057  &	   0.058 &\\
$[$Ti/Fe$]${\sc i}  &     +0.004   &$-$0.004  &$-$0.008 &$-$0.008& 20   & 0.019  &  	0.023 &&     +0.001   &$-$0.001  &$-$0.001 &  +0.022&  3   & 0.074  &	   0.077 &\\
$[$Ti/Fe$]${\sc ii} &     +0.001   &$-$0.002  &  +0.003 &$-$0.022&  9   & 0.028  &  	0.036 &&\\
\hline
$[$Mg/Fe$]$         &   $-$0.005   &  +0.002  &  +0.002 &  +0.005&  3   & 0.063  &  	0.063 &&   $-$0.002   &$-$0.003  &$-$0.002 &  +0.012&  2   & 0.098  &	   0.098 &NGC 6838\\
$[$Si/Fe$]$         &   $-$0.009   &  +0.010  &  +0.009 &  +0.010&  8   & 0.039  &  	0.043 &&   $-$0.006   &  +0.006  &  +0.003 &  +0.022&  9   & 0.046  &	   0.052 &\\
$[$Ca/Fe$]$         &     +0.001   &$-$0.007  &$-$0.005 &$-$0.014& 15   & 0.028  &  	0.033 &&     +0.001   &$-$0.009  &$-$0.003 &$-$0.013&  6   & 0.056  &	   0.059 &\\
$[$Ti/Fe$]${\sc i}  &     +0.004   &$-$0.002  &$-$0.007 &$-$0.026& 17   & 0.026  &  	0.038 &&     +0.002   &$-$0.002  &$-$0.005 &$-$0.001&  4   & 0.069  &	   0.069 &\\
$[$Ti/Fe$]${\sc ii} &     +0.002   &$-$0.005  &$-$0.002 &$-$0.018& 11   & 0.033  &  	0.038 &&\\
\hline
$[$Mg/Fe$]$         &   $-$0.002   &$-$0.002  &$-$0.005 &$-$0.008&  2   & 0.077  &  	0.080 &&   $-$0.002   &  +0.000  &$-$0.002 &  +0.020&  1   & 0.111  &	   0.113 &NGC 7078\\
$[$Si/Fe$]$         &   $-$0.003   &  +0.002  &$-$0.002 &  +0.016&  2   & 0.077  &  	0.079 &&   $-$0.003   &  +0.003  &$-$0.001 &  +0.023&  2   & 0.078  &	   0.082 &\\
$[$Ca/Fe$]$         &   $-$0.001   &$-$0.001  &$-$0.005 &$-$0.008& 13   & 0.030  &  	0.032 &&   $-$0.001   &  +0.000  &$-$0.002 &$-$0.013&  3   & 0.064  &	   0.065 &\\
$[$Ti/Fe$]${\sc i}  &     +0.003   &$-$0.003  &$-$0.007 &$-$0.018& 11   & 0.033  &  	0.038 &&     +0.001   &$-$0.001  &$-$0.004 &  +0.020&  3   & 0.064  &	   0.067 &\\
$[$Ti/Fe$]${\sc ii} &     +0.002   &$-$0.002  &  +0.001 &$-$0.029&  6   & 0.044  &  	0.053 &&\\
\hline
$[$Mg/Fe$]$         &   $-$0.001   &$-$0.003  &$-$0.005 &$-$0.006&  2   & 0.074  &  	0.075 &&   $-$0.002   &  +0.000  &$-$0.001 &  +0.025&  1   & 0.119  &	   0.122 &NGC 7099\\
$[$Si/Fe$]$         &   $-$0.002   &  +0.000  &$-$0.002 &  +0.007&  2   & 0.074  &  	0.075 &&   $-$0.003   &  +0.002  &  +0.000 &  +0.033&  3   & 0.069  &	   0.076 &\\
$[$Ca/Fe$]$         &   $-$0.001   &$-$0.003  &$-$0.005 &$-$0.003& 14   & 0.028  &  	0.029 &&   $-$0.001   &$-$0.001  &$-$0.001 &$-$0.021&  2   & 0.084  &	   0.087 &\\
$[$Ti/Fe$]${\sc i}  &     +0.004   &$-$0.005  &$-$0.008 &$-$0.006& 11   & 0.032  &  	0.034 &&     +0.001   &$-$0.001  &$-$0.002 &  +0.029&  2   & 0.084  &	   0.089 &\\
$[$Ti/Fe$]${\sc ii} &     +0.000   &$-$0.002  &  +0.001 &$-$0.010&  7   & 0.040  &  	0.041 &&\\
\hline
\end{tabular}
\label{t:errabuTOT}
\end{table*}

\FloatBarrier
\newpage
\section{Comparison with the GES, APOGEE, and GALAH surveys}

Comparison with GES (232 stars, six GCs), GALAH (158 stars, eight GCs), and APOGEE
(373 stars, 15 GCs).

\begin{figure*}[h]
\centering
\includegraphics[scale=0.25]{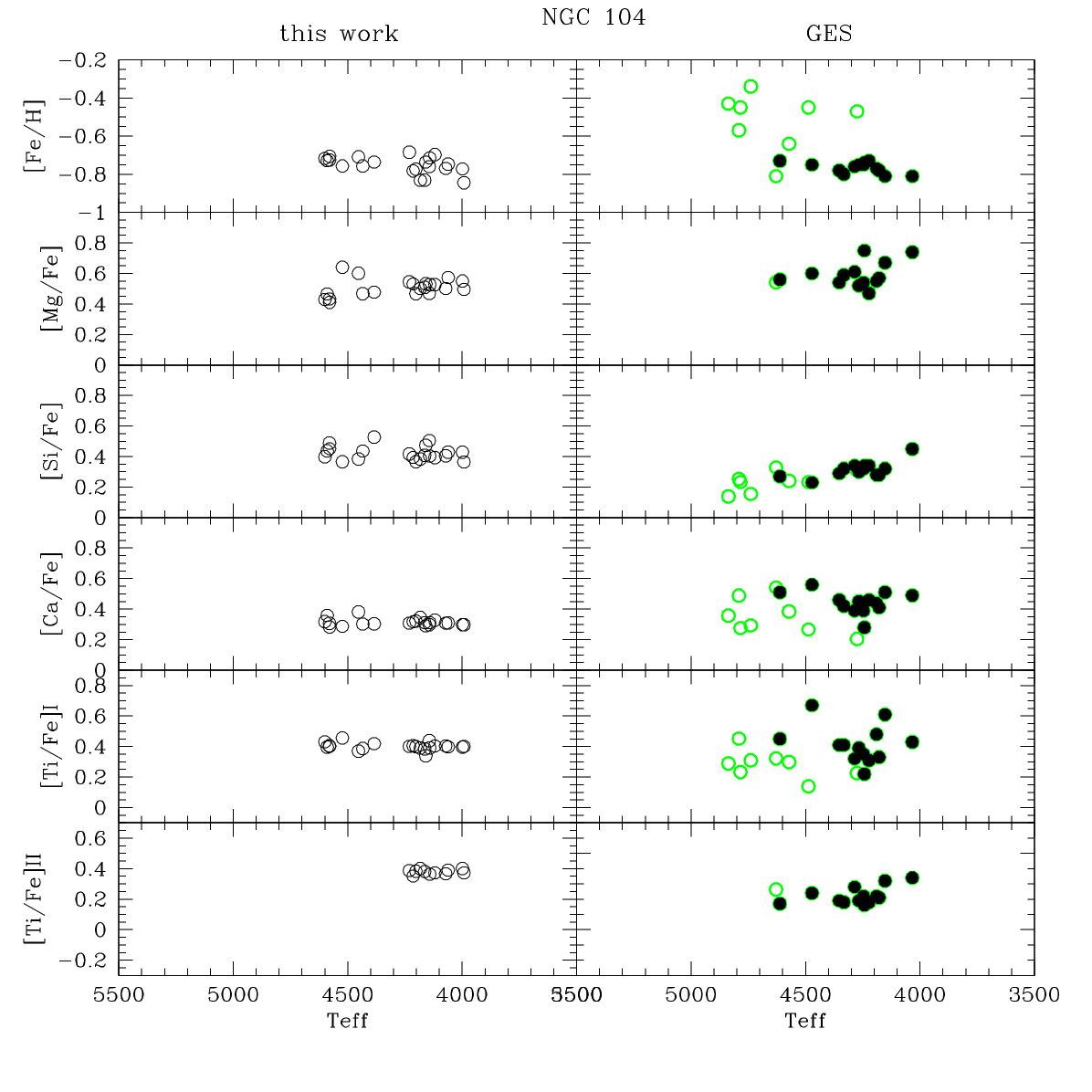}\includegraphics[scale=0.25]{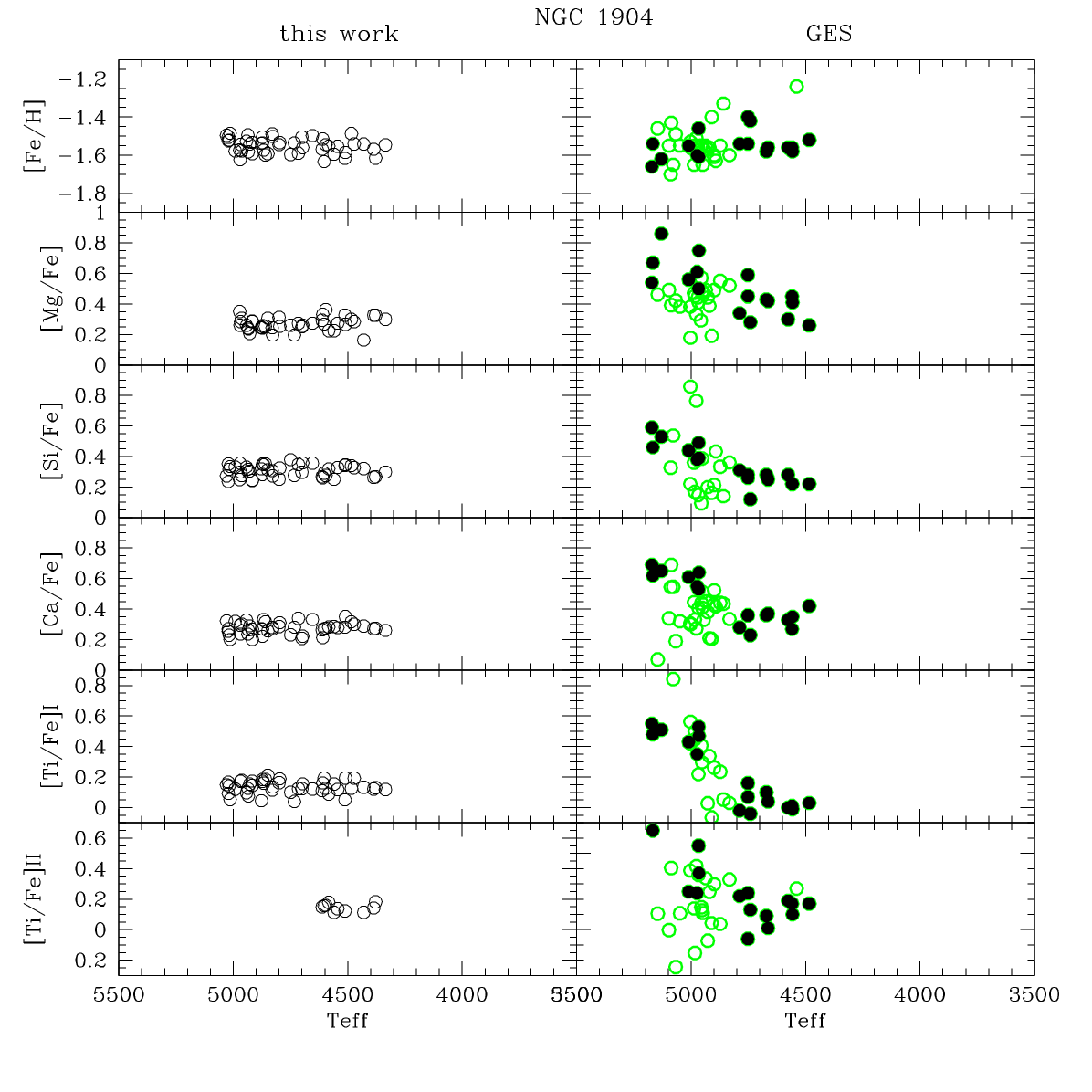}\includegraphics[scale=0.25]{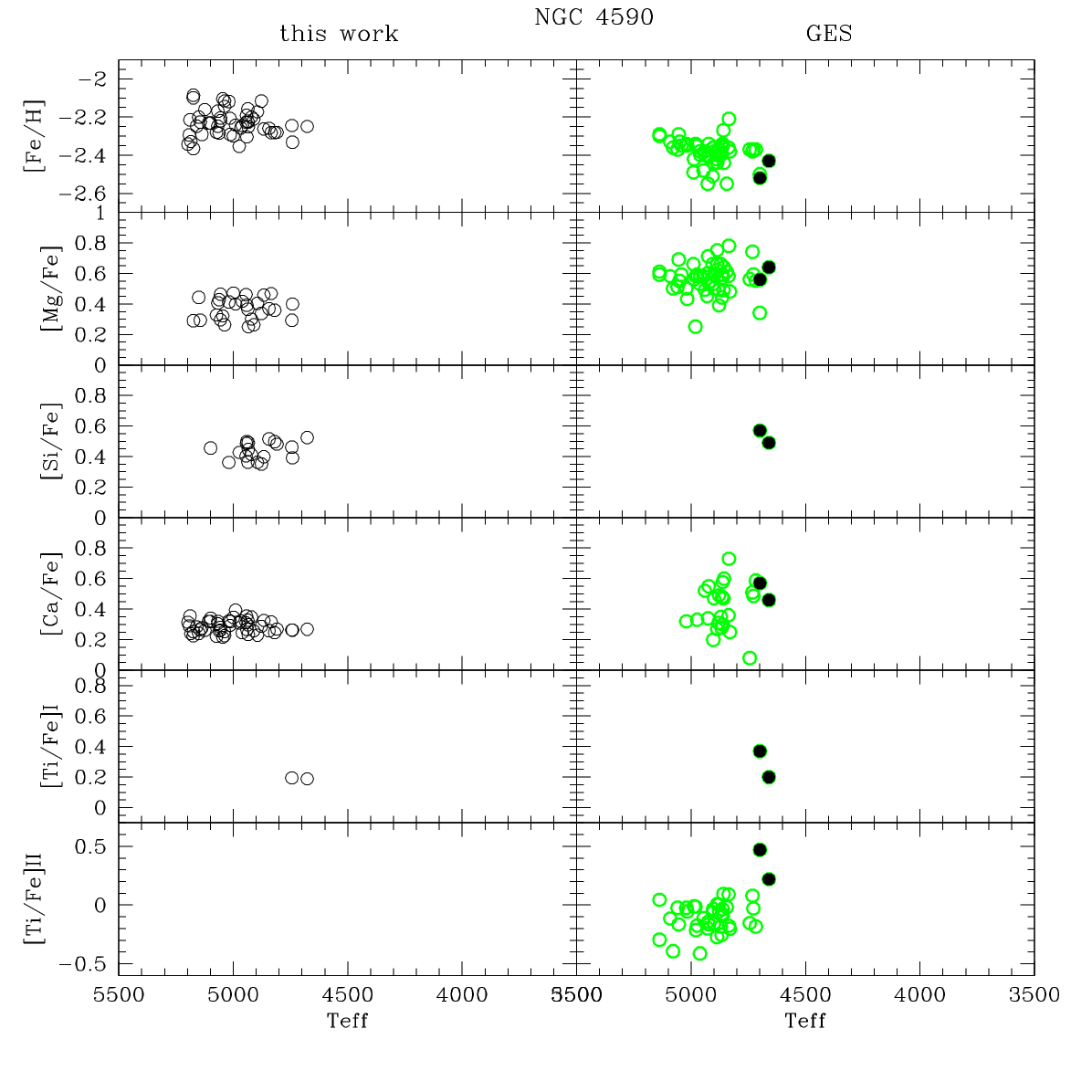}
\includegraphics[scale=0.25]{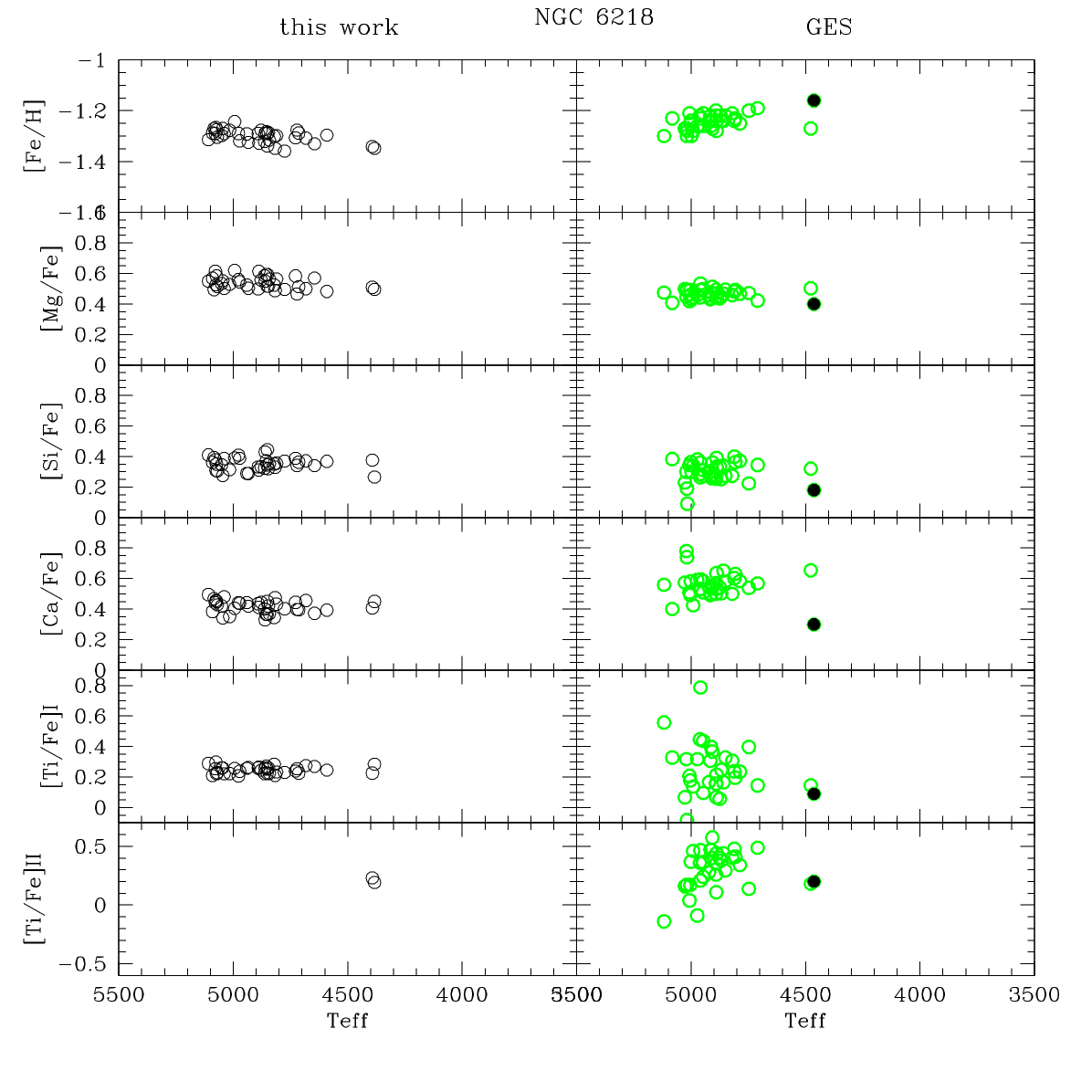}\includegraphics[scale=0.25]{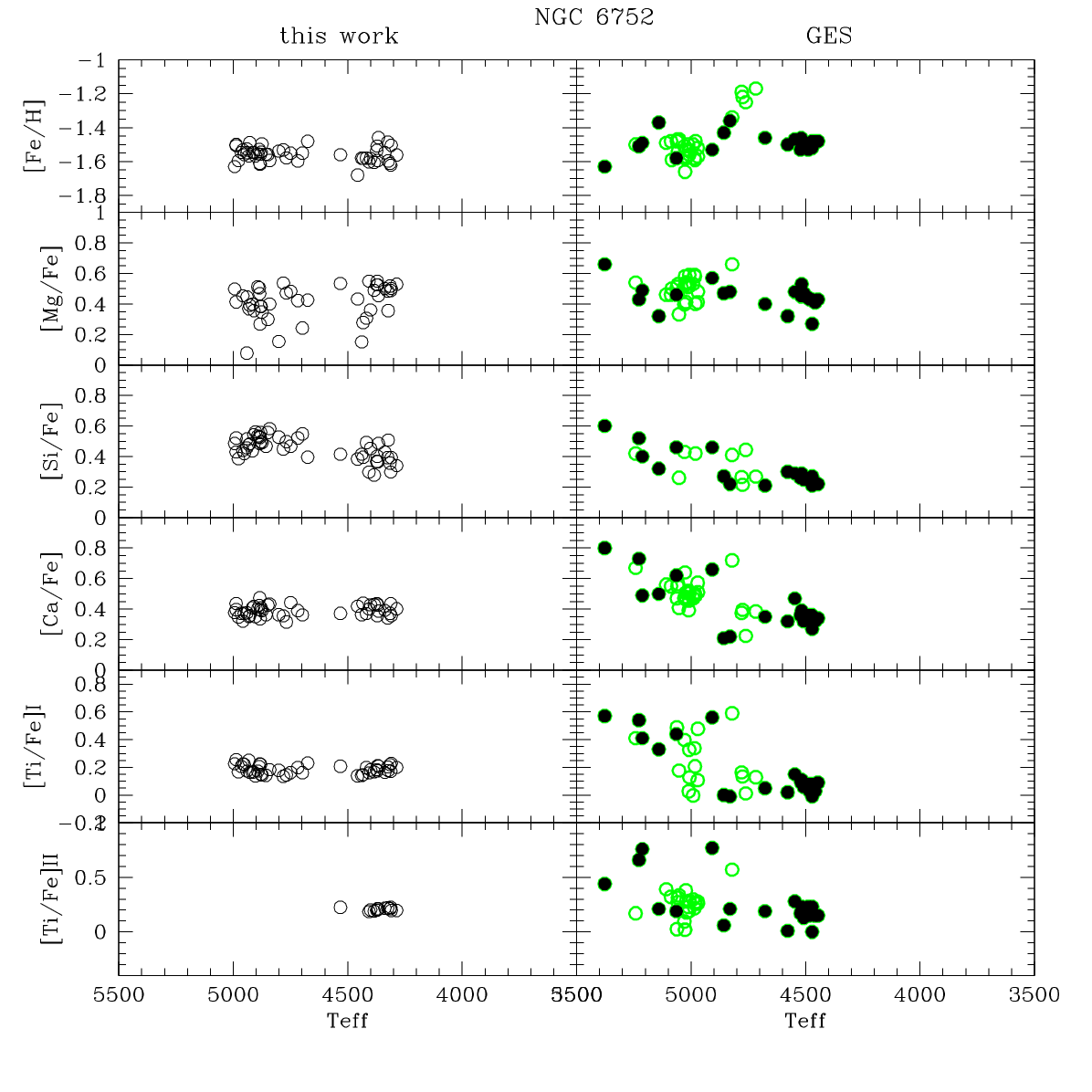}\includegraphics[scale=0.25]{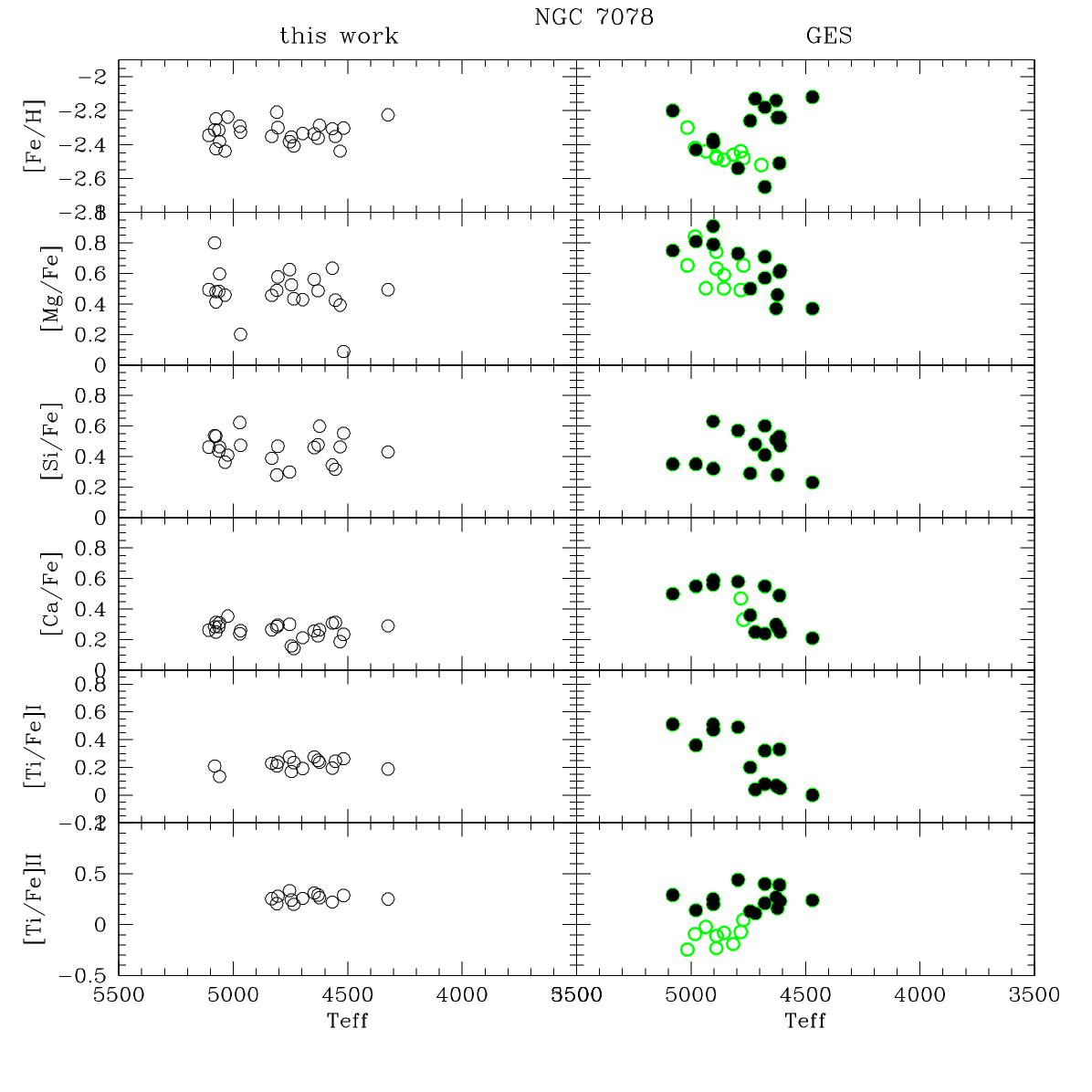}
\caption{Comparison of [Fe/H],[Mg/Fe],[Si/Fe], [Ca/Fe],[Ti/Fe]~{\sc i},and
[Ti/Fe]~{\sc ii} ratios from the present work (left panels) and the GES survey
(right panels). Filled points for GES are stars with UVES U580 spectra.}
\label{f:ges}
\end{figure*}

\begin{figure*}[h]
\centering
\includegraphics[scale=0.23]{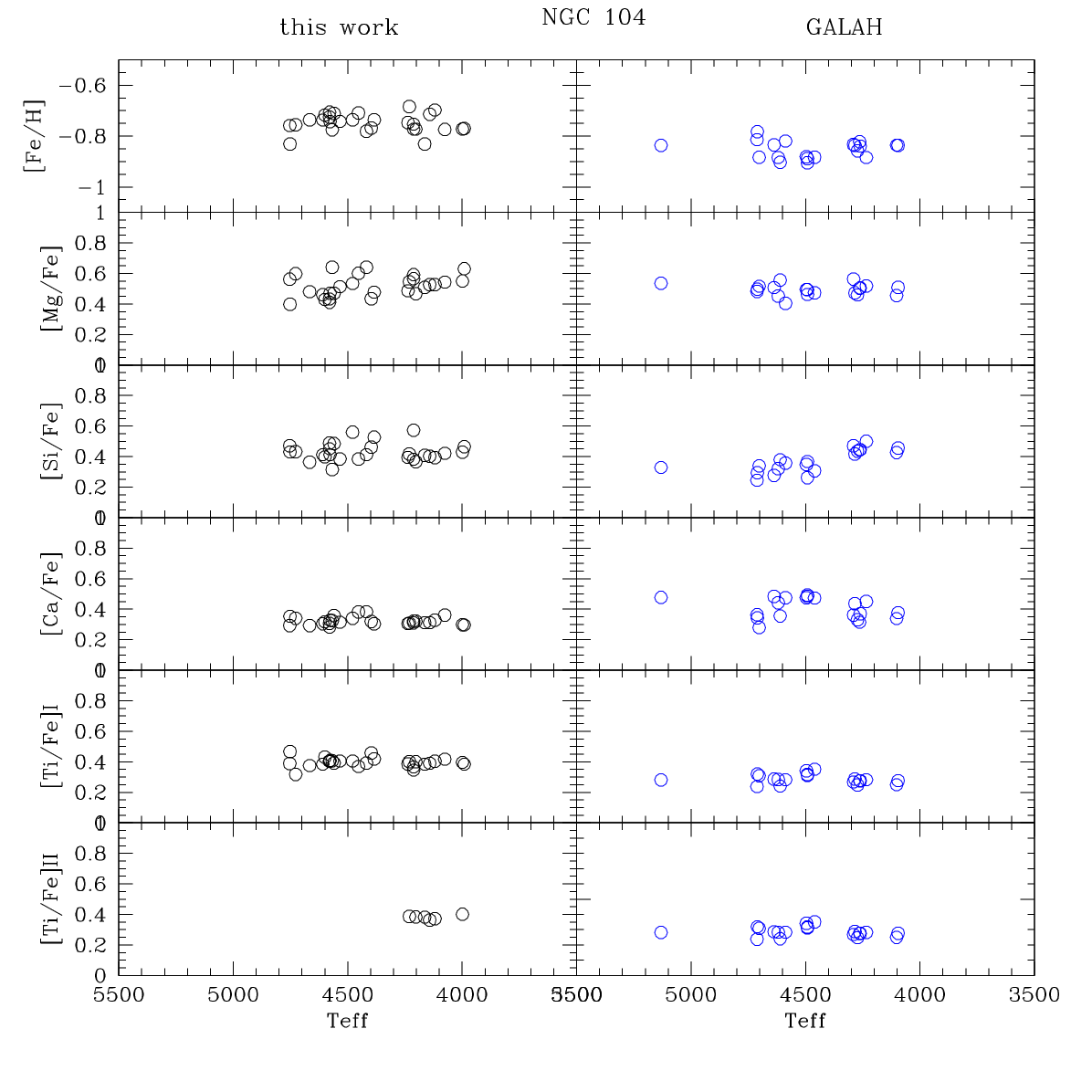}\includegraphics[scale=0.23]{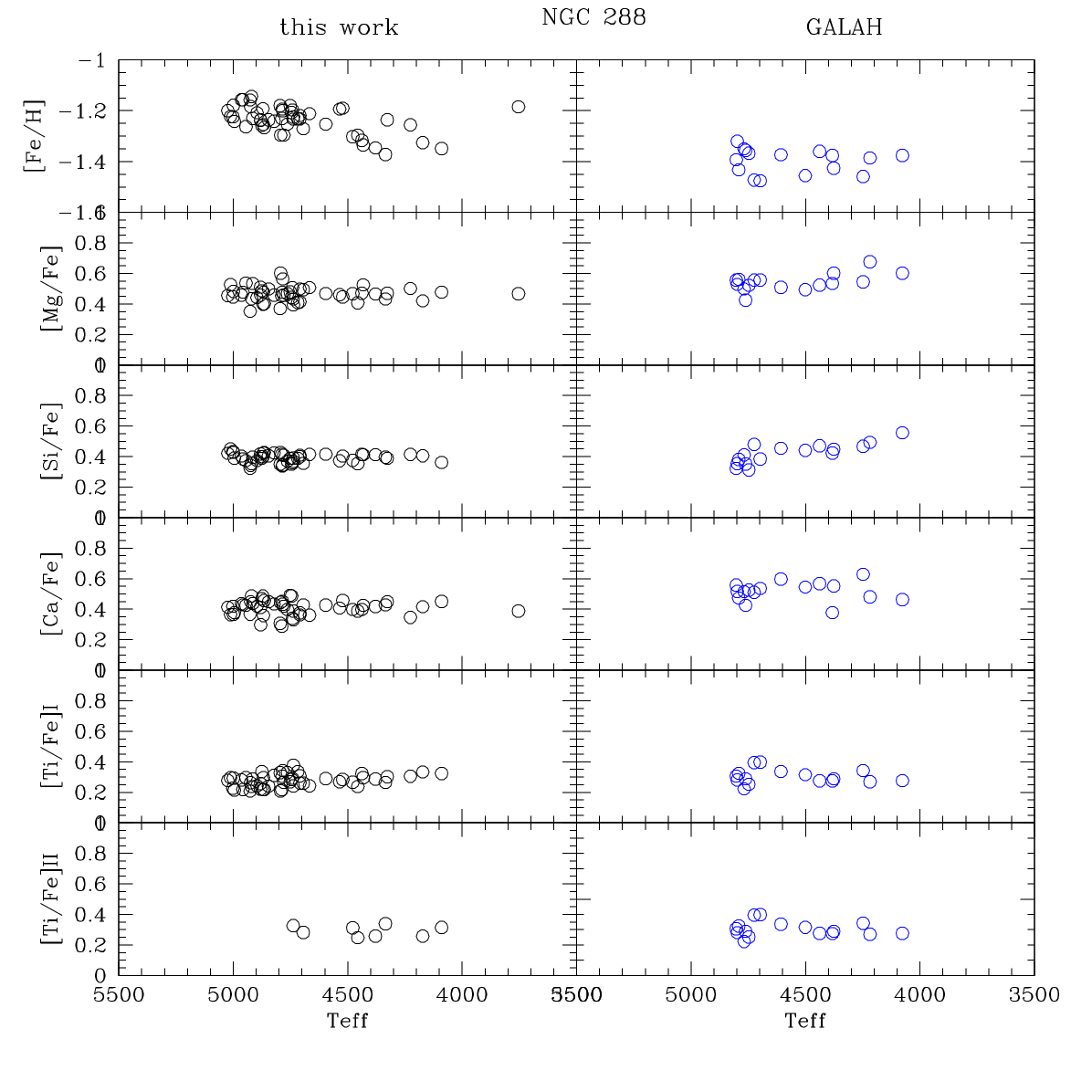}\includegraphics[scale=0.23]{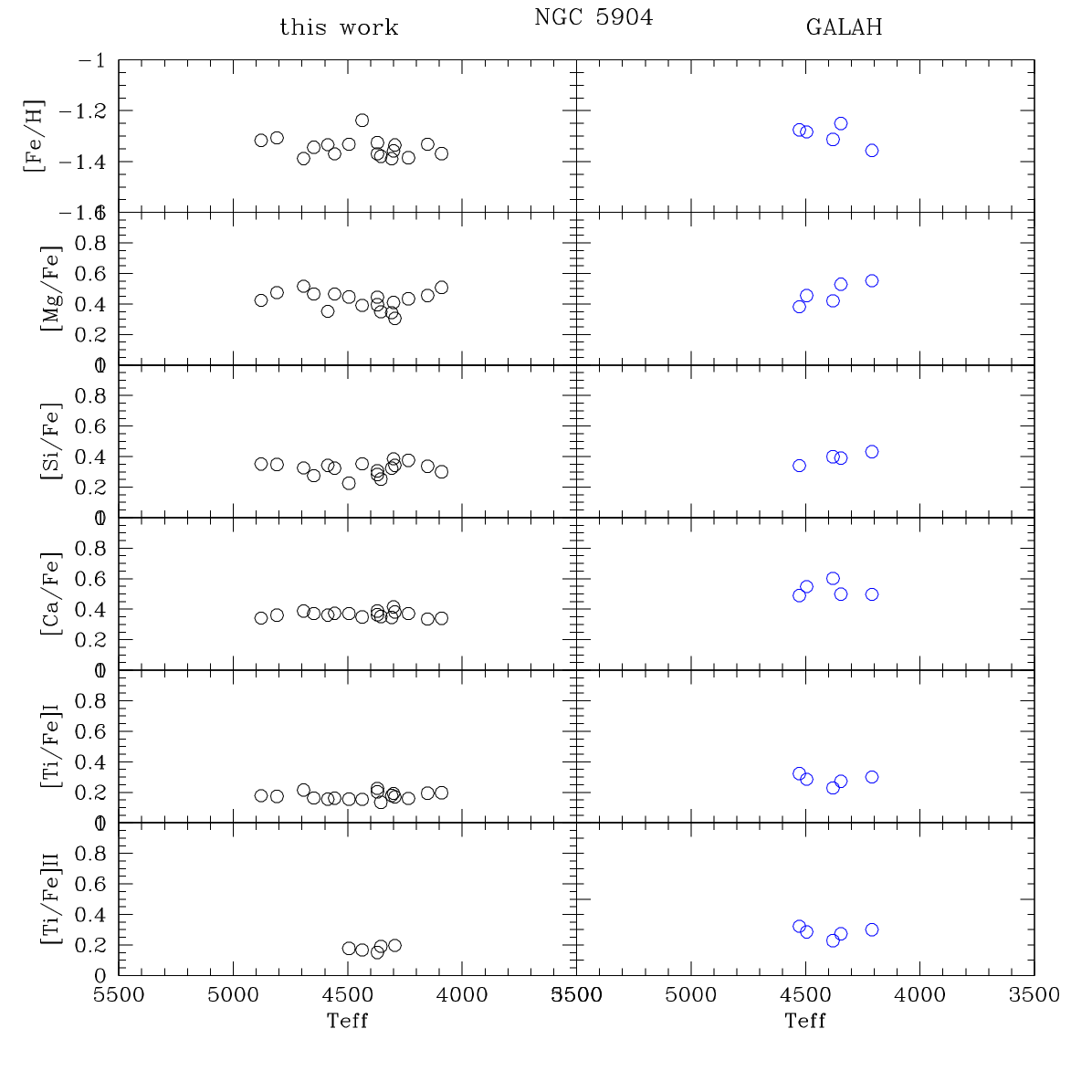}\includegraphics[scale=0.23]{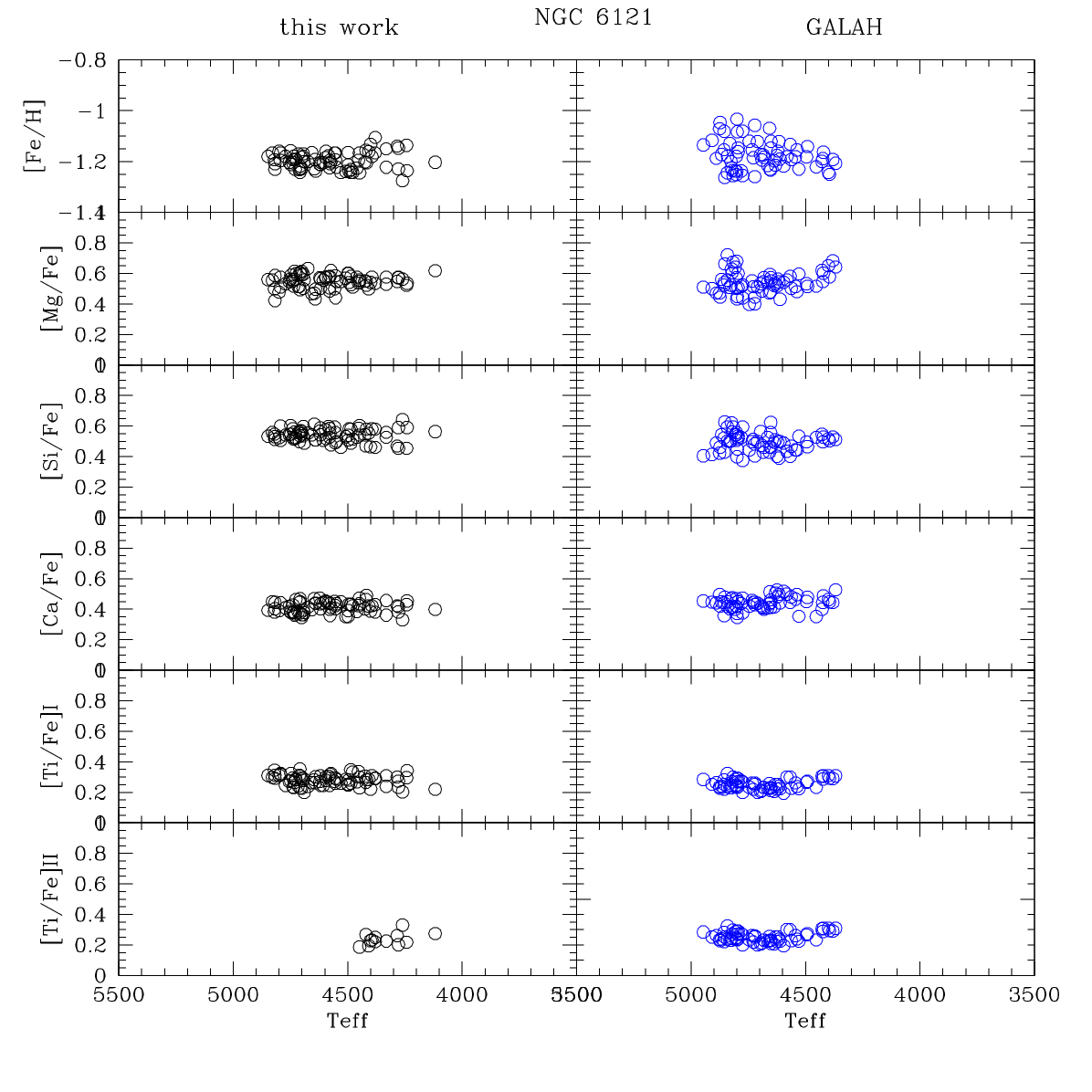}
\includegraphics[scale=0.23]{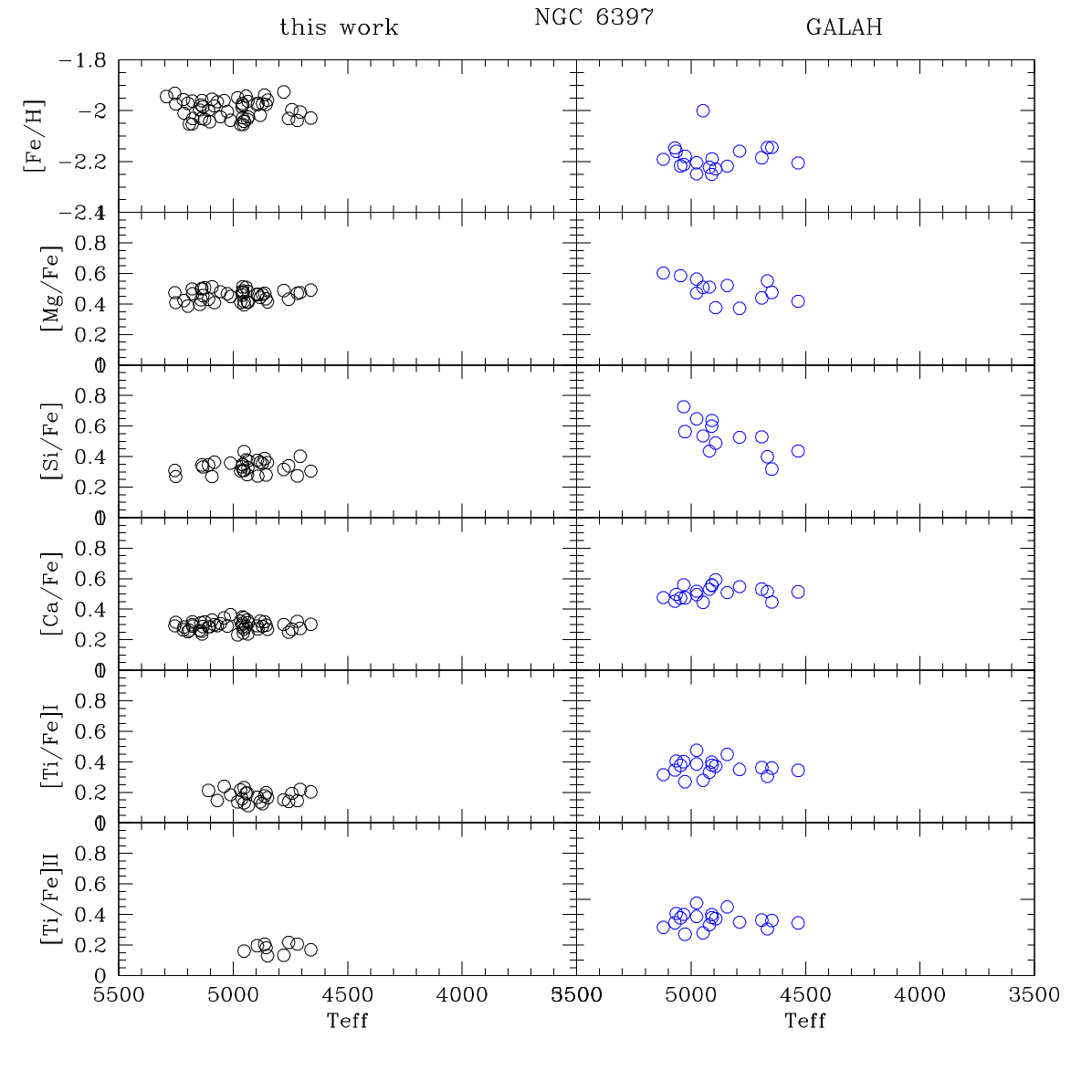}\includegraphics[scale=0.23]{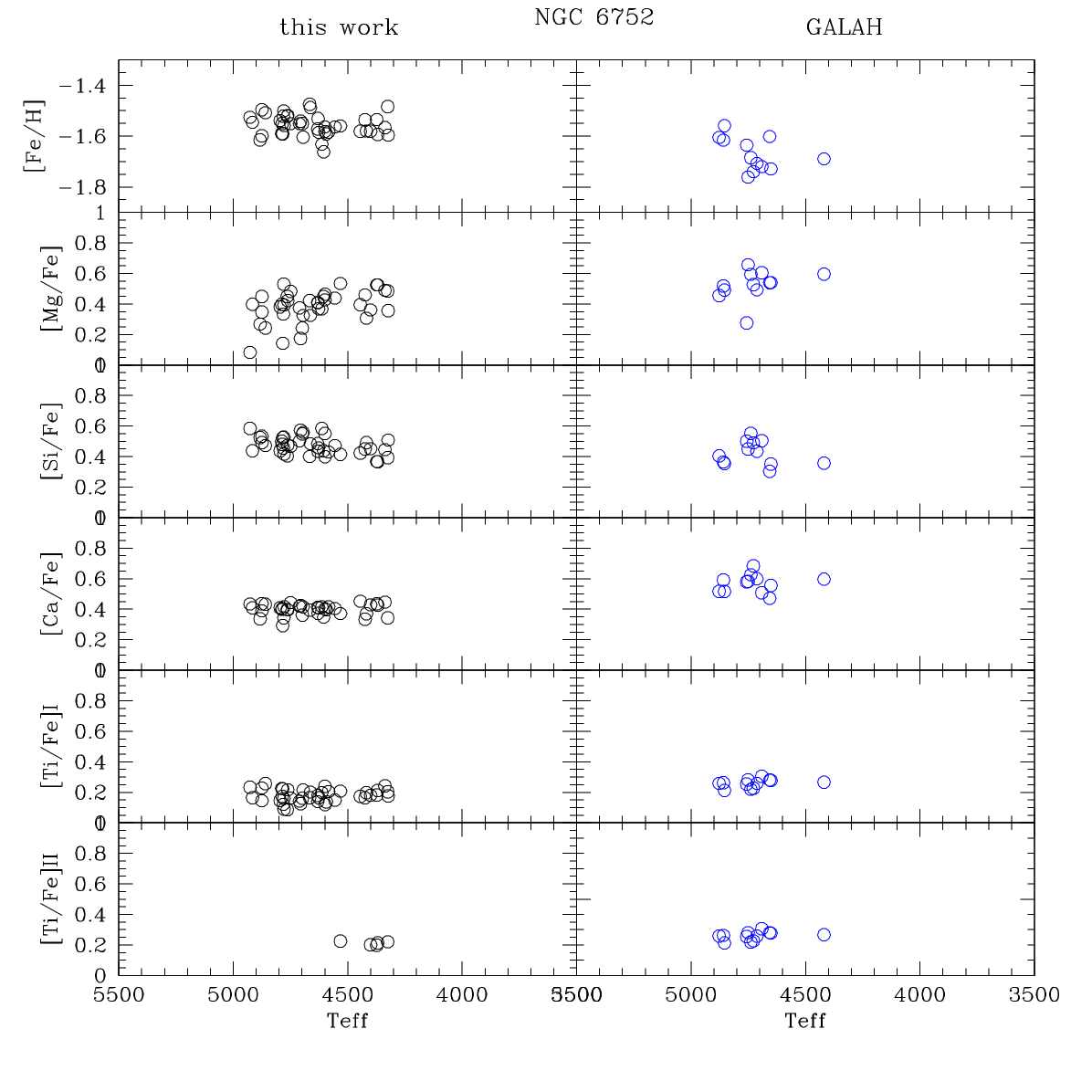}\includegraphics[scale=0.23]{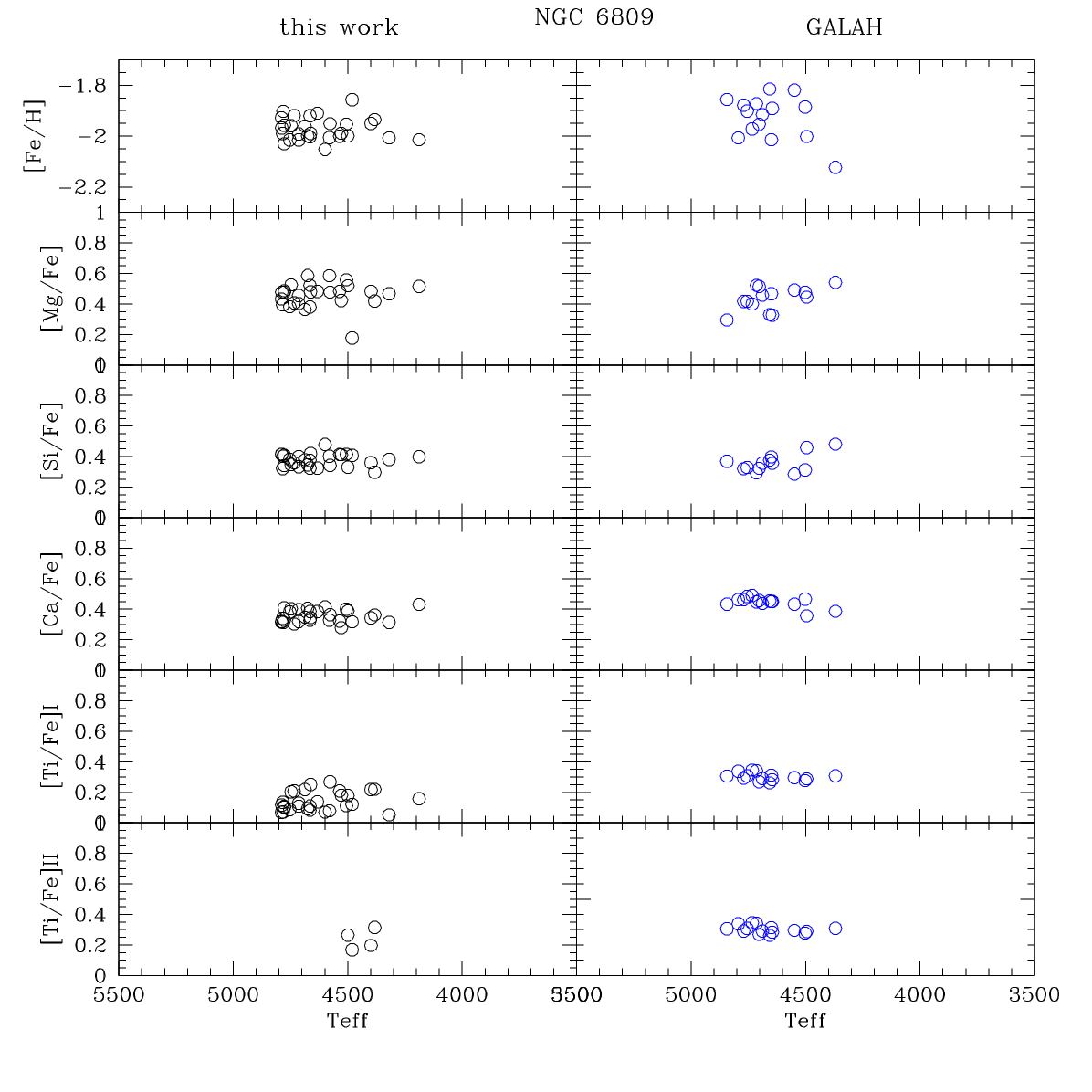}\includegraphics[scale=0.23]{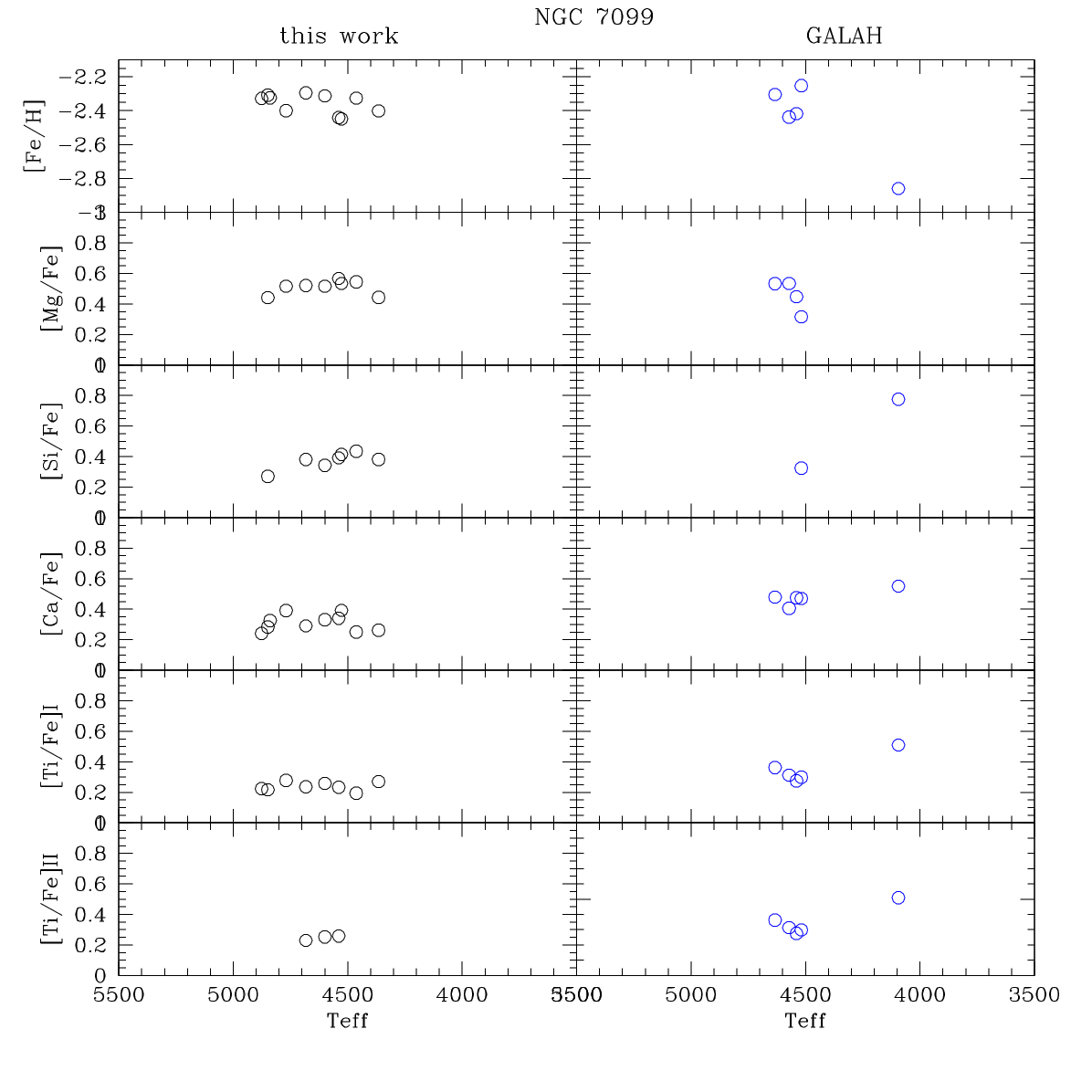}
\caption{Comparison of metallicity [Fe/H] and [Mg/Fe],[Si/Fe],
[Ca/Fe],[Ti/Fe]~{\sc i},and [Ti/Fe]~{\sc ii} ratios from the present work (left
panels) and the GALAH survey (right panels).}
\label{f:gal}
\end{figure*}

\begin{figure*}[h]
\centering
\includegraphics[scale=0.24]{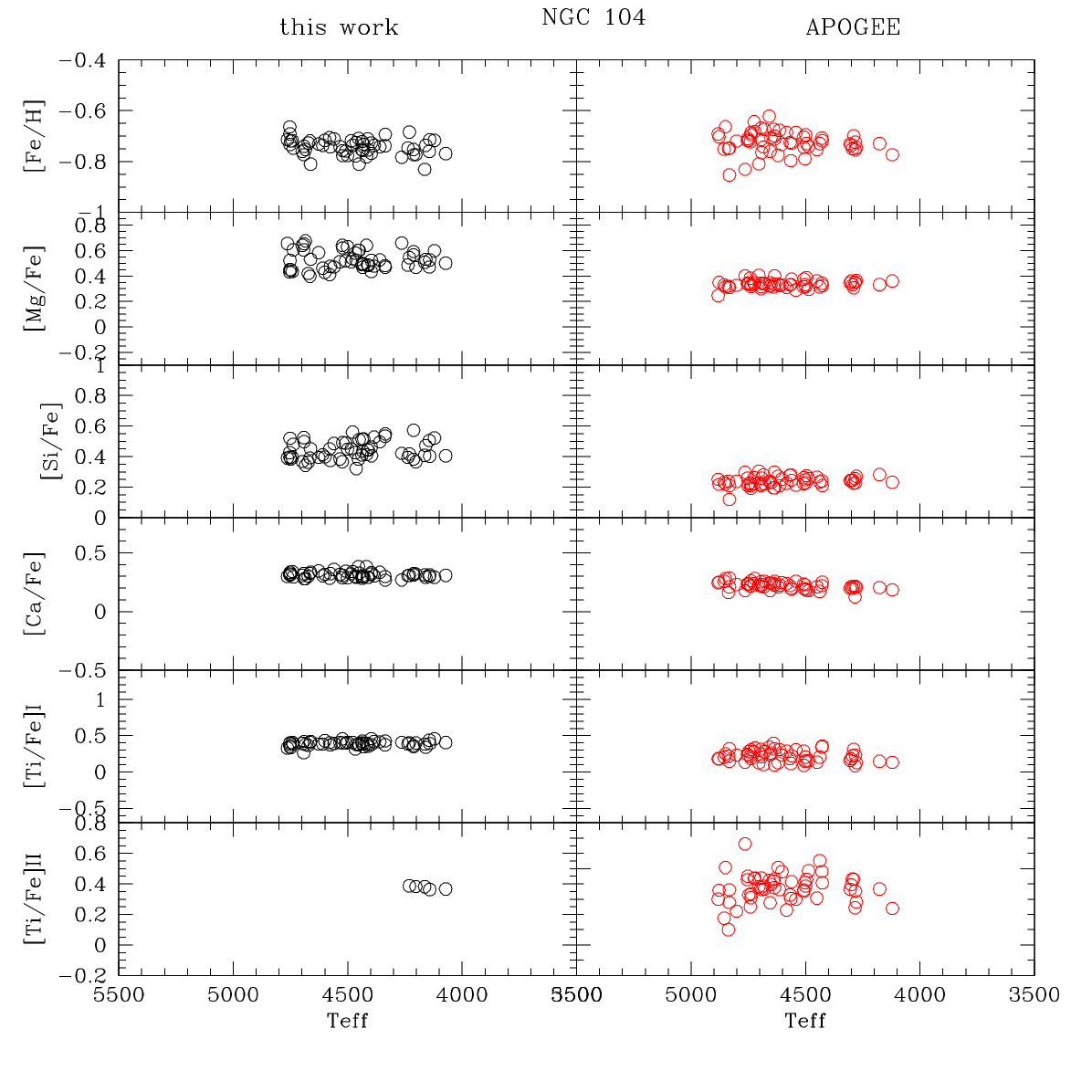}\includegraphics[scale=0.24]{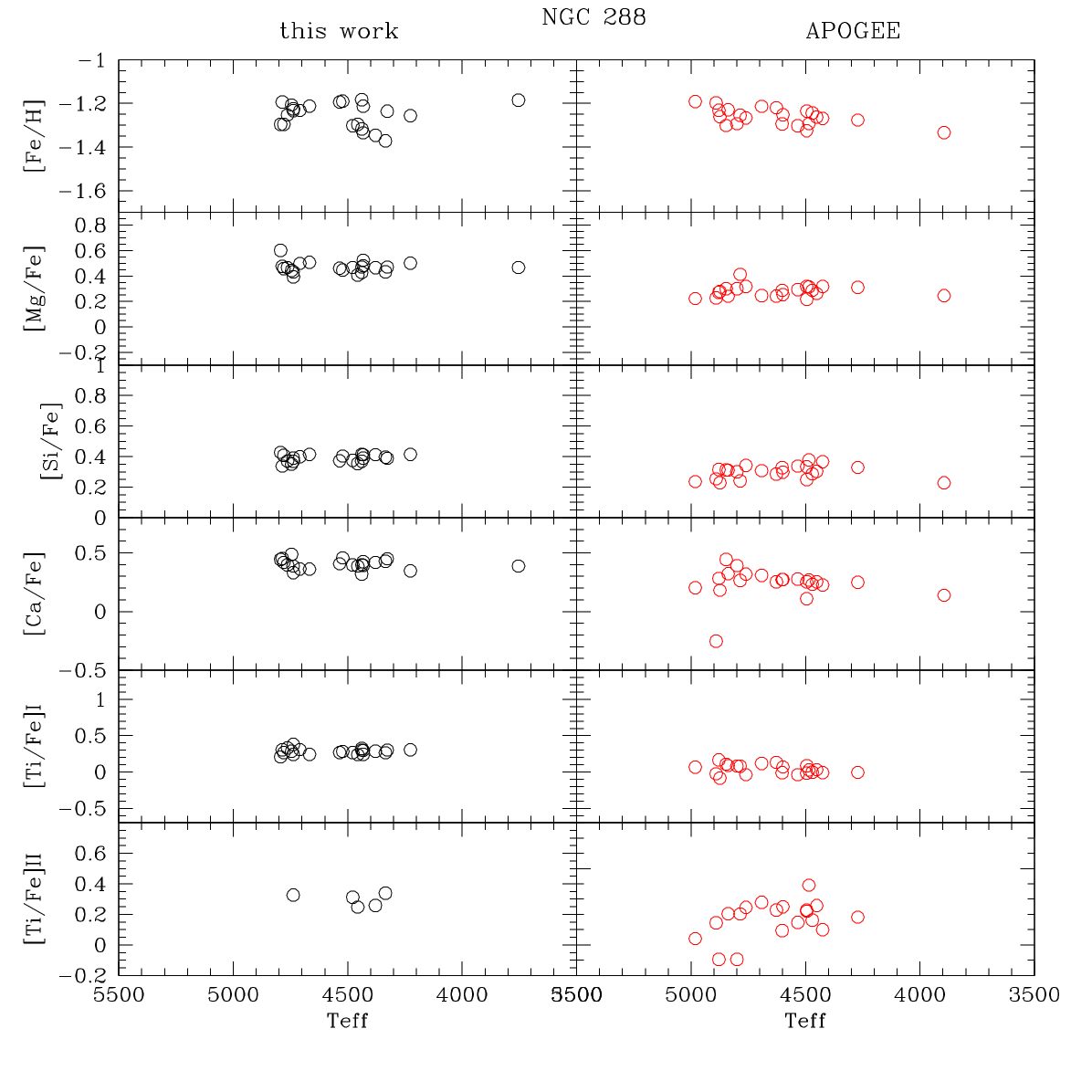}\includegraphics[scale=0.24]{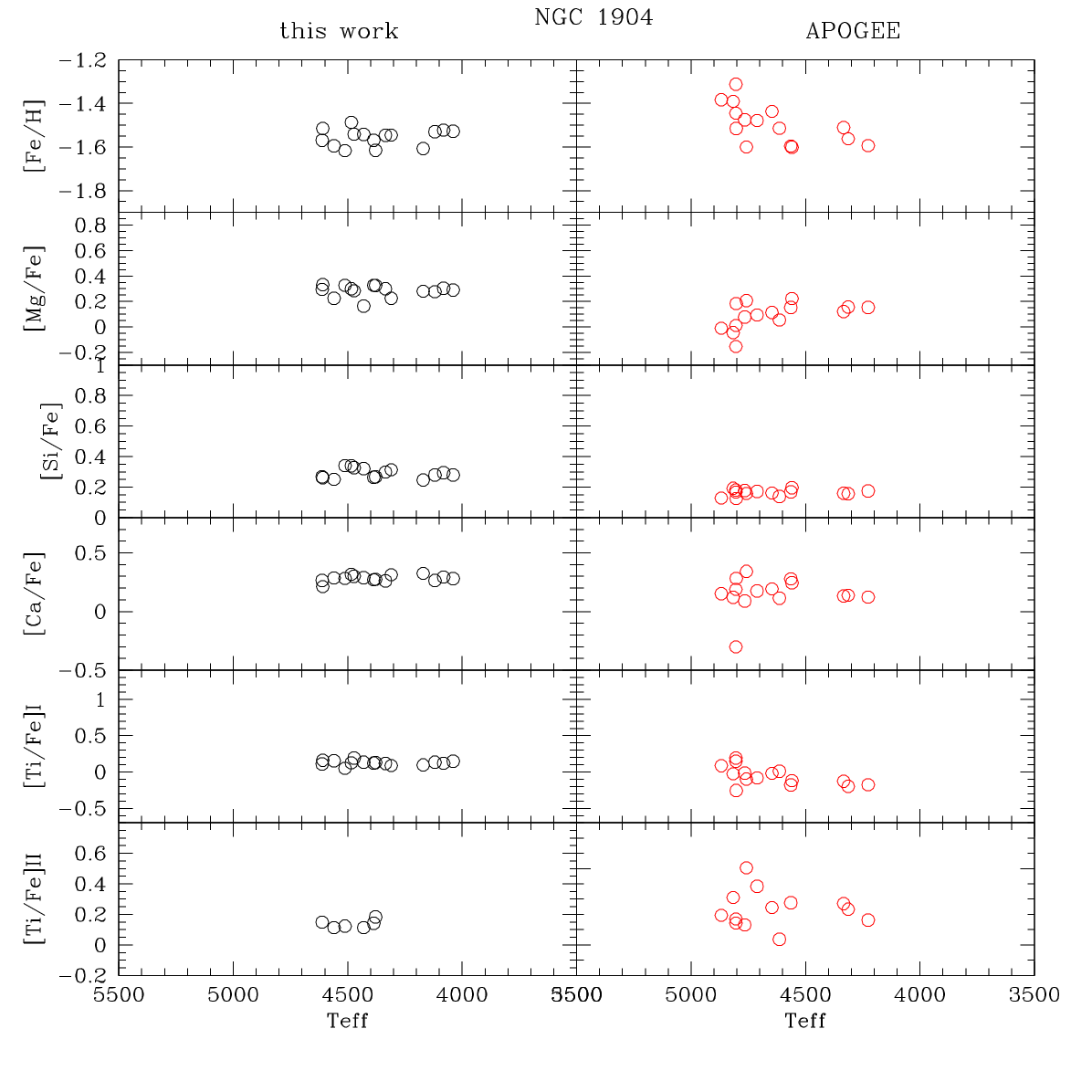}
\includegraphics[scale=0.24]{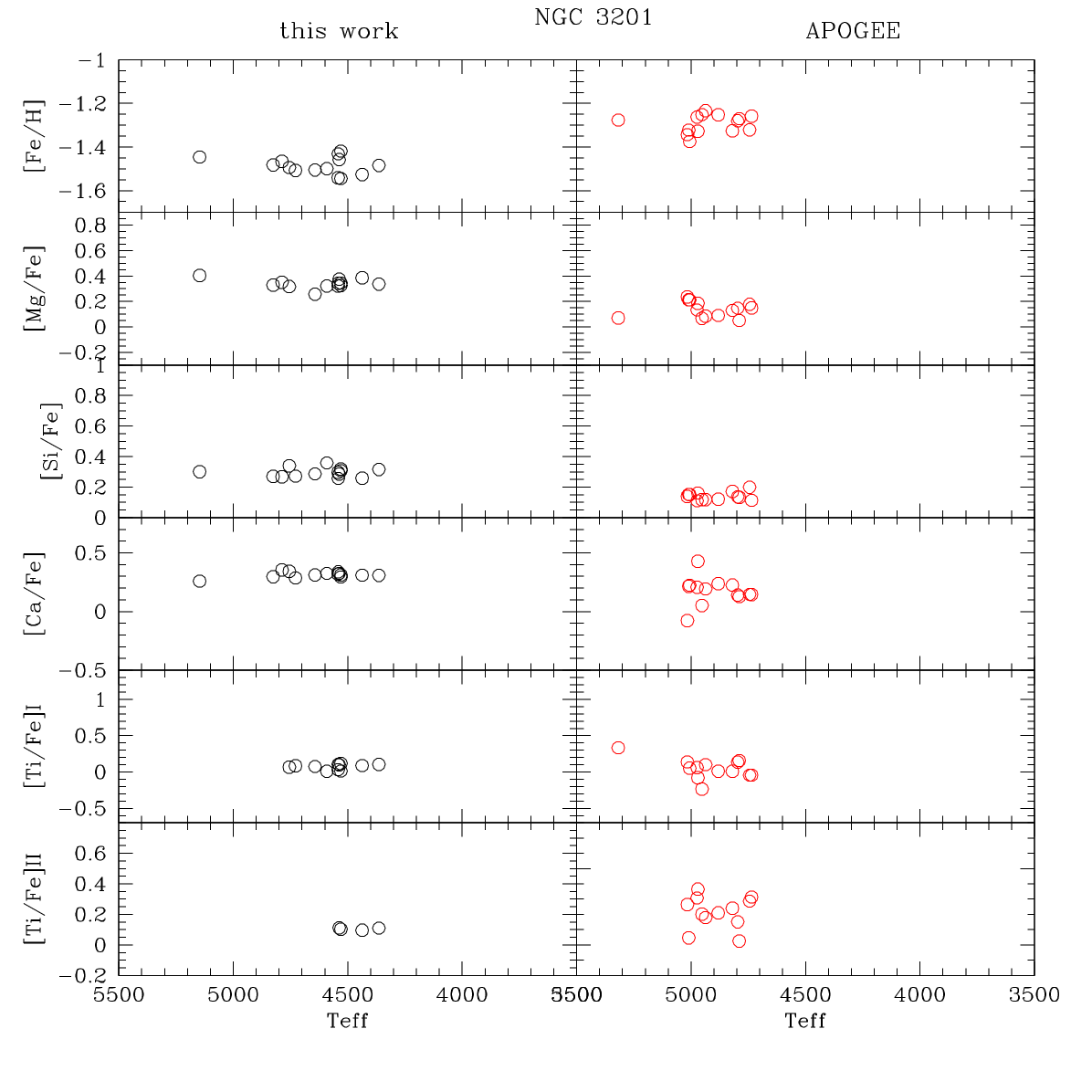}\includegraphics[scale=0.24]{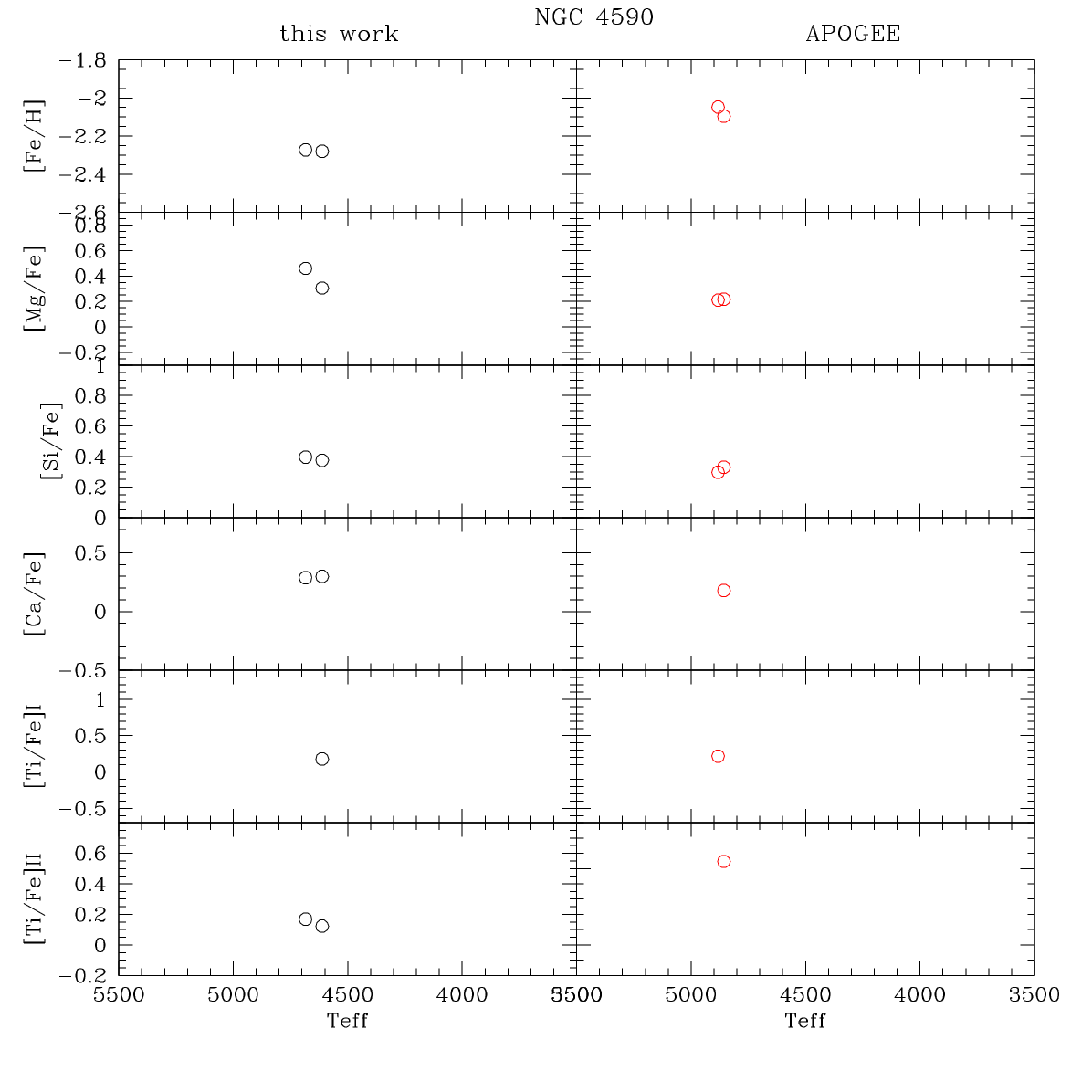}\includegraphics[scale=0.24]{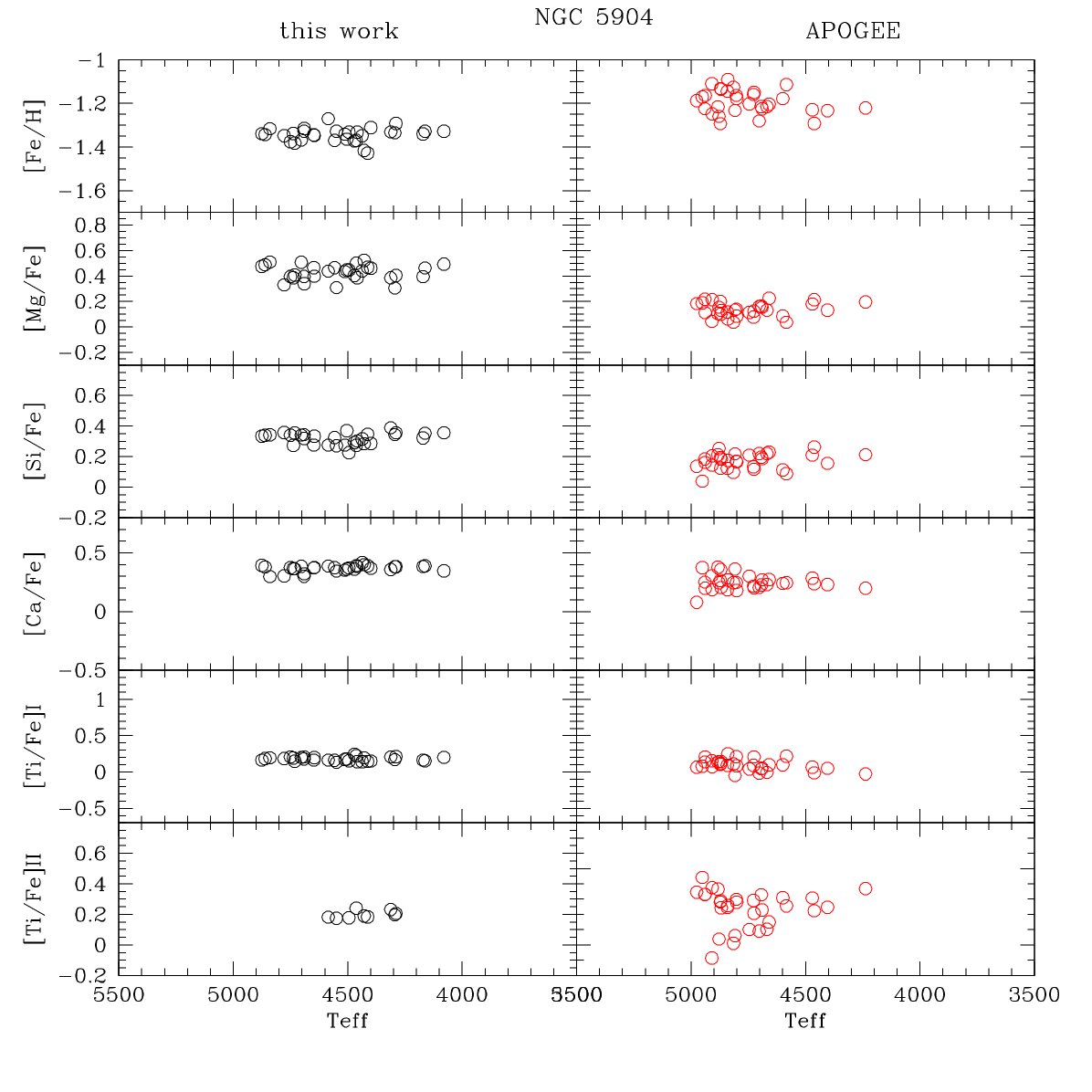}
\includegraphics[scale=0.24]{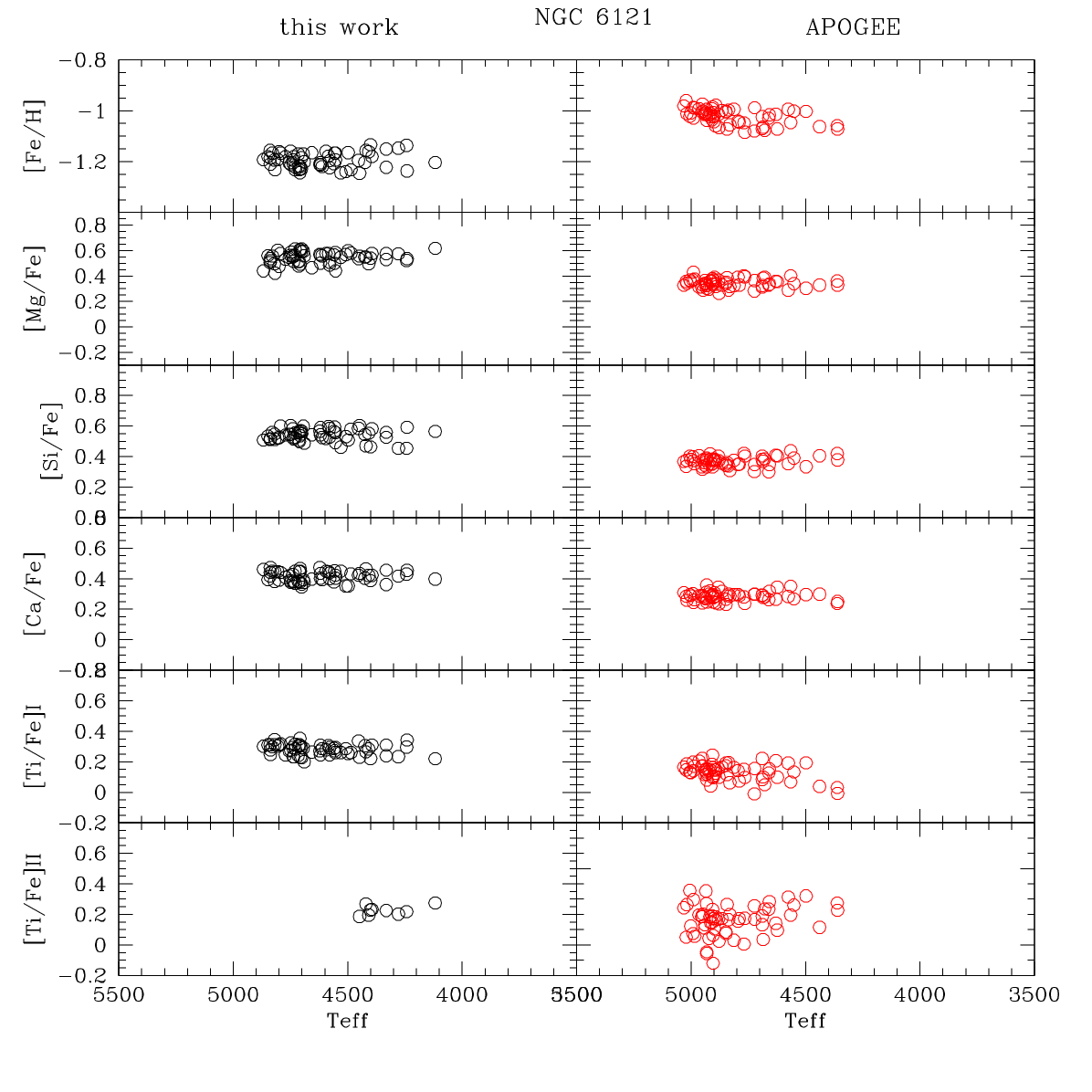}\includegraphics[scale=0.24]{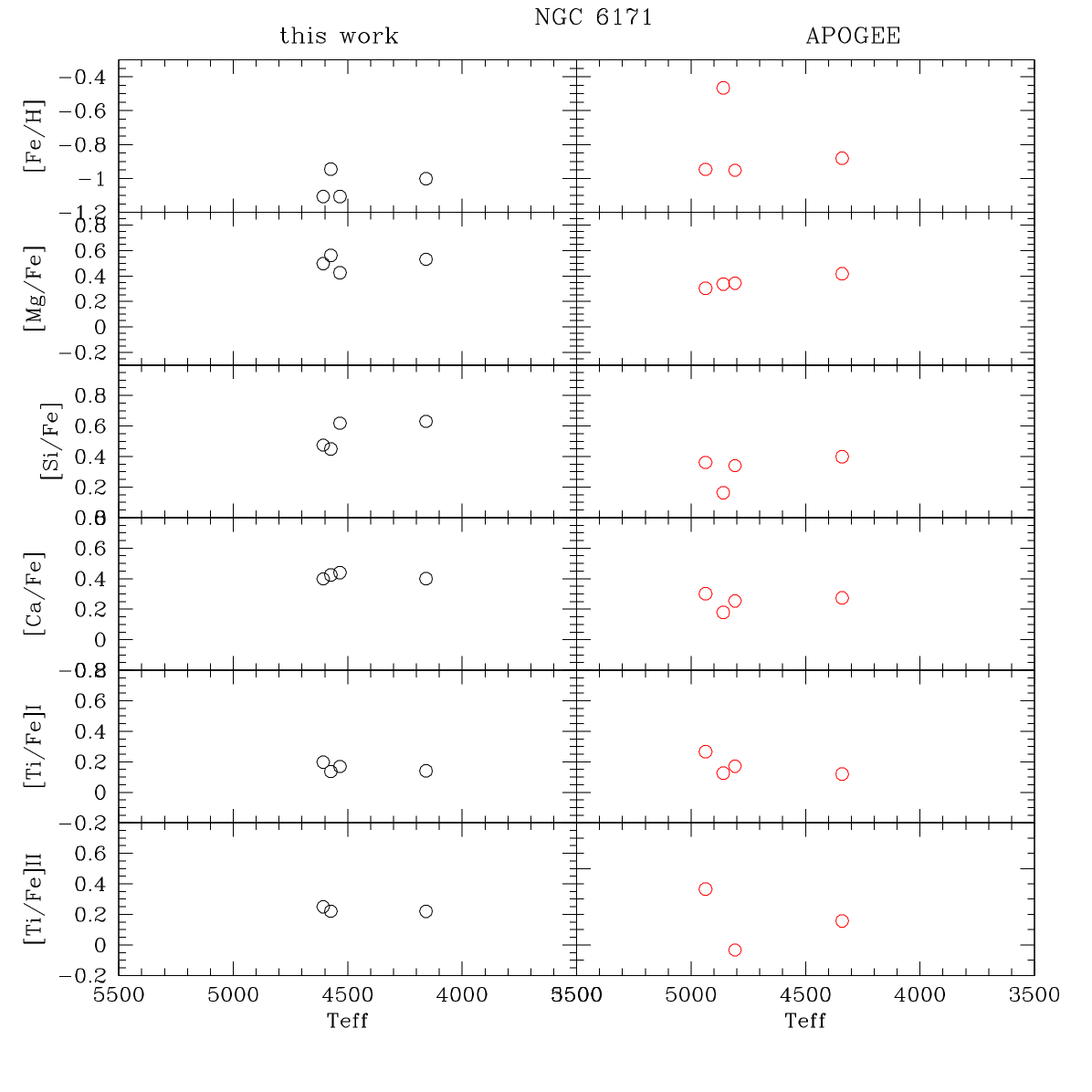}\includegraphics[scale=0.24]{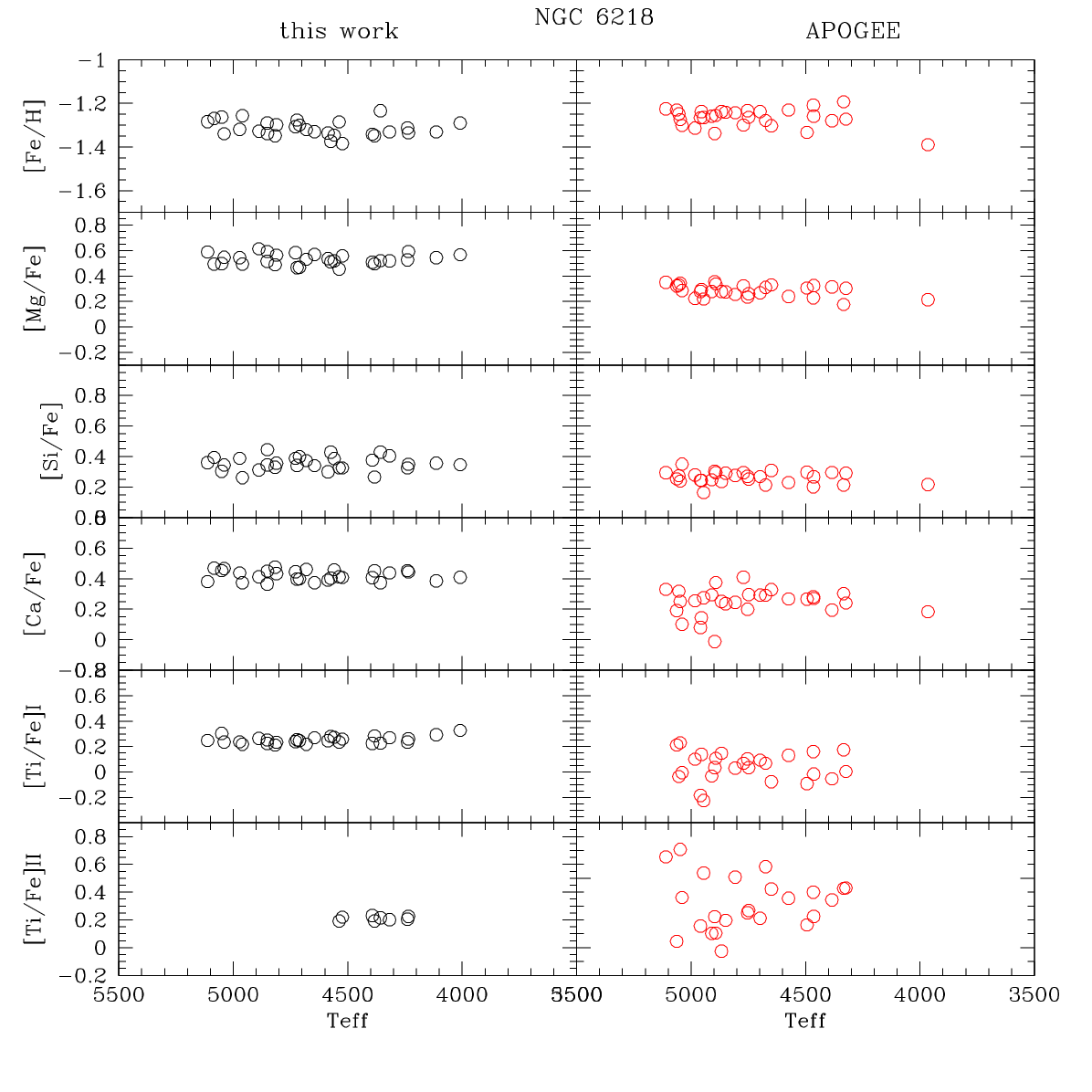}
\includegraphics[scale=0.24]{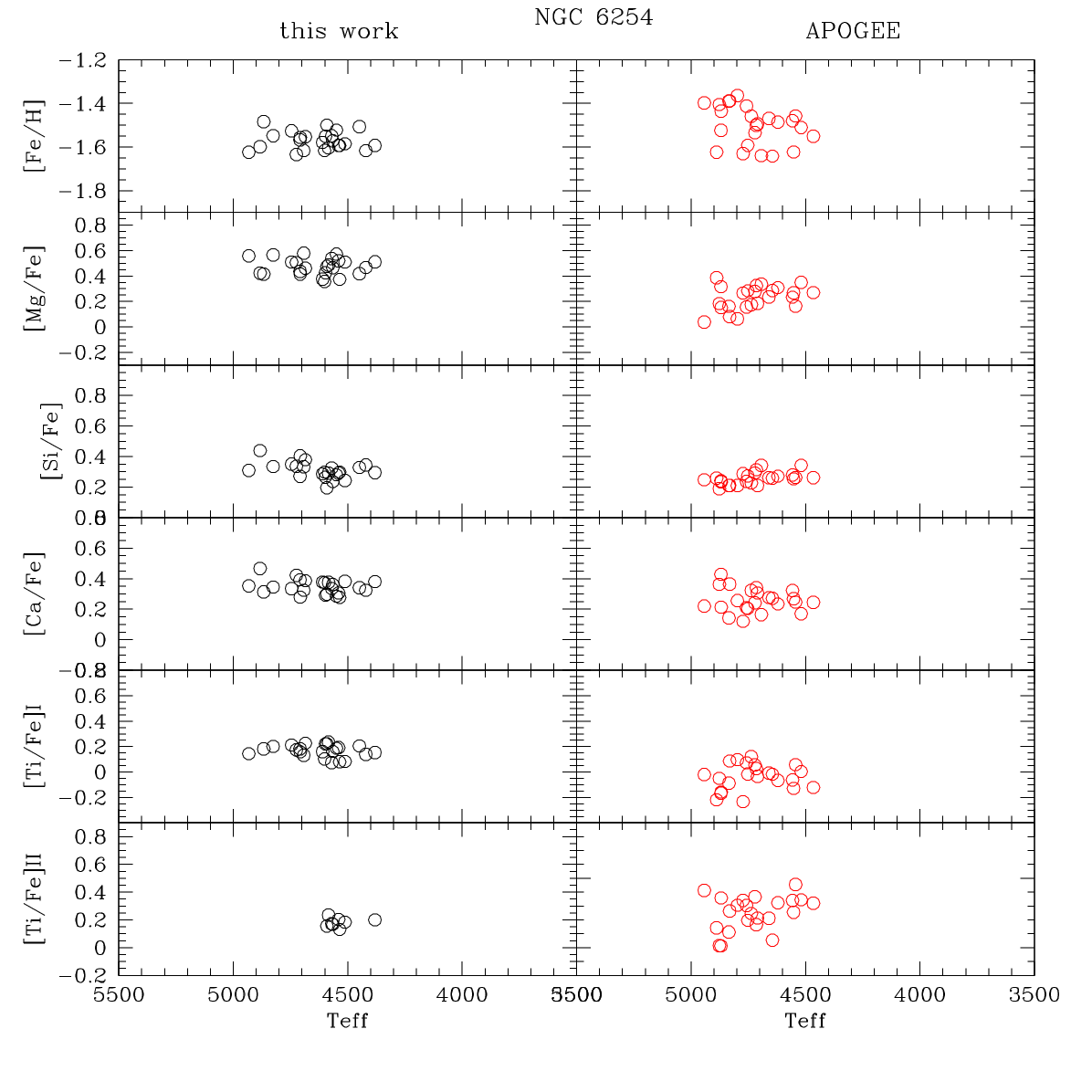}\includegraphics[scale=0.24]{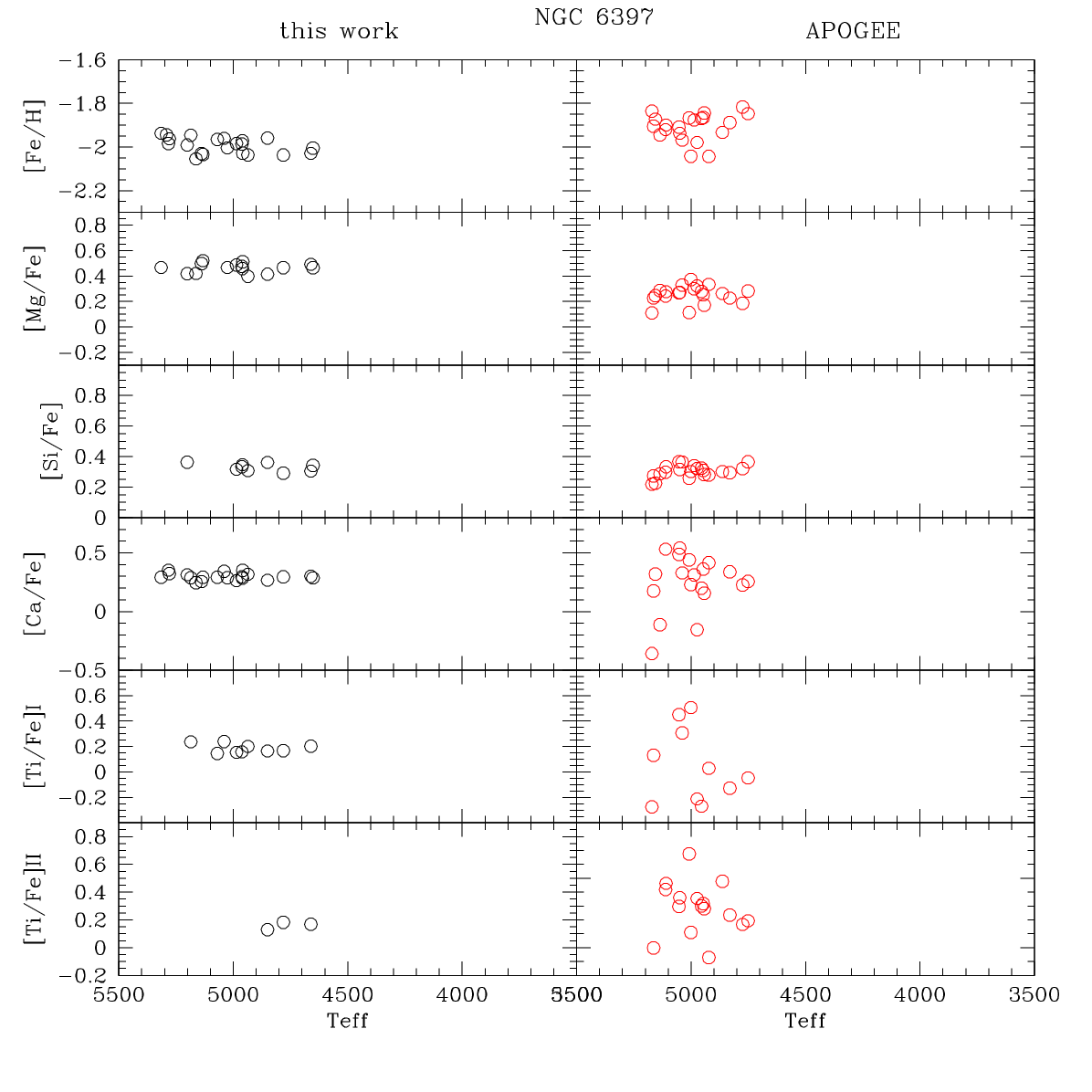}\includegraphics[scale=0.24]{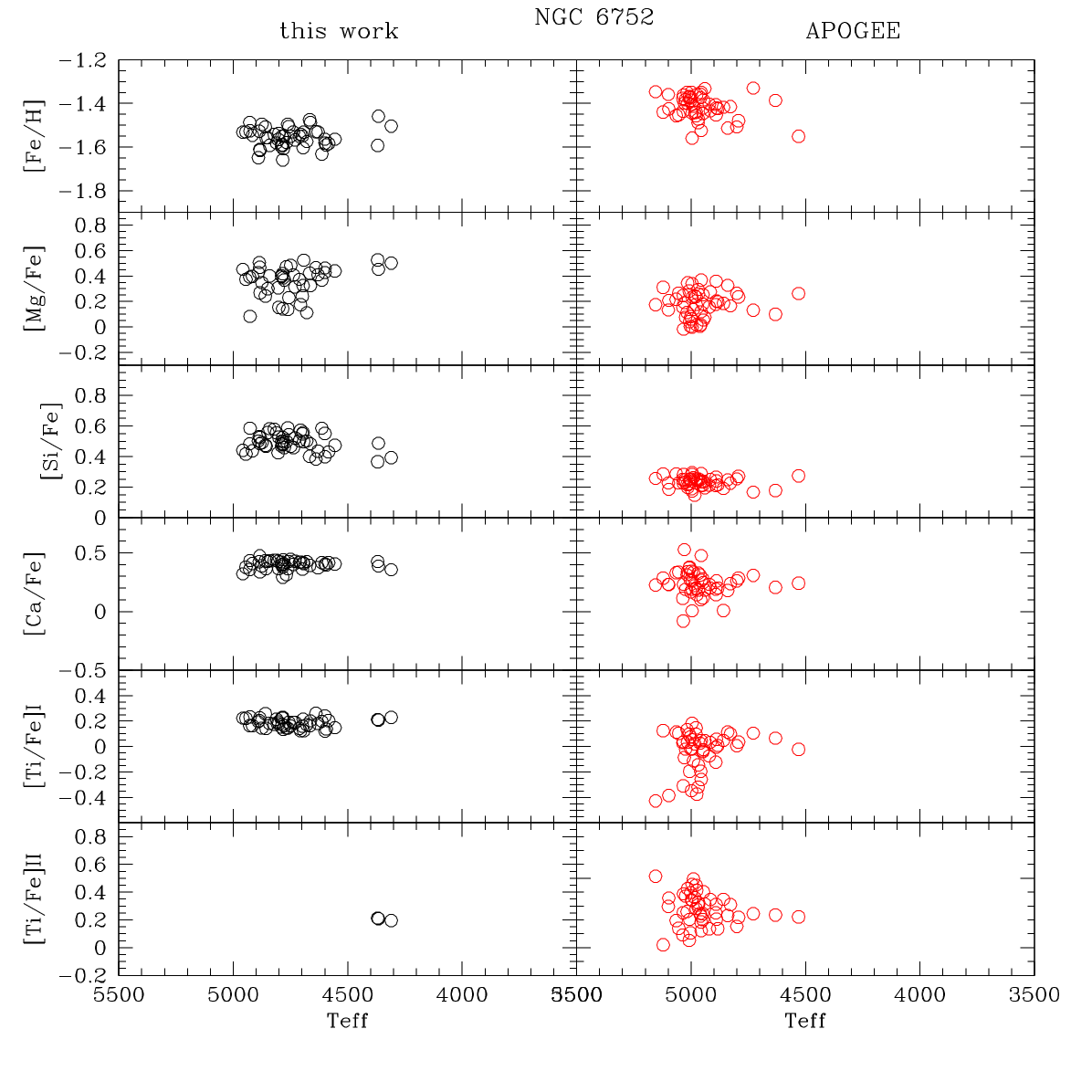}
\includegraphics[scale=0.24]{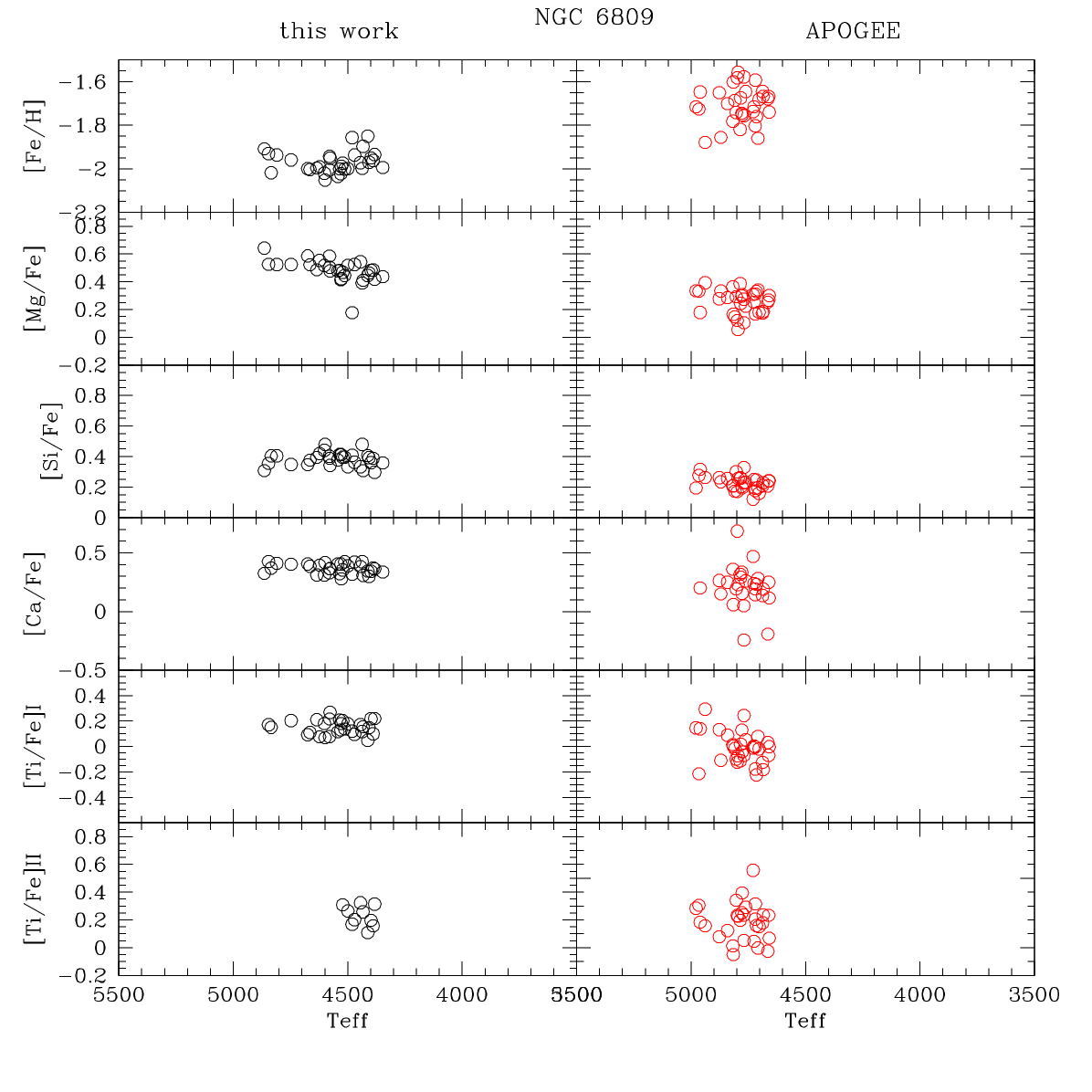}\includegraphics[scale=0.24]{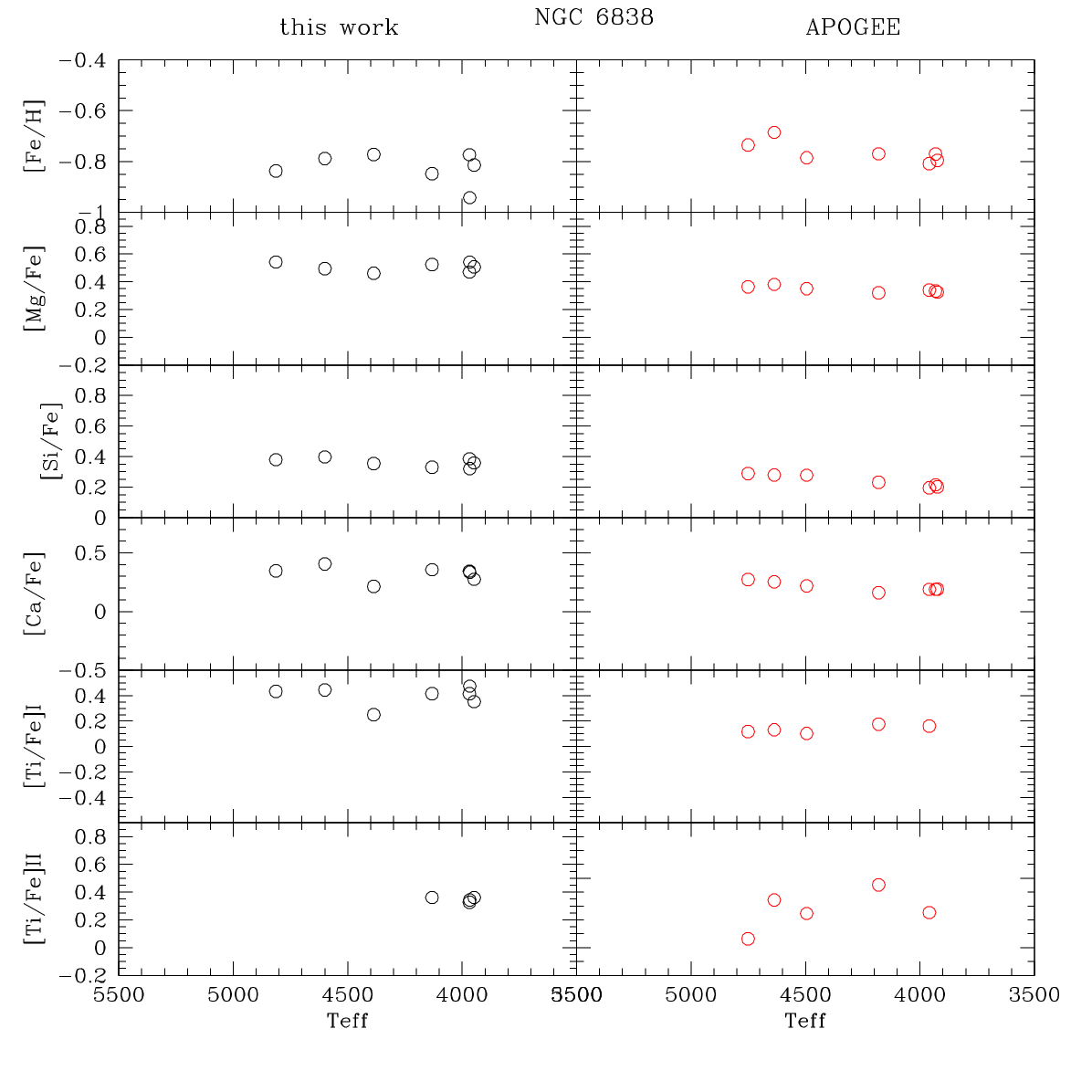}\includegraphics[scale=0.24]{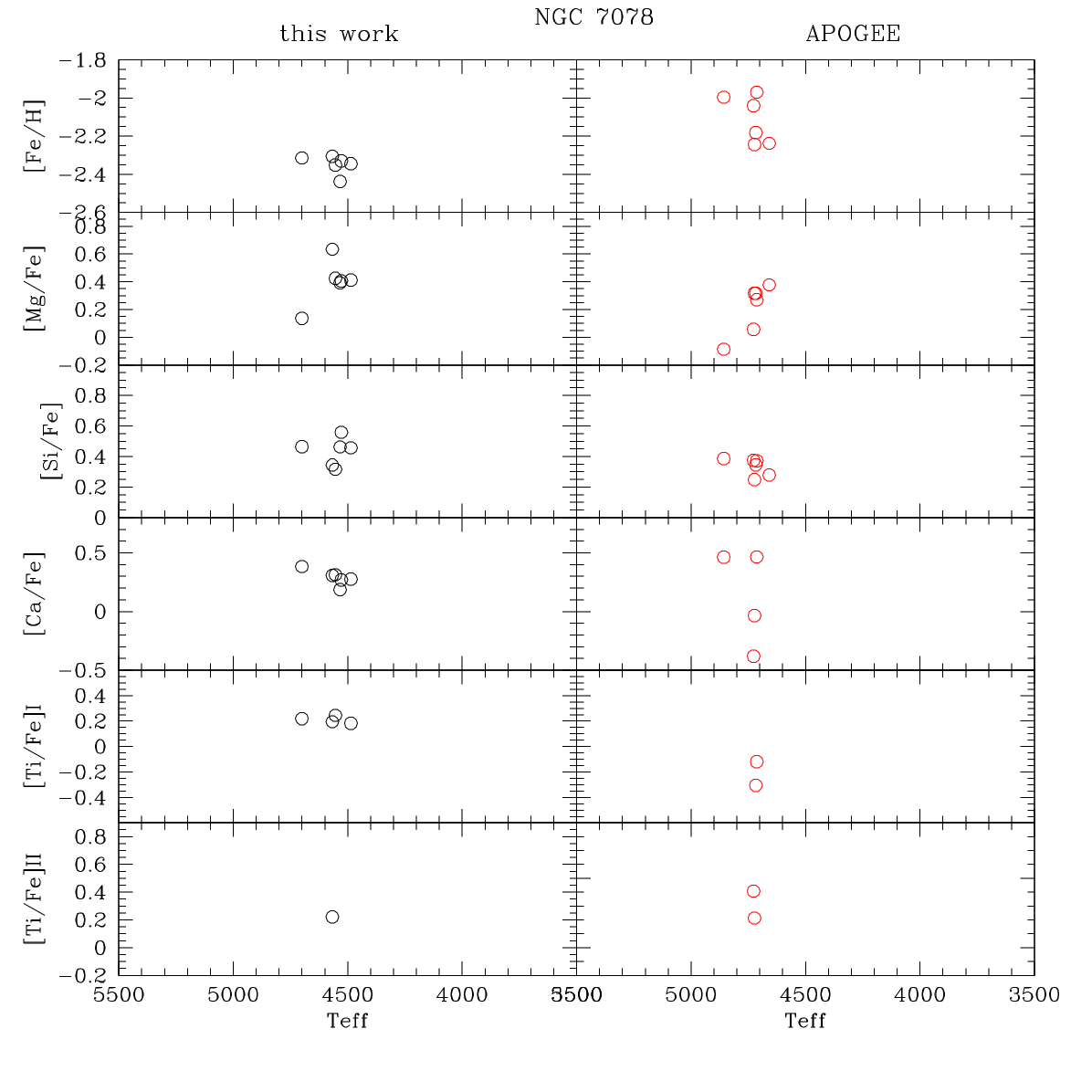}
\caption{Comparison of metallicity [Fe/H] and [Mg/Fe],[Si/Fe],
[Ca/Fe],[Ti/Fe]~{\sc i},and [Ti/Fe]~{\sc ii} ratios from this work (left panels)
and the APOGEE survey (right panels).}
\label{f:apo}
\end{figure*}

\end{appendix}

\end{document}